\documentclass[a4paper,12pt]{article}
\usepackage[utf8]{inputenc}
\usepackage{amsmath}
\usepackage[dvipsnames]{xcolor}
\usepackage[numbers]{natbib}
\usepackage[cm]{fullpage}
\usepackage{graphicx}
\usepackage{subcaption}
\setcitestyle{square}
\numberwithin{equation}{section}
\newcommand\numberthis{\addtocounter{equation}{1}\tag{\theequation}}
\usepackage{setspace}
\usepackage[top=2.75cm,bottom=3.00cm,left=2.5cm,right=2.5cm]{geometry}
\usepackage{overpic}
\usepackage{pgfplots} 
\pgfplotsset{compat=1.8}
\newlength\figureheight 
\newlength\figurewidth 
\usetikzlibrary{arrows.meta}
\providecommand\eps{\varepsilon}
\newcommand\pd[2]{\frac{\partial #1}{\partial #2}}
\newcommand\Pd[2]{\frac{\partial^2 #1}{\partial #2^2}}
\newcommand\f\frac{}{}

\newcommand\p\partial
\newcommand\colour\color

\newcommand\changes[1]{\colour{black} #1 \colour{black}}

\newcommand\FD[3]{
\ifcase #3 
specify deriv
\or 
\frac{#1_{i+1}^{#2} - #1_{i-1}^{#2}}{2\Delta x} 
\or 
\frac{#1_{i+1}^{#2} - 2 #1_{i}^{#2} + #1_{i-1}^{#2}}{\Delta x ^2}
\or
\frac{#1_{i+2}^{#2} - 2 #1_{i+1}^{#2} +2 #1_{i-1}^{#2} - #1_{i-2}^{#2}}{2 \Delta x^3}
\or
\frac{#1_{i+2}^{#2} -4 #1_{i+1}^{#2} +6#1_{i}^{#2} -4 #1_{i-1}^{#2} + #1_{i-2}^{#2}}{\Delta x^4}
\else specify deriv between $1-4$
\fi
}

\usepackage{verbatim}   % for the comment environment
\usepackage{lipsum}

\newif\ifshowtikz
\showtikztrue
\let\oldtikzpicture\tikzpicture
\let\oldendtikzpicture\endtikzpicture
\renewenvironment{tikzpicture}{%
    \ifshowtikz\expandafter\oldtikzpicture%
    \else\comment%   
    \fi
}{%
    \ifshowtikz\oldendtikzpicture%
    \else\endcomment%
    \fi
}
\usepgfplotslibrary{fillbetween}
\usetikzlibrary{intersections}
 \pgfplotsset{
        layers/my layer set/.define layer set={
            bg,
            main,
            ft
        }{
        },
        set layers=my layer set,
    }
    
\usepackage{authblk}
\usepackage{lscape}

\title{Thin film extensional flow of a transversely isotropic viscous fluid}
\author[1,2]{M. J. Hopwood}
\author[1,3]{B. Harding}
\author[1]{J.E.F. Green}
\author[2]{R.J. Dyson}
\affil[1]{\small{School of Mathematical Sciences, University of Adelaide, Adelaide, SA, 5005, Australia}}
\affil[2]{\small{School of Mathematics, University of Birmingham, Edgbaston, Birmingham, B15 2TT, UK}}
\affil[3]{\small{School of Mathematics and Statistics, Victoria University of Wellington, PO Box 600, Wellington, 6140, New Zealand}}
\date{\today}

\begin{document}

\maketitle

\begin{abstract}
 %   Motivated by the aim of understanding the mechanical behaviour of biological fluids which possess a fibrous micro-structure such as cervical mucus, and how this structure affects their function, we consider an existing model for the extensional flow of a thin-two dimensional film of incompressible, transversely isotropic viscous fluid as a first approximation of the `spinnbarkeit test' used as a proxy for fertility. 
     Many biological materials such as cervical mucus and collagen gel possess a fibrous micro-structure. \changes{This micro-structure affects the emergent mechanical properties of the material, and hence the functional behaviour of the system.} We consider the canonical problem of stretching a thin sheet of transversely-isotropic viscous fluid as a simplified version of the spinnbarkeit test for cervical mucus. 
     
     We \changes{propose a novel solution to the model constructed by Green \& Friedman} by \changes{manipulating} the model to a form amenable to \changes{arbitrary Lagrangian--Eulerian} (ALE) techniques. The system of equations, reduced by exploiting the slender nature of the sheet, are solved \changes{numerically} and we discover that the bulk properties of the sheet are controlled by an effective viscosity dependent on the \changes{evolving} angle of the fibres. In addition, we confirm \changes{a} previous conjecture by demonstrating that the centre-line of the sheet need not be flat, and perform a short timescale analysis to capture the full behaviour of the centre-line.
     %\indent This thesis aims to understand how the presence of fibres alters the mechanical behaviour of such materials by considering three canonical examples of thin film flows: the squeezing of a film, and the extensional flows of a sheet or a thread.  The effect of fibres is incorporated via a transversely isotropic fluid stress tensor which models the suspension as a continuum with an evolving single preferred direction, alongside conservation of mass and momentum.  Exploiting the small aspect ratio in each situation, we derive governing equations which we solve via analytical and numerical means. We find throughout that the behaviours of a transversely isotropic fluid are markedly different to that of a Newtonian fluid.
         \end{abstract}

 \section{Introduction}
 Fluids can be classified as isotropic, or anisotropic, depending upon whether the mechanical properties of the fluid are uniform in all directions, or not. Most common fluids such as water are isotropic. However, there are many examples of fluids that arise in biology and industry which contain fibres or elongated particles, for example, collagen gels \cite{green2008extensional}, cervical mucus \cite{cupples2017viscous}, and nematic liquid crystals \cite{cummings2004evolution}. The presence of fibres or particles within the fluid creates an underlying structure that causes the fluid to exhibit anisotropy. The anisotropy of biological fluids mean that they can display interesting behaviours and possess unusual and evolving mechanical properties, which influence  how they perform their particular functions. For example, the anisotropy of cervical mucus is suspected to play a role in fertility, by controlling how easily sperm can migrate through \changes{to the egg} \cite{chretien1978temporary}.
  \\
  
 \indent One of the first models of an anisotropic viscous fluid was formulated by Ericksen \cite{Ericksen1960}. He considered a type of anisotropy known as `transverse isotropy' where the material possesses a single preferred direction which may vary both spatially and temporally; its physical properties are symmetric in all directions normal to this preferred direction. Whilst in general, for a fibrous material, one would need to consider the evolution (in time and space) of a probability distribution \changes{describing the alignment of a fibre} along a particular direction, the Ericksen model simplified the problem by assuming that there is a unique fibre alignment at each point in space. Examples of materials that have been modelled using this approach include fibre--reinforced composites \cite{spencer1997fibre}, entanglements of textile fibres used in the carding process \cite{lee2005continuum} and a number of biological materials -- collagen gels \cite{green2008extensional}, the extra--cellular matrix \cite{dyson2016investigation}, and primary plant cell walls \cite{dyson2010fibre}. A significant number of studies motivated by problems in composites manufacturing include the additional assumption that the fluid is inextensible in the fibre direction \cite{hull1991theory,rogers1989squeezing,spencer1997fibre}; these are termed `ideal' incompressible fibre--reinforced fluids. 
    \\
    
\indent In this paper, we consider the extensional flow of a thin sheet of transversely isotropic viscous fluid. One motivation for studying this problem is that it provides a simplified representation of the `spinnbarkeit' or `spinnability' test, which is applied to cervical mucus as a means of assessing fertility \cite{evans2013cervical}. The test entails taking a sample of mucus and stretching it. Around ovulation, the mucus has a lower pH, a higher concentration of water (which \changes{has} the effect of lowering the viscosity of the mucus), and the fibrous reinforcement takes a more parallel alignment that allows sperm to migrate. At this point the mucus can be stretched the furthest, i.e. the fluid exhibits maximum spinnbarkeit. Conversely, during the most infertile parts of the menstrual cycle, the mucus does not stretch and simply breaks \cite{moghissi1972function,wolf1978human}. Understanding the dynamics of stretching a transversely isotropic sheet can provide some basic insights into how factors such as fibre concentration and alignment influence spinnbarkeit. A second motivation for studying this problem is that is provides a simple prototype situation for investigating how the feedback between the macroscopic fluid flow and the microstructure influences the mechanical behaviour of anisotropic fluids. In particular, as shown in \cite{green2008extensional}, the model can be reduced to one--dimension, with the changing orientation of the fibres specified by a single angle.
\\

\indent Extensional thin film flows of incompressible Newtonian fluids, which arise in a number of industrial and biological applications, have been extensively studied. The problem of a two dimensional thin film was considered by Howell \cite{howell1996models}, who employed an asymptotic expansion of the Navier--Stokes and mass conservation equations in powers of an inverse aspect ratio, to obtain a reduced model (termed the Trouton model) which involves only the leading order longitudinal fluid velocity and sheet thickness. Additionally, the roles of inertia and surface tension were considered, as well as the complementary problem of a fluid thread (i.e. slender cylinder). In order to capture the full behaviours of the centre--line of the fluid of the sheet, a short timescale analysis is carried out \cite{buckmaster1975buckling,howell1994extensional,howell1996models}. These models are not valid for sheets undergoing deformation by bending, and led to the development of a general theory for thin viscous sheets that undergo deformation by stretching, bending, or an arbitrary combination of both, by Ribe \cite{ribe2001bending,ribe2002general}. This work was able to explicitly quantify relations between bending and stretching of the sheet. A complimentary study considered sheets with an inhomogenous viscosity, \cite{pfingstag2011thin}. One of their main results was that `necking' of the sheet, where regions of the sheet thinned faster than others, could be induced by in--plane variations of viscosity throughout the fluid. However, for a transversely isotropic fluid, we are only aware of \changes{three} studies of extensional flow: \changes{ those of \cite{chakraborty2021lockhart,dyson2010fibre},} wherein both works modelled the primary plant cell wall as a thin axisymmetric fibre--reinforced viscous sheet supported between rigid end plates, and that of \cite{green2008extensional}, who used a similar approach to that of Howell \cite{howell1994extensional} to derive a transversely isotropic version of the Trouton model.  The model presented in \cite{green2008extensional} is referred to throughout as the Green \& Friedman model and is of particular relevance to this paper. They presented some analytical results for cases in which the equations simplify (e.g. when some of the anisotropic stress terms are negligible) but did not tackle the general case -- a problem we pursue in this paper.
\\

\indent A number of studies have investigated transversely isotropic fluid flows in other geometries to gain insight into how their behaviour is affected by the microstructure. These include the \changes{studies} of the stability of a transversely isotropic fluid in a Taylor--Couette device with the aim of understanding the behaviour of suspensions of bio-molecules, as well as treating the problem of Rayleigh--Bernard convection \cite{holloway2015linear,holloway2018linear}. It was found in both works that transversely isotropic effects delay the onset of instabilities, primarily through the incorporation of an anisotropic shear viscosity. In the context of modifying transversely isotropic fluid models to incorporate active swimming suspensions, \changes{as in \cite{cupples2017viscous,holloway2018influences},}transversely isotropic effects are also capable of increasing the size of a developing instability, in particular where translation diffusion is neglected. Other studies have focused on prototypical flows, to give more generic insights into fluid-fibre interactions. For example, Phan--Thien and Graham studied both the flow of a transversely isotropic fluid around \changes{a} sphere \cite{phan1991new}, and the squeezing flow of a layer of fluid compressed between two fixed plates \cite{phan1990squeezing} (this latter problem was studied independently for the case of an ideal fibre reinforced fluid \cite{rogers1989squeezing}). 
\\

\indent This paper uses a combination of analysis and numerical simulations to \changes{significantly} extend the previous work of \cite{green2008extensional} to include cases where all of the anisotropic terms in the fluid stress are non-negligible, and the fibre alignment within the sheet may vary with depth. One issue of particular interest was to verify their conjecture that, \changes{in contrast to the Newtonian case,} the centre--line of a transversely isotropic fluid sheet need not always be straight. As preliminary simulation of the model indeed produced results with non-flat centre--lines, we sought to confirm this prediction by additionally investigating the short timescale behaviour of the sheet, similar to the work by \cite{buckmaster1975buckling,howell1996models} on the Newtonian problem. 
\\
\indent The paper is organised as follows. In Section \ref{Section:MathModel} we briefly recap the thin film approach and governing equations as given in \cite{green2008extensional}, as well as introducing a short timescale. \changes{In order to present the first solutions to the full Green \& Friedman model, the equations are manipulated in order become amenable to numerical strategies in section \ref{Section:SectionALE}. We then present the arbitrary Lagrangian--Eulerian techniques we use to solve the model in Section \ref{Section:GreenFriedman},} \changes{and validate the} numerical techniques by comparison with analytical results for short time, and further present new results primarily for \changes{a} passive transversely isotropic fluid. In Section \ref{Section:TIBNTResults} we present results for the behaviour of the fluid on a short timescale. We conclude with a discussion and \changes{suggestions} for future work in Section \ref{Section:Discussion}.

\section{Mathematical model} \label{Section:MathModel}
\begin{figure}[t]
\centering
       \resizebox{0.8\textwidth}{!}{\input{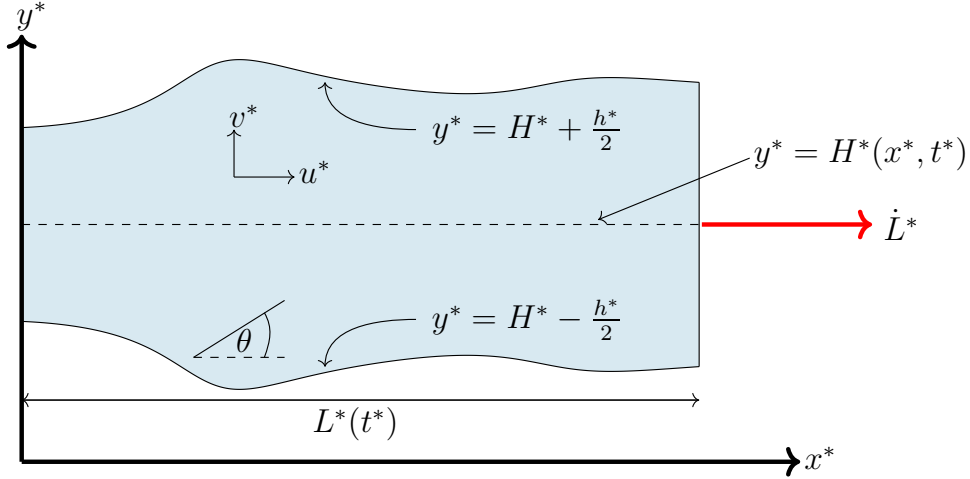}}
        \caption{Schematic of extensional flow of \changes{a} sheet. The fluid is fixed to two plates at $x^*=0,L$, by a no-slip boundary condition. The plate at $x^*=L^*$ is moved at a prescribed speed $\dot{L}^*$. The centre--line of the fluid is given by $H^*(x^*,t^*)$ and the thickness of the film by $h^*(x^*,t^*),$ so that the free boundaries of the film are located at $y^*= H^* \pm \frac{h^*}{2}$. \changes{Note that this figure is not to scale, as we model the thin film behaviour of the sheet.}}
    \label{Schematic}
\end{figure}

We consider the extensional flow of an inertialess thin sheet of incompressible, transversely isotropic, viscous fluid. As shown in Figure \ref{Schematic}, we use the 2D Cartesian coordinates $\left(x^*, y^*\right)$ \changes{to describe the horizontal and vertical axis respectively,} with $t^*$ denoting time (throughout this paper asterisks denote dimensional quantities). The upper and lower boundaries of the fluid sheet are denoted by $y^* = H^{\pm^{*}} = H^{*} \pm \frac{h^{*}}{2}$, where  $H^{*}(x^{*},t^{*})$ is the position of the centre--line and $h^{*}(x^{*},t^{*})$ is the thickness of the fluid sheet. The left-- and right-- hand side boundaries are given by $ x^{*}=0,L^{*}(t^{*})$. The right-hand of the sheet, at $x^{*} = L^{*}$, is pulled in the $x^*$ direction; we prescribe either the speed of pulling, $\dot{L}^*$, or the tension $T^*$ applied to the sheet.
\\

\indent We let $\boldsymbol{u}^* = \left(u^*,v^*\right)$ be the fluid velocities in the $x^*,y^*$ directions respectively, and denoted the stress tensor by $\boldsymbol\sigma^*$. The equations of conservation of fluid mass and momentum are thus
\begin{align}
    \nabla^* \cdot \boldsymbol{u}^* = 0, \label{DimensionFullContinuity}
 \\
    \nabla^* \cdot \pmb\sigma^* = 0. \label{DimensionFullMomentum}
\end{align}
The constitutive law for $\boldsymbol{\sigma}^*$ is
\begin{align}
    \sigma^{*}_{ij}= -p^{*}\delta_{ij} + 2\mu^{*} e^{*}_{ij} + \mu^{*}_{1} a_{i}a_{j} + \mu^{*}_{2}a_{i}a_{j}a_{k}a_{l}e^{*}_{kl} + 2 \mu_{3}^{*}(a_{i}a_{l}e^{*}_{jl} + a_{j}a_{m}e^{*}_{mi}), \label{EricksenConstitutive}
\end{align}
where $p^*$ is the pressure, $\boldsymbol{a}$ is a unit vector describing the orientation of the fibres within the fluid and $\boldsymbol{e}^*$ is the rate of strain tensor. This relationship was derived by Ericksen \citep{Ericksen1960}, as the most general stress tensor that is linear in the rate of strain tensor $\boldsymbol{e}^*$, invariant under the transformation $-\boldsymbol{a} \rightarrow \boldsymbol{a}$, and satisfies $\pmb\sigma^* = \pmb\sigma^{*T}$.
% that is the stress tensor is invariant with respect to a rotation of the fibres through $\pi$ radians (the transformation $\boldsymbol{a} \rightarrow -\boldsymbol{a}$),
The constants $\mu^{*},\mu^{*}_2,\mu^{*}_3$, are all viscosity-like parameters and $\mu^{*}_1$ is the \changes{active} tension in the fibre direction \cite{chakraborty2021lockhart,green2008extensional, spencer1972deformations}. \changes{}

We note first that by setting $\mu^{*}_1=\mu^{*}_2=\mu^{*}_3=0$, one immediately recovers the stress tensor for an incompressible, isotropic Newtonian fluid, with $\mu^*$ as the familiar dynamic (shear) viscosity.
The physical interpretations of $\mu_2^*$ and $\mu_3^*$ can be identified by considering three deformations of a $2D$ sheet of fibres in a Cartesian plane, as illustrated in  \cite{dyson2010fibre,green2008extensional,holloway2015linear}. In the same way as previous work, we interpret $\mu_2^*$ as the anisotropic extensional viscosity, and $\mu_3^*$ the anisotropic shear viscosity. Additionally, we observe that there is no velocity component to the $\mu_1^*$ term, indicating that there exists stress in the fluid even when at rest \changes{and that this stress must be a tensile stress as no stress is induced in the fibres when the fluid is compressed.}
\\

\indent In addition to the constitutive law for $\pmb\sigma^{*}$, we require an equation  governing the evolution of the fibre direction. We use the form given by Green \& Friedman (for derivation, see \citep{green2008extensional}), which results from allowing the fibres to advect with the fluid flow 
\begin{align}
  \pd{\boldsymbol{a}}{t^*} + (\boldsymbol{u}^* \cdot \nabla^*) \boldsymbol{a}+\zeta^*\boldsymbol{a} = (\boldsymbol{a} \cdot \nabla^*) \boldsymbol{u}^*, \label{KinematicConditionOnA}
\end{align}
where
    \begin{align}
\zeta^*(x^*,y^*,t^*) = \boldsymbol{a} \cdot (\boldsymbol{a} \cdot \nabla^* \boldsymbol{u}^*), \label{ZetaStar}
\end{align}
is the fractional rate of extension of the fibres. The first two terms are the convective derivative of the fibre orientation, the third is the fractional rate of extension of the fibre in the direction of the fibres, all of which balances with effect of the fibres upon the flow velocity field on the right hand side. Equation \eqref{KinematicConditionOnA} corresponds to the specific case of Ericksen's equation in the long fibre limit \cite{cupples2017viscous,Ericksen1960,green2008extensional}. Since $\boldsymbol{a}$ is a unit vector and our model is two dimensional we write $\boldsymbol{a} = \left(\cos\theta,\sin\theta\right)$, where $\theta\left(x^{*},y^{*},t^{*}\right)$ is the angle the fibre direction makes with the $x^*$-axis.
\\

In order to close \changes{the} model, we must impose suitable boundary and initial conditions. At the ends of the sheet, we set 
\begin{align*}
    &u^*\left(0,y^*,t^*\right) = 0, &u^*\left(L^*,y^*,t^*\right) = \dot{L}^*,
    \\
    &H^*\left(0,t^*\right) =  0,  &H^*\left(L^*,t^*\right) = 0.
\end{align*}
On the upper and lower free surfaces, we apply a no-stress boundary condition 
\begin{align*}
   \boldsymbol{\sigma^*} \cdot \hat{\boldsymbol{n}} = 0;\text{ on } y^* = H^{*} \pm \frac{1}{2}h^*,
\end{align*}
together with the usual kinematic condition 
\begin{align}
    v^{*} = \frac{\p H^{*}}{ \p t^*} \pm \frac{1}{2}\frac{\p h^{*}}{\p t^*} + u^{*}\left(  \frac{\p H^{*}}{ \p x^*} \pm \frac{1}{2}\frac{\p h^{*}}{\p x^*}\right);\text{ on } y = H^{*} \pm \frac{1}{2}h^*. \label{KinematicConditionOnV}
\end{align}
Initial conditions must also be prescribed, however, since the number of initial conditions required depends upon the timescale considered, this will be discussed later.

\subsection{Thin film approximation}% for $t^* \sim L_0/U$}
We now introduce the assumption that the sheet is thin, which allows considerable simplification of the governing equations. Full details of the derivation can be found in \cite{green2008extensional}, but for the sake of completeness, we recapitulate the main points here. We let $L_0$ and $h_0$ be the initial length and typical initial thickness of the fluid sheet, respectively, and let $U$ be a typical value for the velocity \changes{of} the fluid at the pulled boundary. We then introduce the parameter $\varepsilon = h_0/L_0 \ll 1$, which is the initial inverse aspect ratio of the sheet. We are interested in the behaviour of the sheet as it undergoes significant changes in length, and so consider the timescale $t^* \sim L_0/U$. Following \cite{green2008extensional,howell1994extensional} we nondimensionalise as follows:
\begin{align*}
    \left(x^*,y^*\right) = \left(xL_{0},\varepsilon yL_{0}\right), \quad \left(u^*,v^*\right) &= \left(uU,\varepsilon v U\right), \quad
   p^{*} = \dfrac{\mu^* U}{L_{0}}p, \quad t^*=\frac{L_0}{U}t,
   \\
\left(H,L,h\right)=& \left(\varepsilon L_0 H^{*},  L_0 L^{*} , \varepsilon L_0 h^*  \right).
\end{align*}
These scalings introduce the dimensionless \changes{material parameters}
\begin{align*}
    \mu_1 = \frac{\mu_1^{*} L}{\mu^{*} U},\quad \mu_2 = \frac{\mu_2^{*}}{\mu^{*}}, \quad\mu_3 = \frac{\mu_3^{*}}{\mu^{*}}.
\end{align*}
At this point, we exploit the thin geometry of the sheet by expanding all of the dependent variables as power series in terms of the inverse aspect ratio $\varepsilon$, that is
\begin{align*}
    u = u^{(0)} + \varepsilon u^{(1)} +...
\end{align*}
with similar expressions for the other dependent variables (here, unlike in Howell \cite{howell1994extensional}, we encounter terms involving odd powers of $\varepsilon$). After some lengthy algebra, full details of which are given in \citep{green2008extensional}, we obtain a system of one dimensional equations for the quantities $h^{(0)},H^{(0)},u^{(0)},u^{(1)},v^{(0)}, \theta^{(0)}$. As a consequence of the analysis, it is found that the leading order longitudinal velocity satisfies $u^{(0)} = u^{(0)}(x,t)$ only (i.e. the flow is extensional). Conservation of mass yields:
\begin{align}
    \pd{h^{(0)}}{t} +& \pd{}{x}\left(h^{(0)}u^{(0)}\right) = 0. \label{Cons Mass}
\end{align}
Taking a depth-averaged force balance over the sheet leads to the following equation for \changes{$u^{(0)}$}
\begin{align}
    \pd{}{x} \int\limits_{H^{(0)^{-}}}^{H^{(0)^{+}}}   4\left(1+\mu_3\right)\pd{u^{(0)}}{x} +& \mu_1\cos2\theta^{(0)} + \mu_2\left(\cos^2 2\theta^{(0)} \pd{u^{(0)}}{x} + \f{1}{4}\sin 4\theta^{(0)} \pd{u^{(1)}}{y} \right)\mathrm{d}y = 0. \label{Original u integral}
    \end{align}
Consideration of the momentum equations and associated no-stress boundary conditions on the upper and lower boundaries of the fluid at higher order gives an equation governing the centre--line of the fluid, $H^{(0)}$
\begin{multline}
      \pd{}{x} \int\limits_{H^{(0)^{-}}}^{H^{(0)^{+}}} \pd{}{x} \int\limits_{H^{(0)^{-}}}^{y}  4\left(1+\mu_3\right)\pd{u^{(0)}}{x} + \mu_1\cos2\theta^{(0)} 
      \\
      + \mu_2\left(\cos^2 2\theta^{(0)} \pd{u^{(0)}}{x} + \f{1}{4}\sin 4\theta^{(0)} \pd{u^{(1)}}{y} \right)\mathrm{d}s\mathrm{d}y  = 0. \label{Original H integral}
\end{multline}
We note first that $s$ is a dummy variable, we are integrating over the second argument of the functions in the integrand twice, so that the quantities in the integrand of equation \eqref{Original H integral} are $\theta^{(0)}\left(x,s,t\right), u^{(1)}\left(x,s,t\right)$ (to emphasise this, the dummy variable is not used in \citep{green2008extensional}). Additionally, we note that equation \eqref{Original H integral} is a particularly unusual form of an integro-differential equation, wherein one of the variables of integration appears in the limit of one of the integrals. The equation \eqref{KinematicConditionOnA} yields
\begin{align}
    \pd{\theta^{(0)}}{t} + u^{(0)}\pd{\theta^{(0)}}{x} + v^{(0)}\pd{\theta^{(0)}}{y} =& -2\sin \theta^{(0)} \cos \theta^{(0)} \pd{u^{(0)}}{x} - \sin^2\theta^{(0)} \pd{u^{(1)}}{y}, \label{Theta Equation}
\end{align}
as an evolution equation governing the behaviour of the fibre director field. The leading order transverse velocity \changes{is found by integrating the incompressiblity condition and applying the kinematic condition \eqref{KinematicConditionOnV}}
\begin{align}
    v^{(0)} =& \pd{H^{(0)}}{t} + \pd{}{x} \left(H^{(0)} u^{(0)}\right) - y \pd{u^{(0)}}{x}, \label{Transverse Velocity}
\end{align}
\changes{consideration of the $x$-momentum equation at $\mathcal{O}\left(\varepsilon\right)$ with its associated boundary condition supplies} 
\begin{equation}
    \pd{u^{(1)}}{y} = -\mu_1 \f{2\sin 2\theta^{(0)}}{4+4\mu_3 + \mu_2\sin^2 2\theta^{(0)}} - \mu_2 \f{\sin 4\theta^{(0)}}{4+4\mu_3 + \mu_2\sin^2 2\theta^{(0)}}\pd{u^{(0)}}{x}, \label{UOne}
\end{equation}
as the next-order correction term for $u$. In this form, one can observe that in the case of $\mu_1 = \mu_2 =\mu_3 =0$, we return to the Trouton model for a Newtonian fluid. In the case of $\mu_1=\mu_2=0$, the fluid behaves very similarly to a Newtonian fluid, with the only modification being a modified Trouton ratio, which effectively only changes the tension required to be applied to the sheet to achieve the same behaviour as a Newtonian fluid. In these cases, the model can be solved by means of a Lagrangian transformation, as detailed in \cite{green2008extensional,howell1994extensional}. \\

\indent The boundary conditions for the \changes{leading order} longitudinal velocity are then
 \begin{align}
     u^{(0)}(0,t) = 0, \quad u^{(0)}(L,t) = \dot{L}.
 \end{align}
We assume that the sheet is being extended in the $x$ direction and that the end points of the centre--line remain \changes{fixed to} $y=0$, thus
\begin{align}
    H^{(0)}(0,t) = H^{(0)}(L,t) = 0.
\end{align}
The kinematic boundary condition yields
\begin{align}
    v^{(0)} = \frac{\p H^{(0)}}{ \p t} \pm \frac{1}{2}\frac{\p h^{(0)}}{\p t} + u^{(0)}\left(  \frac{\p H^{(0)}}{ \p x} \pm \frac{1}{2}\frac{\p h^{(0)}}{\p x}\right);\text{ on } y = H^{(0)} \pm \frac{1}{2}h^{(0)}. \label{Kinematic}
\end{align}
We will need to prescribe initial conditions for the thickness, $h^{(0)},$ and the fibre direction, $\theta^{(0)}$ in the sheet. As in the Newtonian problem (see \cite{howell1994extensional}), we do not need to prescribe an initial condition for $H^{(0)}$ as we are unable to satisfy an arbitrary initial condition for $H^{(0)}$, as noted in \cite{green2008extensional}. 
In order to study the behaviour of sheets that do not initially obey equation \eqref{Original H integral} we must consider a shorter timescale than $\dfrac{L_0}{U}$. We turn to such a timescale in the next subsection. %To this end, we follow a similar path to work by Howell \cite{howell1994extensional}, and consider the behaviour of a sheet on a timescale of $\varepsilon\changes{^2}\dfrac{L_0}{U}$.

\subsection{Short timescale model} \label{Section:TIBNT}
Similarly to the Newtonian problem, we have a singular perturbation problem for $H^{(0)}$ in $t$, the outer solution of which is determined by the solution of \eqref{Original H integral}. As suspected by Green \& Friedman \cite{green2008extensional}, we discover that the centre--line is not necessarily straight (unlike the Newtonian case). In order to study the evolution of a sheet with an arbitrary initial condition for $H^{(0)}$, we follow a similar approach to Howell and Buckmaster \textit{et al.} \cite{buckmaster1975buckling,howell1994extensional} by considering a timescale of $\varepsilon^2\dfrac{L_0}{U}$\\

\indent We note from \eqref{Kinematic}, that in addition to rescaling time, we must also rescale the velocity $v$; hence, we introduce
\begin{equation}
 \tau = \f{t}{\eps^2}, \quad   v = \f{V}{\eps^2}. 
\end{equation}
\indent As in the previous section, we exploit the slender geometry of the sheet by expanding variables as a power series of $\varepsilon$. At leading order the continuity equation gives
\begin{align}
    \pd{V^{(0)}}{y} =0, \label{LeadingCont}
\end{align}
which, combined with the kinematic condition at the same order
\begin{align}
    V^{(0)} = \pd{H^{(0)}}{\tau} \pm \f{1}{2} \pd{h^{(0)}}{\tau}; \text{ on } y=H^\pm,
\end{align}
tells us
\begin{align}
    V^{(0)} = \pd{H^{(0)}}{\tau}, \qquad \pd{h^{(0)}}{\tau} = 0,
\end{align}
since if $V^{(0)}$ is independent of $y$, it must take the same value on the upper and lower free surfaces. Hence, we have an expression for $V^{(0)}$ in terms of $H^{(0)}$, and note that there is no thinning of the sheet on this timescale. Additionally we obtain the equation for $\theta^{(0)}$
\begin{equation}
     \pd{\theta^{(0)}}{\tau} + \pd{H^{(0)}}{\tau}\pd{\theta^{(0)}}{y} = 0. \label{FirstTheta}
\end{equation}
Consideration of the $x$-momentum equation and the associated no-stress boundary condition at $\mathcal{O}\left(\varepsilon\right)$ yields a result for $u^{(0)}$:
\begin{align}
    u^{(0)} = \bar{u}\left(x,\tau\right) + \left(H^{(0)} - y \right)\frac{\p^2 H^{(0)}}{\p \tau \p x},
\end{align}
where $\bar{u}(x,t)$ is an as yet unknown function arising from integration. This is precisely the same result as for a Newtonian fluid \cite{howell1994extensional}. Consideration of the $y$-momentum equation at the same order yields an identity \changes{which is already satisfied.} Next, the $y$-momentum equation at $\mathcal{O}\left(\varepsilon^2\right)$ yields an expression for pressure
\begin{multline}
    -\pd{p^{(0)}}{y} +  \Pd{V^{(0)}}{x} + \Pd{V^{(2)}}{y} + \mu_1 \pd{}{y}\left(\sin^2\theta^{(0)}\right) + \mu_2 \pd{}{y}\left( \sin^2\theta^{(0)} \cos^2\theta^{(0)} \pd{ u^{(0)}}{ x} \right.
    \\
    + \left. \cos\theta^{(0)}\sin^3\theta^{(0)}\left(\pd{u^{(1)}}{y} + \pd{V^{(1)}}{x} \right)  + \sin^4\theta \pd{V^{(2)}}{y} \right)
    \\
    + 2\mu_3 \pd{}{y}\left(2\sin^2\theta^{(0)}\pd{V^{(2)}}{y} + \cos\theta^{(0)}\sin\theta^{(0)}\left(\pd{u^{(1)}}{y} + \pd{V^{(1)}}{x} \right)  \right) = 0, \label{OE2YM}
\end{multline}
with the associated boundary condition
\begin{multline}
    -p^{(0)} + 2\pd{V^{(2)}}{y} + \mu_1 \sin^2\theta^{(0)} + \mu_2\left( \sin^2\theta^{(0)} \cos^2\theta^{(0)} \pd{ u^{(0)}}{ x} \right.
    \\
   \left. + \cos\theta^{(0)}\sin^3\theta^{(0)}\left(\pd{u^{(1)}}{y} + \pd{V^{(1)}}{x} \right) + \sin^4\theta \pd{V^{(2)}}{y} \right) 
    \\
    + 2\mu_3\left( 2\sin^2\theta^{(0)} \pd{V^{(2)}}{y} + \sin\theta^{(0)}\cos\theta^{(0)}\left(\pd{u^{(1)}}{y} + \pd{V^{(1)}}{x} \right)\right) = 0; \text{ on } y=H^{(0)^\pm}. \label{OE2YBDY}
\end{multline}
A number of terms involving $H^{(1)}$ arise in the calculation of \eqref{OE2YBDY}, these terms are multiplied by a term that is identically zero and are thus omitted. We can directly integrate \eqref{OE2YM} and apply \eqref{OE2YBDY} to obtain an equation for pressure
\begin{multline}
    p^{(0)} = -2 \pd{u^{(0)}}{x} + \mu_1 \sin^2\theta^{(0)} + \mu_2\left( \sin^2\theta^{(0)} \cos^2\theta^{(0)} \pd{ u^{(0)}}{ x} \right.
    \\ \left.    +\cos\theta^{(0)}\sin^3\theta^{(0)}\left(\pd{u^{(1)}}{y} + \pd{V^{(1)}}{x} \right) \right.
    \\
    \left. - \sin^4\theta \pd{u^{(0)}}{x} \right)  + 2\mu_3\left( - 2\sin^2\theta^{(0)} \pd{u^{(0)}}{x} + \sin\theta^{(0)}\cos\theta^{(0)} \left(\pd{u^{(1)}}{y} + \pd{V^{(1)}}{x} \right)\right), \label{Pressure}
\end{multline}
where $\mathcal{O}\left(\varepsilon^2\right)$ continuity has been used to eliminate the $V^{(2)}$ terms. This result for pressure is essentially a Newtonian pressure enhanced by the presence of the fibres; should we choose $\mu_1 = \mu_2 = \mu_3 = 0$, we recover the Newtonian pressure. We have also introduced $u^{(1)}$ and $V^{(1)}$ terms, which must now \changes{be eliminated.} The $\mathcal{O}\left(\varepsilon^2\right)$ $x$-momentum equation is 
\begin{multline}
    \Pd{u^{(1)}}{y} + \mu_1 \pd{}{y}\left(\cos\theta^{(0)}\sin\theta^{(0)}\right) + \mu_2 \pd{}{y}\left( \cos^3\theta^{(0)}\sin^2\theta^{(0)}\pd{u^{(0)}}{x} \right. 
    \\
    + \cos^2\theta^{(0)}\sin^2\theta^{(0)} \left(\pd{u^{(1)}}{y} + \pd{V^{(1)}}{x}\right)
    + \left. \cos\theta^{(0)} \sin^3\theta^{(0)} \pd{V^{(2)}}{y} \right) 
    \\
    + \mu_3 \pd{}{y} \left(\pd{u^{(1)}}{y} + \pd{V^{(1)}}{x} \right) = 0,
\end{multline}
with the associated boundary condition
\begin{multline}
    \pd{u^{(1)}}{y} + \pd{V^{(1)}}{x} + \mu_1 \cos\theta^{(0)}\sin\theta^{(0)} + \mu_2\left(  \cos^3\theta^{(0)}\sin^2\theta^{(0)}\pd{u^{(0)}}{x} \right.
    \\
    +\left. \cos^2\theta^{(0)}\sin^2\theta^{(0)} \left(\pd{u^{(1)}}{y} + \pd{V^{(1)}}{x}\right) \right. + \left. \cos\theta^{(0)} \sin^3\theta^{(0)} \pd{V^{(2)}}{y} \right)
    \\
    + \mu_3 \left(\pd{u^{(1)}}{y} + \pd{V^{(1)}}{x} \right) = 0, \text{ on } y= H^{(0)^{\pm}}.
\end{multline}
Combining these yields the following compatibility condition
\begin{align}
   \left( \pd{u^{(1)}}{y} + \pd{V^{(1)}}{x} \right) \left( 1 + \mu_2  \cos^2\theta^{(0)}\sin^2\theta^{(0)} + \mu_3\right) + \mu_1  \cos\theta^{(0)}\sin\theta^{(0)} \nonumber \\
   + \mu_2 \pd{u^{(0)}}{x} \left(\cos^3\theta^{(0)}\sin\theta^{(0)} - \cos\theta^{(0)} \sin^3\theta^{(0)}\right) = 0. \label{Compat}
\end{align}
We note that in \changes{the} analysis, the terms $V^{(1)}$,$u^{(1)}$ only appear together. We can thus view \eqref{Compat} as an expression which allows us to eliminate \changes{both} $V^{(1)}$ and $u^{(1)}$. Hence, we can now write leading order pressure in terms of $\theta^{(0)}$ and $u^{(0)}$ only. \\

\indent In order to close the model, we must go to yet higher orders in order to obtain equations for $\bar{u}$ and $H^{(0)}$. Our approach is similar to Green \& Friedman, \cite{green2008extensional}: we integrate the relevant equations over the depth of the sheet and apply the no-stress boundary conditions at $y=H^{(0)^{\pm}}$. The expressions involved are rather cumbersome, but substituting for previously-determined quantities produces considerable simplification; full details can be found in Appendix \ref{Appendix:BNTDeriv}. We finally obtain the following equation for $\bar{u}$
%In our approach, a number of higher order terms in the boundary conditions will be eliminated by substitution of previously obtained quantities. At $\mathcal{O}\left(\varepsilon^3\right)$, equation \eqref{Compat} also appears as part of the boundary conditions. Using this equation to cancel out numerous terms greatly simplifies those conditions. We leave the details to the appendix, and finally obtain
\begin{equation}
    \pd{}{x}\int \limits_{H^{(0)^{-}}}^{H^{(0)^{+}}} \dfrac{\mu_1 \cos 2\theta^{(0)} + \left(4+4\mu_3 + \mu_2\right)\left( \pd{\bar{u}}{x} +\left(H^{(0)} -  y \right)\frac{\p H^{{(0)}^3}}{\p x^2 \p \tau} + \frac{\p H^{(0)}}{\p x}\frac{\p^2 H^{(0)} }{\p x \p \tau} \right)  }{4+4\mu_3 + \mu_2  \sin^2 2\theta^{(0)}} \mathrm{d}y  = 0.\label{UTransformedIntro}
\end{equation}
The equation for $H^{(0)}$ is
\begin{multline}
    \Pd{}{x} \left(~ \int\limits_{H^{(0)^{-}}}^{H^{(0)^{+}}} \int \limits_{H^{(0)^{-}}}^{y} \dfrac{\mu_1 \cos 2\theta^{(0)} + \left(4+4\mu_3 + \mu_2\right)\left( \pd{\bar{u}}{x} +\left(H^{(0)} -  y \right)\frac{\p H^{{(0)}^3}}{\p x^2 \p \tau} \right)}{4+4\mu_3 + \mu_2  \sin^2 2\theta^{(0)}} \mathrm{d}s \mathrm{d}y \right) 
   \\
    = \frac{\p}{\p x^2}\left( H^{(0)^{+}} \right) \left(\int\limits_{H^{(0)^{-}}}^{H^{(0)^{+}}} \dfrac{\mu_1 \cos 2\theta^{(0)} + \left(4+4\mu_3 + \mu_2\right)\left( \pd{\bar{u}}{x} +\left(H^{(0)} -  y \right)\frac{\p H^{{(0)}^3}}{\p x^2 \p \tau} + \frac{\p H^{(0)}}{\p x}\frac{\p^2 H^{(0)} }{\p x \p \tau} \right) }{4+4\mu_3 + \mu_2  \sin^2 2\theta^{(0)}}  \mathrm{d}y \right). \label{HTransformedIntro}
\end{multline}
We have now derived a system of equations for $\theta^{(0)}, \bar{u}$, and $H^{(0)}$, namely \eqref{FirstTheta},\eqref{UTransformedIntro},\eqref{HTransformedIntro} respectively. 
We note that \eqref{HTransformedIntro} now includes a $\dfrac{\p^5 H^{(0)}}{\p x^4 \p \tau}$ term. We must prescribe two more boundary conditions for $H^{(0)}$, in addition to the those discussed in the previous section. We set
\begin{align}
    \pd{H^{(0)}}{x}(0,\tau) = \pd{H^{(0)}}{x}(L,\tau) = 0.
\end{align}
\indent Much like the Green \& Friedman model, the equations \eqref{FirstTheta},\eqref{UTransformedIntro},\eqref{HTransformedIntro} must be solved numerically. The approach to the short time-scale problem is more straightforward than we use for the Green \& Friedman model (the details of which we will present in Section \ref{Section:SectionALE}).

%Despite equations \eqref{UTransformed}, \eqref{HTransformed} appearing to be  significantly more complex than \eqref{U Eqn Rewrite}, \eqref{H Equation Liebniz}, at each time-step $\theta$ is known and $\dfrac{\p \bar{u}}{\p x}$ depends on only $x,\tau$ and can be removed from the integral. Hence the integrals are pre-computable at each time step, and we have two equations we must simultaneously solve for $\bar{u},H$. Our strategy is as follows.
%Given an initial condition for $\theta, H$, we simultaneously solve for $\bar{u}$ at the current time step, and $H$ at the next time step using a FTCS finite difference scheme. Then we use equation \eqref{FirstThetaS} to update $\theta$ to the next time step using an upwinding method. We repeat until we reach the desired time. The details of the discretisation are included in Appendix \ref{AppendixBNTDisc}.

\section{Arbitrary Lagrangian--Eulerian Methods} \label{Section:SectionALE}
Although the Green \& Friedman system, \eqref{Cons Mass}--\eqref{UOne} and the short timescale system, \eqref{FirstTheta}, \eqref{UTransformedIntro}--\eqref{HTransformedIntro} are both significant simplifications compared to the full two-dimensional problem, they are too complex to allow significant analytical progress and must be solved numerically. 
In this section, we reformulate the Green \& Friedman and short-timescale equations into an arbitrary Lagrangian--Eulerian (ALE) formulation. ALE methods involve the construction of a reference domain together with mappings from this domain to both the Lagrangian and Eulerian descriptions of the flow. Unlike numerical techniques based on either a purely Lagrangian description, where the nodes of the computational mesh follow an associated material particle throughout the motion, or upon a purely Eulerian description, where the computational mesh is fixed and the motion of the continuum is with respect to the grid, ALE methods allow freedom in moving the mesh in a way that is not necessarily fixed to a material particle. This can provide accurate solutions when modelling greater distortions of a flow problem than can ordinarily be handled by numerical techniques upon a Lagrangian description, with more resolution than is often attainable by a purely Eulerian description \citep{donea2017arbitrary}. \\
\\
\indent Our approach largely follows \citep{donea2017arbitrary}, but we outline the details here for completeness. We introduce Lagrangian, Eulerian, and reference domain variables which we denote by  $\boldsymbol{X} = \left(X,Y\right)$, $\boldsymbol{x} = \left(x,y\right)$ and $\boldsymbol{x'} = \left(x',y'\right),$ respectively. Converting between Eulerian and Lagrangian descriptions of flow fields is well established, and has been employed in Newtonian extensional flow problems \cite{green2008extensional, howell1994extensional, wylie2016evolution}. We define the map $\pmb\varphi$ from the Lagrangian to the Eulerian descriptions such that 
\begin{equation*}
    \left(\boldsymbol{x},t\right) = \left(\pmb\varphi\left(\boldsymbol{X},t\right),t\right),
\end{equation*}
where the material velocity ${\boldsymbol{\upsilon}}$ is given by 
\begin{equation}
   \pd{\pmb\varphi}{t} = \boldsymbol{\upsilon}\left(\boldsymbol{X},t\right), \label{MaterialVelocity}
\end{equation}
so that the familiar material time derivative of an arbitrary scalar field (e.g pressure) is 
\begin{align}
    \frac{Df\left(\boldsymbol{X},t\right)}{Dt} = \pd{f\left(\boldsymbol{x},t\right)}{t} + \pd{f\left(\boldsymbol{x},t\right)}{\boldsymbol{x}}\pd{\pmb\varphi}{t} = \pd{f}{t} + \left( \boldsymbol{\upsilon} \cdot \nabla_{\boldsymbol{x}}\right)f.
\end{align}
Now, it remains only to define the map from the reference domain to the Eulerian domain, which we denote by ${\pmb\Phi}$. This satisfies 
\begin{equation*}
    \left(\boldsymbol{x},t\right) = \left({\pmb\Phi}\left(\boldsymbol{x'},t\right), t\right).
\end{equation*}
The mesh velocity, $\boldsymbol{u}_{\text{mesh}}$, is given by
\begin{equation*}
    \boldsymbol{u}_{\text{mesh}}\left(\boldsymbol{x'},t\right) = \pd{{\pmb\Phi}}{t}.
    \end{equation*}
We obtain the relations between physical quantities, such as pressure, in the reference and Eulerian descriptions in the same way we do between Lagrangian and Eulerian descriptions. In particular, the time derivative of an arbitrary scalar field $f$ is
\begin{align}
    \pd{f\left(\boldsymbol{x'},t\right)}{t} &= \pd{f\left(\boldsymbol{x},t\right)}{t} +\left(\boldsymbol{u}_{\text{mesh}} \cdot \nabla_{\boldsymbol{x}}\right)f, \label{ALERelationA}
\end{align}
so that the transformation from the reference to the Eulerian description behaves very much like a familiar transformation between Lagrangian and Eulerian frames. For convenience, we rewrite \eqref{ALERelationA} as
\begin{align}
    \pd{f\left(\boldsymbol{x'},t\right)}{t} - \frac{\p \boldsymbol{x'}}{\p \boldsymbol{x}}\left(\boldsymbol{u}_{\text{mesh}} \cdot \nabla_{\boldsymbol{x'}}\right)f = \pd{f\left(\boldsymbol{x},t\right)}{t}. \label{ALERelation}
\end{align}
As a final note, if $\pmb\Phi = \pmb\varphi$, then the reference description is the same as the Lagrangian description and thus $\boldsymbol{u}_{\text{mesh}} = \boldsymbol{\upsilon}$. If $\pmb\Phi = \boldsymbol{x'}$, then the reference description is the same as the Eulerian description, and $\boldsymbol{u}_{\text{mesh}} = \boldsymbol{0}$. For further explanation of ALE techniques, we direct the reader to \citep{donea2017arbitrary}.

\subsection{Employing ALE upon the \changes{Green \& Friedman} model}
\changes{Henceforth we drop the superscript notation on leading order terms. We first employ ALE techniques upon the equations governing the $t \sim \dfrac{L_0}{U}$ timescale, equations \eqref{Cons Mass} -- \eqref{Kinematic}. }We define the reference domain, $D_{\text{ref}}$, to be
\begin{equation}
    \left(x',y'\right) \in D_{\text{ref}} = [0,1] \times \left[-\frac{1}{2}, \frac{1}{2}\right]. \label{RefDom}
\end{equation}
 We define the mapping from the reference variables to the Eulerian variables $\pmb\Phi: D_{\text{ref}} \rightarrow  \left[0,L\right] \times \left[H^{-},H^{+}\right]$ such that 
\begin{equation}
    \left(x,y\right) = \pmb\Phi \left(x',y'\right) = \left(Lx',H\left(\frac{L x'}{L(0)},t\right)+ h\left(\frac{L x'}{L(0)},t\right)y' \right). \label{ALEDefn}
\end{equation}
This choice of map maintains equidistant spacing in the mesh in both horizontal and vertical directions. Therefore, our discretisation of \changes{$D_{\text{ref}}$} can be a simple equidistant grid.
As already discussed, the mesh velocity is readily obtained by differentiating the mapping $\pmb\Phi$ with respect to time. Written in terms of Eulerian variables, we have
\begin{multline}
   \pd{\pmb\Phi}{t} =  \mathbf{u}_{\text{mesh}} = \left( \dot{L} \frac{x}{L}, \dot{L} \frac{x}{L}\left(\pd{H}{x}(x,t) + \frac{y-H(x,t)}{h(x,t)} \pd{h}{x}(x,t) \right) + \pd{H}{t}(x,t) \right. \\ \left.
   + \frac{y-H(x,t)}{h(x,t)} \pd{h}{t}(x,t) \right).
\end{multline}
\\
\indent We now account for the introduction of the moving mesh by using \eqref{ALERelation} and \eqref{ALEDefn}, we first modify the equation for $\theta$, \eqref{Theta Equation}. Using \eqref{UOne} to eliminate $u^{(1)}$, this equation becomes
\begin{equation}
\pd{\theta}{t} + \left(\frac{u-u_{\text{mesh}}}{L}\right)\pd{\theta}{x'} + \left(\frac{v-v_{\text{mesh}}}{h}\right)\pd{\theta}{y'} = \frac{\sin2\theta\left(2\mu_1 \sin^2\theta - \frac{1}{L}\left(1+2\mu_2\sin^2\theta \right)\pd{u}{x'} \right)}{4+4\mu_3+\mu_2 \sin^2 2\theta}. \label{Theta Equation ALE} 
\end{equation}
This form of \eqref{Theta Equation ALE} allows us yet further simplification. As explicitly demonstrated in Appendix \ref{AppendixTheta}, this equation corresponds to advection purely in a horizontal direction on the reference domain, which significantly eases implementation. We note that this also decouples $\theta$ from $H$, and since $v$ does not arise in any other equation, we need not prescribe an initial $H$ and can simply compute the transverse velocity on demand. To summarise the model in the ALE form, we have
\begin{align}
  \pd{h}{t} &+ \left(\frac{u-u_{\text{mesh}}}{L}\right) \pd{h}{x'} + \frac{1}{L}\pd{u}{x'}=0, \label{h ALE}
  \\
        \pd{\theta}{t} &+ \left(\frac{u-u_{\text{mesh}}}{L}\right)\pd{\theta}{x'}  =  \frac{\sin2\theta\left(2\mu_1\sin^2\theta - \f{1}{L}\left(4+4\mu_3+2\mu_2\sin^2\theta \right)\pd{u}{x'} \right)}{4+4\mu_3+\mu_2 \sin^2 2\theta}, \label{Theta Equation Rewrite Summary}
        \\
    \pd{}{x'} &\int\limits_{-\frac{1}{2}}^{\frac{1}{2}} \f{\mu_1\cos2\theta + \f{1}{L}\left(4+4\mu_3 + \mu_2\right) \pd{u}{x'} }{ 4+4\mu_3+\mu_2\sin^2 2\theta} h \mathrm{d}y' = 0, \label{U ALE}
    \\
   \Pd{}{x'}&\int\limits_{-\frac{1}{2}}^{\frac{1}{2}} \int\limits_{-\frac{1}{2}}^{y'} \frac{\mu_1 \cos 2\theta + \frac{1}{L}\left(4+4\mu_3 + \mu_2 \right) \pd{u}{x'}}{4+4\mu_3+\mu_2\sin^2 2\theta}h^2 \mathrm{d}s' \mathrm{d}y' \nonumber
   \\
   &\qquad  =  \left(\Pd{H}{x'} + \frac{1}{2}\Pd{h}{x'} \right) \int\limits_{-\frac{1}{2}}^{\frac{1}{2}} \frac{\mu_1 \cos 2\theta + \frac{1}{L}\left(4+4\mu_3 + \mu_2\right) \pd{u}{x'}}{4+4\mu_3+\mu_2\sin^2 2\theta}h\mathrm{d}y', \label{H ALE}
\end{align}
for $\left(x',y'\right) \in D_{\text{ref}}$. The associated boundary and initial conditions now become
\begin{align}
    &u(0,t) = 0,  &u(1,t) = 1,
    \\
    &H(0,t) = 0, &H(1,t) = 0,
\\
&h(x',0) = h_{i}\left(x'\right), &\theta\left(x',y',0\right) = \theta_{i}\left(x',y'\right).
\end{align}
We exclude the equation for the transverse velocity $v$, since this quantity is readily calculated from \eqref{Transverse Velocity} at any $(x',y',t)$ once the model \eqref{h ALE}--\eqref{H ALE} is solved.
\\

\indent In order to solve this model we use the following algorithm. Given initial conditions, equation \eqref{U ALE} is solved to obtain $u\left(x',0\right)$, using the trapezoidal rule and a finite difference approximation. We note that this particular equation can be approached in either ALE or Eulerian variables, since applying the trapezoidal rule to solve the integral \eqref{U ALE} effectively decouples $H$ from the system (see Appendix C). Once solved, equations \eqref{h ALE} and \eqref{Theta Equation Rewrite Summary} can be used to update $h,\theta$ to the next time--step, at which point the process is repeated until the required end time is reached. The quantities $v,H$ can then be computed at any time using  \eqref{Transverse Velocity} and \eqref{H ALE}, respectively, as well as any other quantities of interest. In our implementation, we solved equations \eqref{h ALE}--\eqref{H ALE} in the ALE form given above. Since the ALE framework acts like a simple substitution on the integral equations, it is equally possible to discretise and solve equations \eqref{U ALE} and \eqref{H ALE} in their Eulerian forms, and indeed the discretisation is similar. We include the discretisations of equations \eqref{U ALE},\eqref{H ALE} in Appendix \ref{AppendixDiscretisation}.

\subsection{Validation of numerical method}
\begin{figure}[ht]
    \centering
   \begin{overpic}[width=0.49\textwidth]{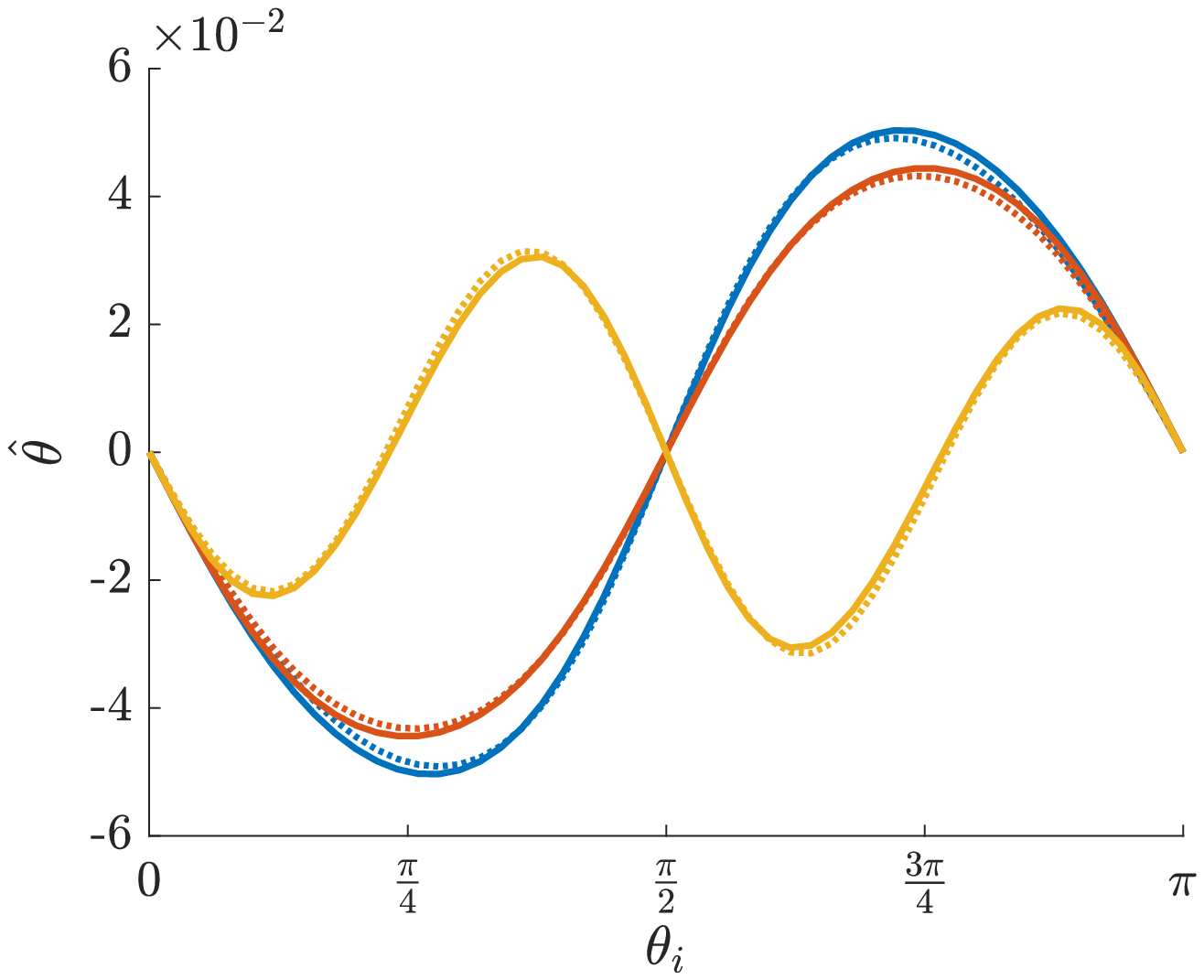}
      \put (3, 73) {a)}
     \put(40,15){\vector(0,1){50}}
\put(28,66){$\mu_1$ increasing}
  \end{overpic}
  \begin{overpic}[width=0.49\textwidth]{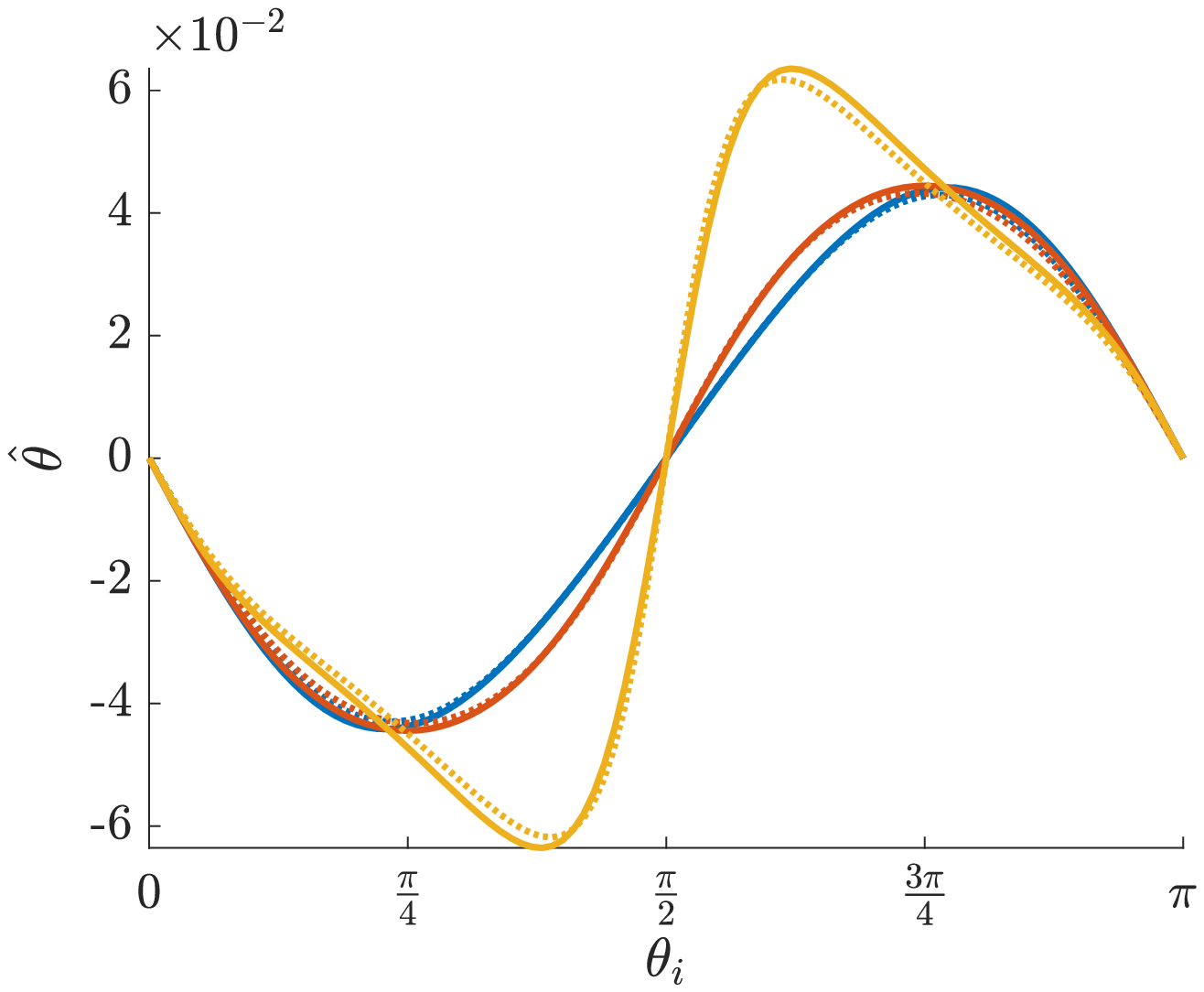}
      \put (3, 73) {b)}
   %   \linethickness{2pt}
\put(62,42){\vector(-1,1){15}}
\put(28,58){$\mu_2$ increasing}
  \end{overpic}
  \\
  \centering
     \begin{overpic}[width=0.49\textwidth]{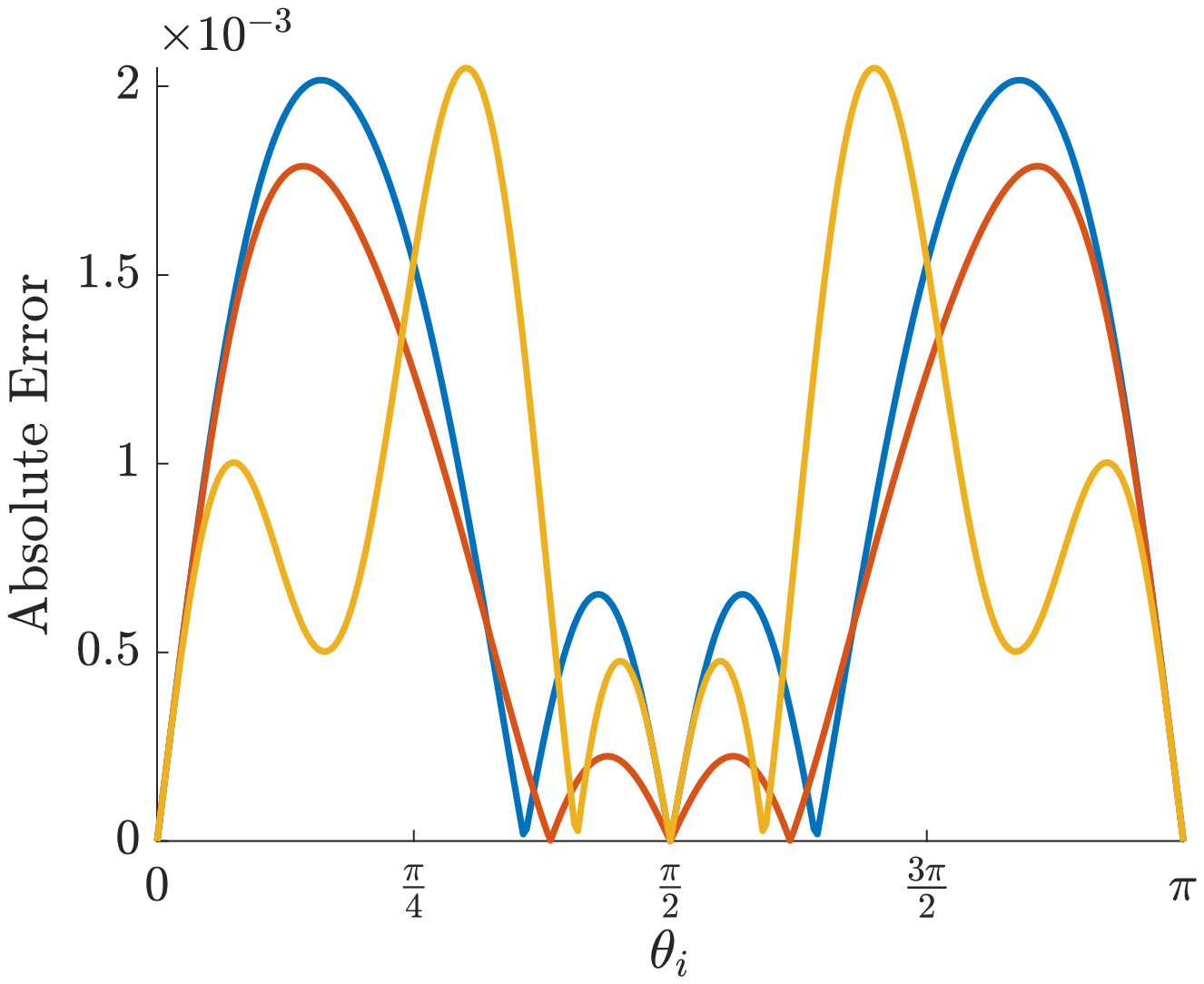}
      \put (3, 73) {c)}
     \put(40,20){\vector(1,2){10}}
\put(40,42){$\mu_1$ increasing}
  \end{overpic}
  \begin{overpic}[width=0.49\textwidth]{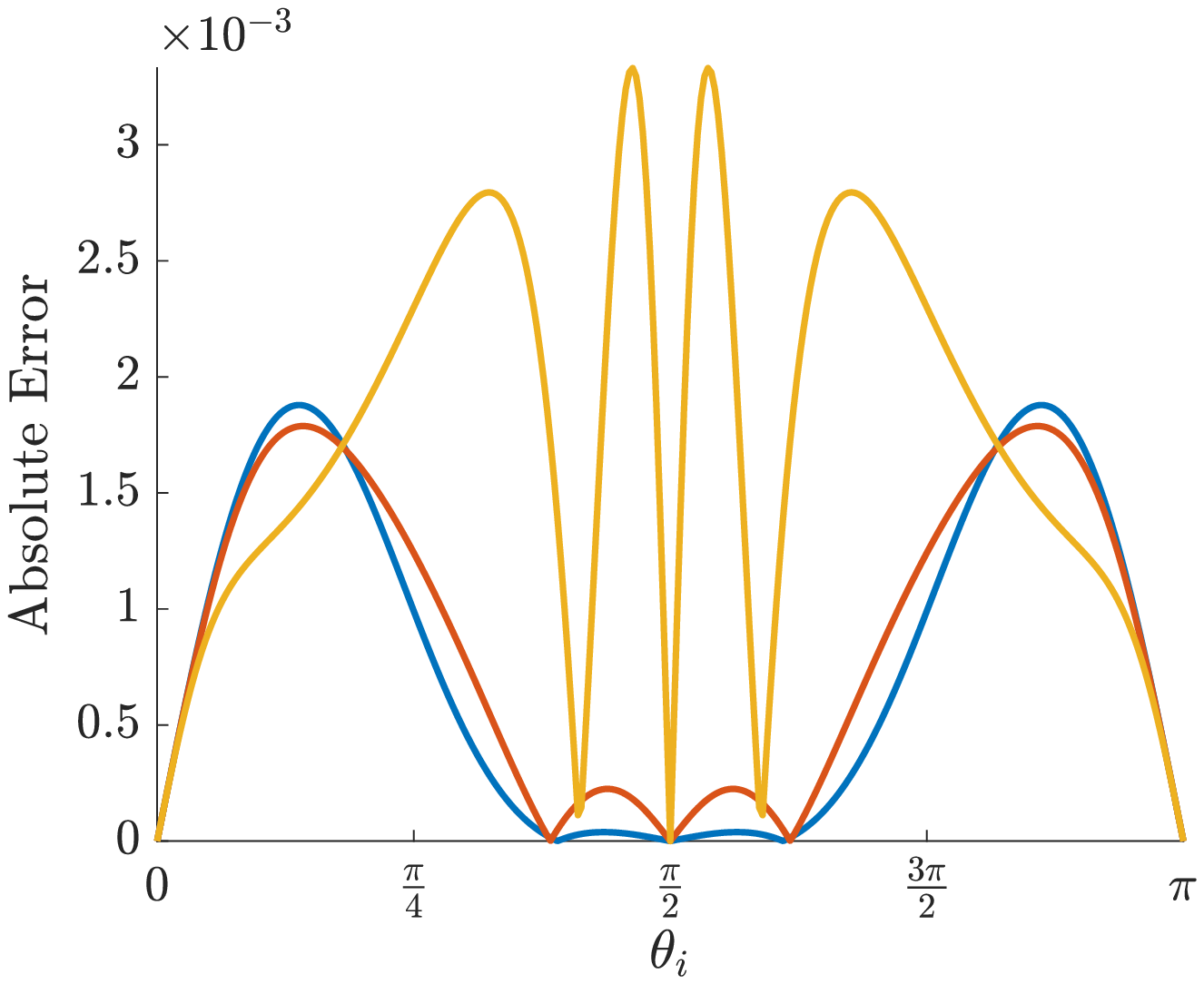}
      \put (3, 73) {d)}
   %   \linethickness{2pt}
\put(30,30){\vector(0,1){38}}
\put(15,70){$\mu_2$ increasing}
  \end{overpic}
    \caption{Comparison of the analytical result for $\hat\theta$ at $t' = 0.05$, with $\theta_i \in [0,\pi]$ for a) $\mu_1 =0,1,10$ (blue, red, yellow respectively) with $\mu_2=\mu_3=1$ fixed and b) $\mu_2 = 0,1,10$ (blue, red, yellow respectively) with $\mu_1=\mu_3=1$ fixed.
    The analytical results from \eqref{SheetThetaHat} are solid lines, the dotted lines are numerical results obtained by solving \eqref{h ALE}--\eqref{H ALE}. The absolute errors between the numerical and analytical results are given for c) varied $\mu_1$ and d) varied $\mu_2$.}
\label{ShortTimeNumericsValidation}
\end{figure}

In order to validate \changes{the} approach, we consider the early--time behaviour of the sheet for which an analytical expression is available \citep{green2008extensional}. Introducing the short timescale $t' = \delta^{-1} t$, where $\varepsilon \ll \delta \ll 1$, assuming constant initial film thickness, $h_i$, and alignment angle, $\theta_i$, and performing a Taylor expansion on the variables so that
\begin{align*}
    h = h_i + \hat{h}t' \delta + \mathcal{O}\left(\delta^2\right), \\
    \theta = \theta_i + \hat{\theta} t' \delta +\mathcal{O}\left(\delta^2\right), 
\end{align*}
where $\hat{h},\hat{\theta}$ are the changes in thickness and fibre direction respectively, it is possible to derive an analytical expression for the evolution of the fibre angle in the sheet for constant initial conditions for thickness and fibre direction, see \cite{green2008extensional}. The result is:
\begin{equation}
\hat\theta =  - 2\sin 2\theta_{i} + \frac{2\sin^2 \theta_{i}}{4+ 4\mu_3 + \mu_2 \sin^2 2\theta_i}\left[ \mu_1 \sin 2\theta_i +\frac{\mu_2}{2} \sin4\theta_i \right], \label{SheetThetaHat}
\end{equation}
where $t' \hat\theta $ is the total change in the fibre angle over the short time $\tau$, and $\theta_i$ is the constant initial condition. In \cite{green2008extensional}, the authors note that a consequence of this analysis is that so long as $\mu_1, \mu_2, T$ are all sufficiently small, the fibres will align with the direction of pulling as long as $\theta_{i} \neq \frac{\pi}{2},\frac{3\pi}{2}$. In Figure \ref{ShortTimeNumericsValidation} we plot the numerical solution of \changes{\eqref{h ALE} -- \eqref{H ALE} against equation \eqref{SheetThetaHat} up to $t = 0.05$.} The maximum absolute error in Figure \ref{ShortTimeNumericsValidation}b is 0.0033, and occurs at $\theta_i = 1.4556,1.6860$, for $\mu_2 =10$.

\subsection{Employing ALE \changes{to the short timescale}}
We again define the reference domain and mapping from the reference to the Eulerian domain by \eqref{RefDom} and \eqref{ALEDefn}. In Section 2.2 we determined that on the short timescale there \changes{is} no extension or thinning of the sheet, which gives the simpler mesh velocity
\begin{align}
\mathbf{u}_{\text{mesh}} = \left(0,\pd{H}{t}(x,t) \right).
\end{align}
The fibre director equation on the short timescale, \eqref{FirstTheta}, becomes
\begin{align}
    \frac{\p \theta}{\p \tau} = 0, \label{ALETheta}
\end{align}
hence there is no evolution in $\theta$ in the ALE framework on this timescale. Next, equation \eqref{UTransformedIntro} gives,
\begin{multline}
     \pd{}{x'}\int \limits_{-\frac{1}{2}}^{\frac{1}{2}} \dfrac{\mu_1 \cos 2\theta + \left(4+4\mu_3+\mu_2\right) \pd{\bar{u}}{x'}}{4+4\mu_3+\mu_2\sin^2 2\theta} h\mathrm{d}y'  +  \frac{1}{L}\pd{}{x'}\left(\pd{H}{x'}\dfrac{\p^2 H}{\p x' \p \tau}\int \limits_{-\frac{1}{2}}^{\frac{1}{2}} \dfrac{\left(4+4\mu_3+\mu_2\right)}{4+4\mu_3+\mu_2 \sin^2 2\theta} h\mathrm{d}y' \right.
    \\
    \left. - \dfrac{\p^3 H}{\p x'^2 \p \tau}\int \limits_{-\frac{1}{2}}^{\frac{1}{2}} \dfrac{\left(4+4\mu_3+\mu_2\right) y'}{4+4\mu_3+\mu_2 \sin^2 2\theta} h^2\mathrm{d}y'\right) =0, \label{UTransformedBNTSec}
\end{multline}
with the momentum equation \eqref{HTransformedIntro} becoming
\begin{multline}
    \Pd{}{x'} \left( \int\limits_{-\frac{1}{2}}^{\frac{1}{2}} \int \limits_{-\frac{1}{2}}^{\tilde{y}} \dfrac{\mu_1 \cos 2\theta + \left(4+4\mu_3+\mu_2\right) \frac{1}{L}\pd{\bar{u}}{x'}}{4+4\mu_3+\mu_2 \sin^2 2\theta} h^2 \mathrm{d}\tilde{y}' \mathrm{d}\tilde{y}
    \right. 
    \\ 
    \left. + \pd{H}{x'}\frac{\p^2 H}{\p x' \p \tau}  \int\limits_{-\frac{1}{2}}^{\frac{1}{2}} \int \limits_{-\frac{1}{2}}^{\tilde{y}}\dfrac{4+4\mu_3+\mu_2}{4+\mu_3+\mu_2 \sin^2 2\theta} h^2 \mathrm{d}\tilde{y}' \mathrm{d}\tilde{y} \right.
    \left. 
    -\frac{\p^3 H}{\p x'^2 \p \tau}  \int\limits_{-\frac{1}{2}}^{\frac{1}{2}} \int \limits_{-\frac{1}{2}}^{\tilde{y}} \dfrac{(4+4\mu_3+\mu_2)\tilde{y}'}{4+4\mu_3+ \mu_2 \sin^2 2\theta} h^3 \mathrm{d}\tilde{y}' \mathrm{d}\tilde{y}  \right)     \\
    = \left(\Pd{H}{x'} +\frac{1}{2}\Pd{h}{x'}\right)\left(\int\limits_{-\frac{1}{2}}^{\frac{1}{2}}  \dfrac{\mu_1 \cos 2\theta + \left(4+4\mu_3+\mu_2\right)\frac{1}{L}\pd{\bar{u}}{x'}}{4+4\mu_3+\mu_2 \sin^2 2\theta} h \mathrm{d}\tilde{y} \right.
    \\\left. 
    + \pd{H}{x'}\frac{\p^2 H}{\p x' \p \tau}  \int\limits_{-\frac{1}{2}}^{\frac{1}{2}} \dfrac{4+4\mu_3+\mu_2}{4+4\mu_2+\mu_2 \sin^2 2\theta} h \mathrm{d}\tilde{y} \right.
  +
    \left. - \frac{\p^3 H}{\p x'^2 \p \tau}  \int\limits_{-\frac{1}{2}}^{\frac{1}{2}} \dfrac{\left(4+4\mu_3+\mu_2\right)\tilde{y}}{4+4\mu_3 + \mu_2 \sin^2 2\theta} h^2  \mathrm{d}\tilde{y}  \right). \label{HTransformedBNTSec}
\end{multline}
As the equation for $\theta$, \eqref{ALETheta}, tells us that there is no evolution of $\theta$ in the ALE framework, we now need only to solve \eqref{UTransformedBNTSec}, \eqref{HTransformedBNTSec} for $\bar{u}$ and $H$ respectively. As $\dfrac{\p \bar{u}}{\p x}$ depends on only $x',\tau$, it can be removed from the integrals, the integral coefficients are pre--computable at each time step, which significantly eases the numerical implementation.
\\

\indent The strategy is as follows: given an initial condition for $\theta, H$, we simultaneously solve for $\bar{u}$ at the current time step, and $H$ at the next time step using a forward time centred space finite difference discretisation. We repeat this until we reach the desired time. We give the discretisations of equations \eqref{FirstTheta},\eqref{UTransformedBNTSec},\eqref{HTransformedBNTSec} and details of the construction of the resulting linear system in Appendix \ref{AppendixBNTDisc}.

\section{Results \changes{on the extensional flow timescale}} \label{Section:GreenFriedman}
We first examine the behaviours of the sheet on the $t\sim\frac{L_0}{U}$ timescale. Green \& Friedman \changes{considered} special cases \changes{where} further analytical insight into the behaviour of the model \changes{is possible}. They considered two cases: \changes{where the sheet is not extending} $(\dot{L}= 0)$, in which the fibres tend to align parallel to the $y$-axis, and the case of an extensional flow with $\mu_1=\mu_2=0$. In this case, the equations imply that $H\equiv 0$, and are amenable to further progress via Lagrangian transformation. It is found that the fibres align with the direction of extension (i.e parallel to the $x$-axis) \cite{green2008extensional}. \changes{Throughout the next section we assume that $\mu_1 = 0$ in order to examine the contributions of the anisotropic viscosities $\mu_2, \mu_3$ to the behaviour of the sheet.} % Throughout the next section we assume that $\mu_1 = 0$.
%-------------------------------%
\subsection{Solutions for initially-uniform transversely-isotropic sheets with $\mu_1 = 0$} \label{Section:IUTI}
We now aim to study the effect of the \changes{anisotropic} viscosities $\mu_2,\mu_3$ upon the fluid as the sheet is stretched. We first turn our attention to transversely isotropic sheets that have an initially constant thickness and $\mu_1 = 0$. %, which for simplicity we set $h_i = 1$.
For clarity of exposition, we introduce the quantity
 \begin{align}
     G\left(x',y',t\right) = h G_{1} + \frac{h}{L}\pd{u}{x'}G_{2} =  \int\limits_{-\frac{1}{2}}^{y'}\frac{\mu_1 \cos 2\theta + \left(4+4\mu_3 + \mu_2\right) \frac{1}{L} \pd{u}{x'}}{4+4\mu_3+\mu_2 \sin^2 2\theta}h \mathrm{d}y',
 \end{align} 
 so that
 \begin{align}
    G_1(x',y',t) = \int\limits_{-\frac{1}{2}}^{y'}\frac{\mu_1 \cos 2\theta }{4+4\mu_3+\mu_2 \sin^2 2\theta} \mathrm{d}y', \\
      G_2(x',y',t) = \int\limits_{-\frac{1}{2}}^{y'}\frac{\left(4+4\mu_3 + \mu_2\right) }{4+4\mu_3+\mu_2 \sin^2 2\theta} \mathrm{d}y'.
\label{G2Defn}     
\end{align}
The definitions of $G_{1}$ and $G_{2}$ are readily obtained in the \changes{physical domain} by performing the inverse of the transformation $\pmb\Phi$ \changes{and reintroducing dimensional parameters}.
We start by considering sheets with no tension in the fibre direction when the fluid is at rest, that is, $\mu_1 = 0$, in which case the equation for $u$ in ALE variables, \eqref{U ALE}, becomes
\begin{align}
    \pd{}{x'} \left( h \pd{u}{x'} G_2\left(x',\frac{1}{2},t\right)\right) = 0. \label{IntroG2}
\end{align}
This is of the same form as the longitudinal momentum equation in the Trouton model, with $G_{2}$ playing the role of a spatially-varying viscosity (setting $\mu_2 = \mu_3 = 0$ gives $G_2 = 4$, the Trouton ratio for a Newtonian thin sheet \cite{howell1994extensional}).  We interpret $G_{2}$ as a heterogeneous, time-dependent, `effective viscosity'. As we shall show, we see the effect of $G_{2}$ is to induce `necking' in the sheet (the sheet undergoes thinning at a greater rate in some areas of the sheet than others, generating a `neck'). This behaviour has been observed in Newtonian fluids which possess an inhomogenous viscosity.
\\

\indent We also note that if the fibre angle is independent of $x'$, or if $\mu_1 = \mu_2 = 0$ \cite{green2008extensional}, then $G_2$ as it appears in equation \eqref{IntroG2} does not possess $x'$ dependence. Hence, a fluid that does not possess tension in the fibre direction when at rest, and has a uniform fibre direction will behave like a Newtonian fluid, with a modified viscosity. In this case, the centre--line of the fluid is always a straight line, and the model can be solved by transforming to Lagrangian variables, as detailed in \cite{green2008extensional, howell1994extensional}.
\\

\begin{figure}[ht]
    \centering
   \begin{overpic}[width=0.49\textwidth]{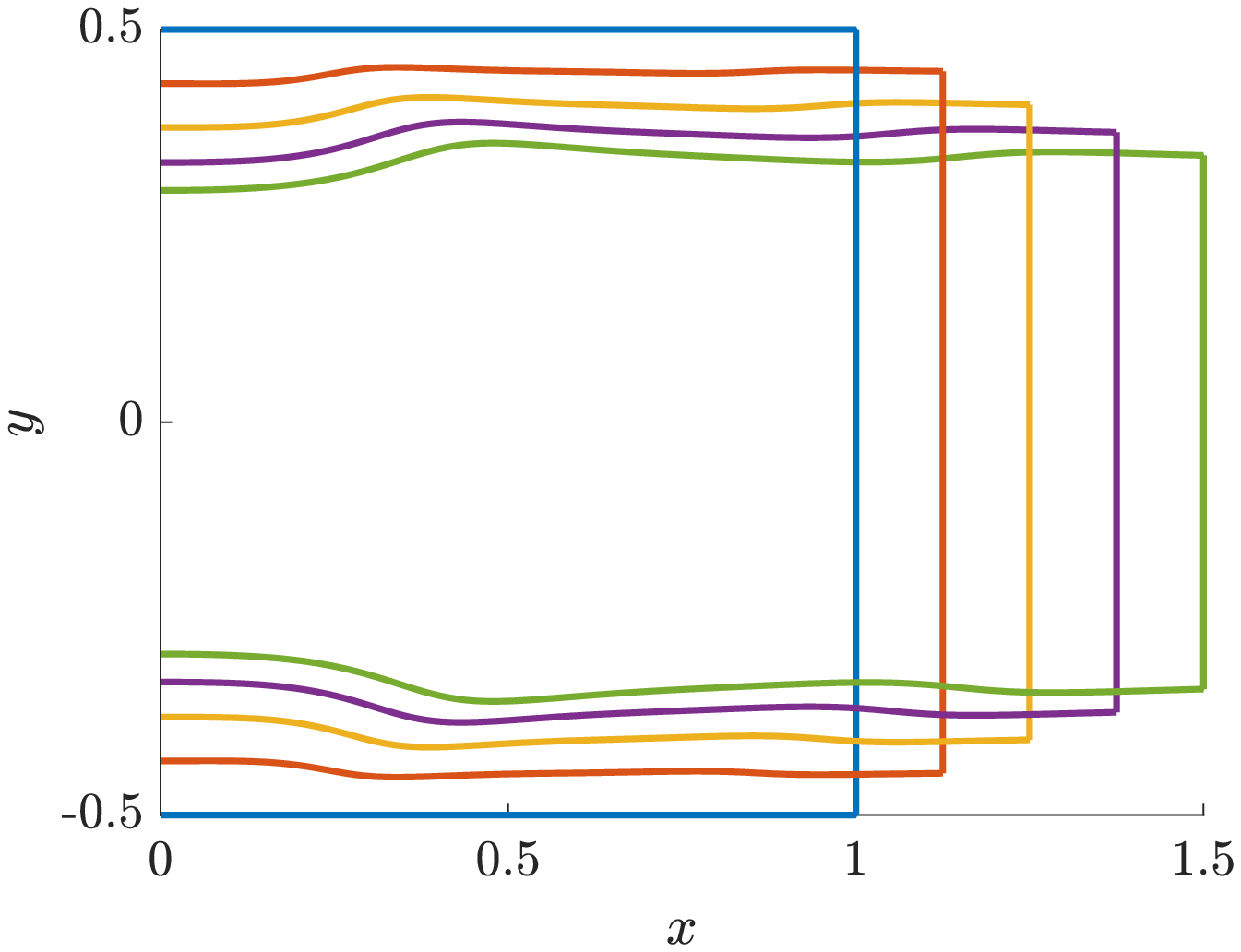}
      \put (3, 73) {a)}
     \put(50,9){\vector(0,1){15}}
\put(35,25){$t$ increasing}
  \end{overpic}
  \begin{overpic}[width=0.49\textwidth]{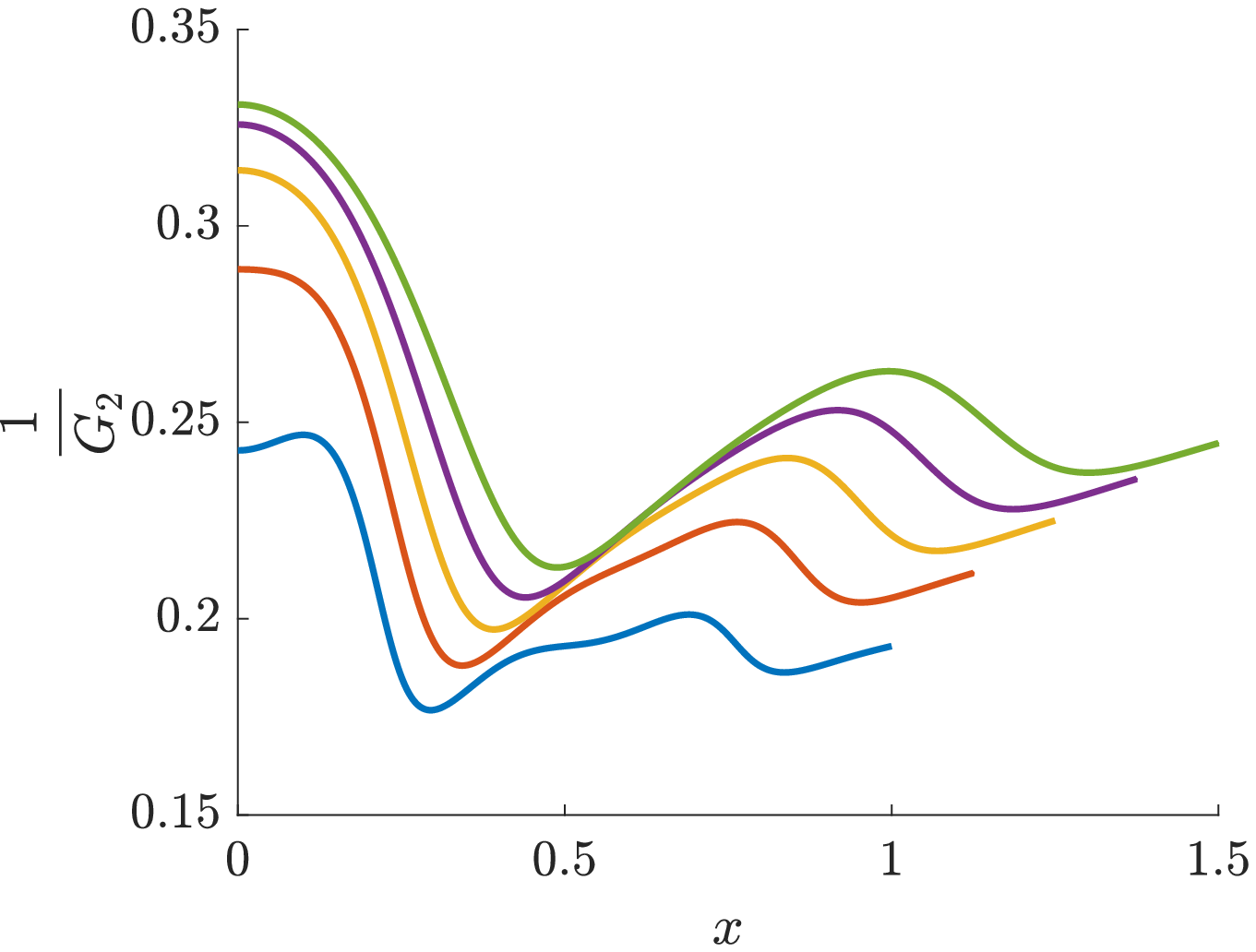}
      \put (3, 73) {b)}
   %   \linethickness{2pt}
\put(52,20){\vector(0,1){22}}
\put(38,44){$t$ increasing}
  \end{overpic}
    \caption{The evolution of a) the thickness of the sheet and b) the function $\dfrac{1}{G_{2}}$ at the times $t=0,0.125,0.25,0.375,0.5$, pulling at $L(t) = 1+t$, with the initial conditions  $h\left(x',0\right) = 1, \theta\left(x',y',0\right) = \cos\left(4\pi x' y'\right) - 0.1, \mu_1 = \mu_3 = 0, \mu_2 = 5$.}
\label{ShortTimeCosine}
\end{figure}

\indent A first integral of \eqref{IntroG2} yields, 
\begin{align}
    h\pd{u}{x'}G_{2}\left(x',\frac{1}{2},t\right)  = T\left(t\right), \label{G2UX}
\end{align}
where $T$ is the tension applied to the sheet. Directly from \eqref{G2UX} we see that $G_2$ may induce non-linear behaviours in $u$. For a Newtonian fluid, or a transversely isotropic fluid where only $\mu_3 \neq  0$, choosing an initial condition of constant thickness across the sheet yields that $u$ must be linear in $x'$. In that case, as shown by Howell, $u=\dot{L}x$ for all time, and $h$ is a function of $t$ only \cite{howell1994extensional} (using a similar approach, the same result was shown for a fluid with $\mu_1 = \mu_2 = 0,$ \cite{green2008extensional}). However, for a transversely isotropic sheet with $\mu_2 \neq 0$, the existence of the trigonometric terms inside $G_{2}$ \changes{prohibits} this. When $\mu_2 > 0 $, with $h\left(x',0\right)$ constant, if $\theta(x',y',0)$ depends upon $x'$, this will result in $u$ becoming nonlinear in $x'$, and by \eqref{h ALE}, this will cause $h$ to become spatially non-uniform. In Figure \ref{ShortTimeCosine}, we illustrate the early evolution of the thickness of the sheet and the behaviour of ${1}/{G_{2}}$, for $h\left(x',0\right) = 1, \theta\left(x',y',0\right) = \cos\left(4\pi x' y'\right) - 0.1, \mu_1 = \mu_3 = 0, \mu_2 = 5$. Notice that the sheet thickness immediately becomes non-uniform, and the location of the peaks and troughs in $1/G_{2}$ correlates with the locations of local minima and maxima of the thickness of the sheet \changes{when $t\neq0$.}
\subsubsection{Effect of varying the extensional and shear viscosities $\mu_2,\mu_3$ with $\mu_1 = 0.$} 
\begin{figure}
    \begin{overpic}[width=0.49\textwidth]{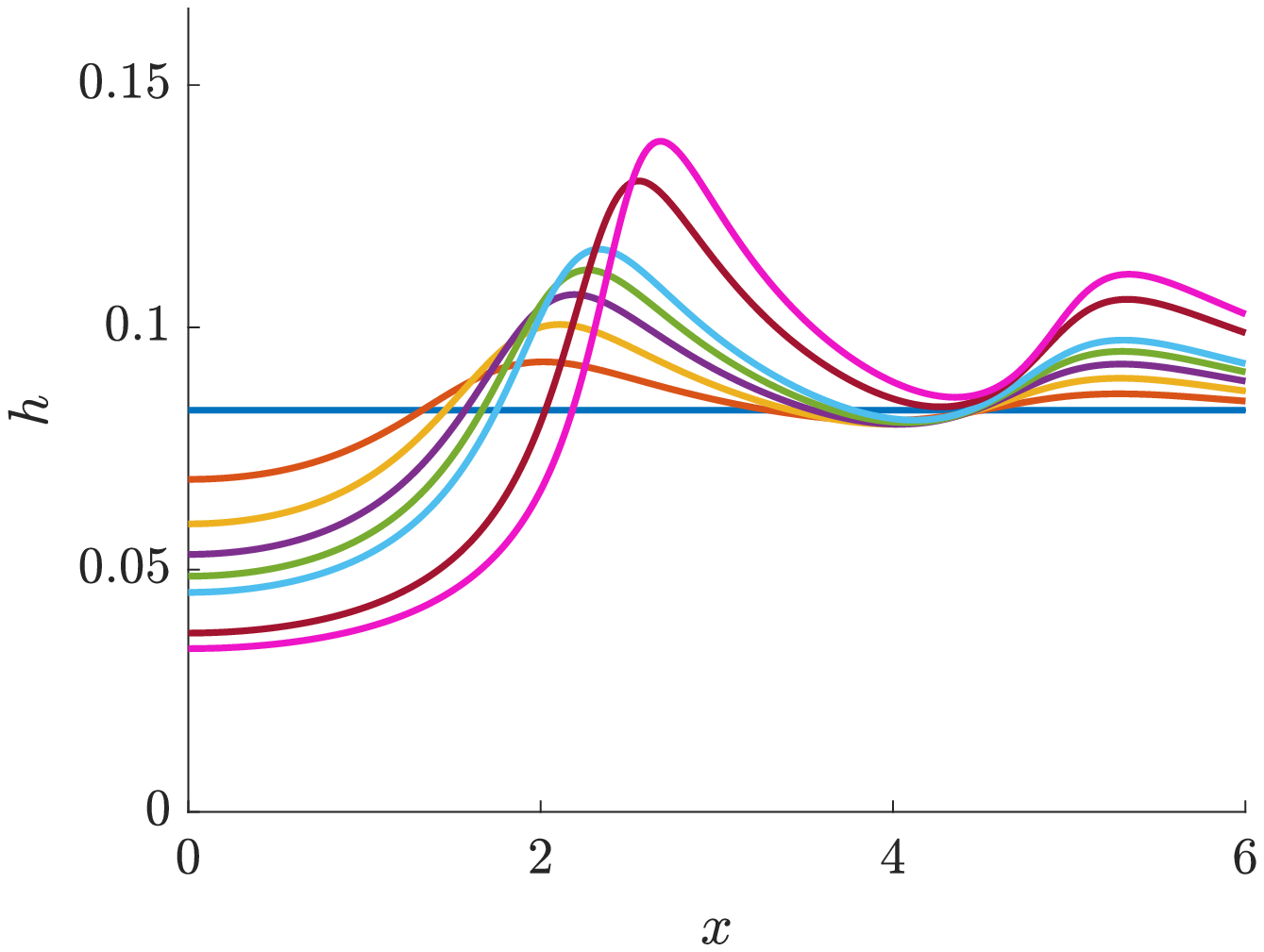}
      \put (3, 73) {a)}
     \put(50,35){\vector(0,1){30}}
\put(48,66){$\mu_2$ increasing}
  \end{overpic}
  \begin{overpic}[width=0.49\textwidth]{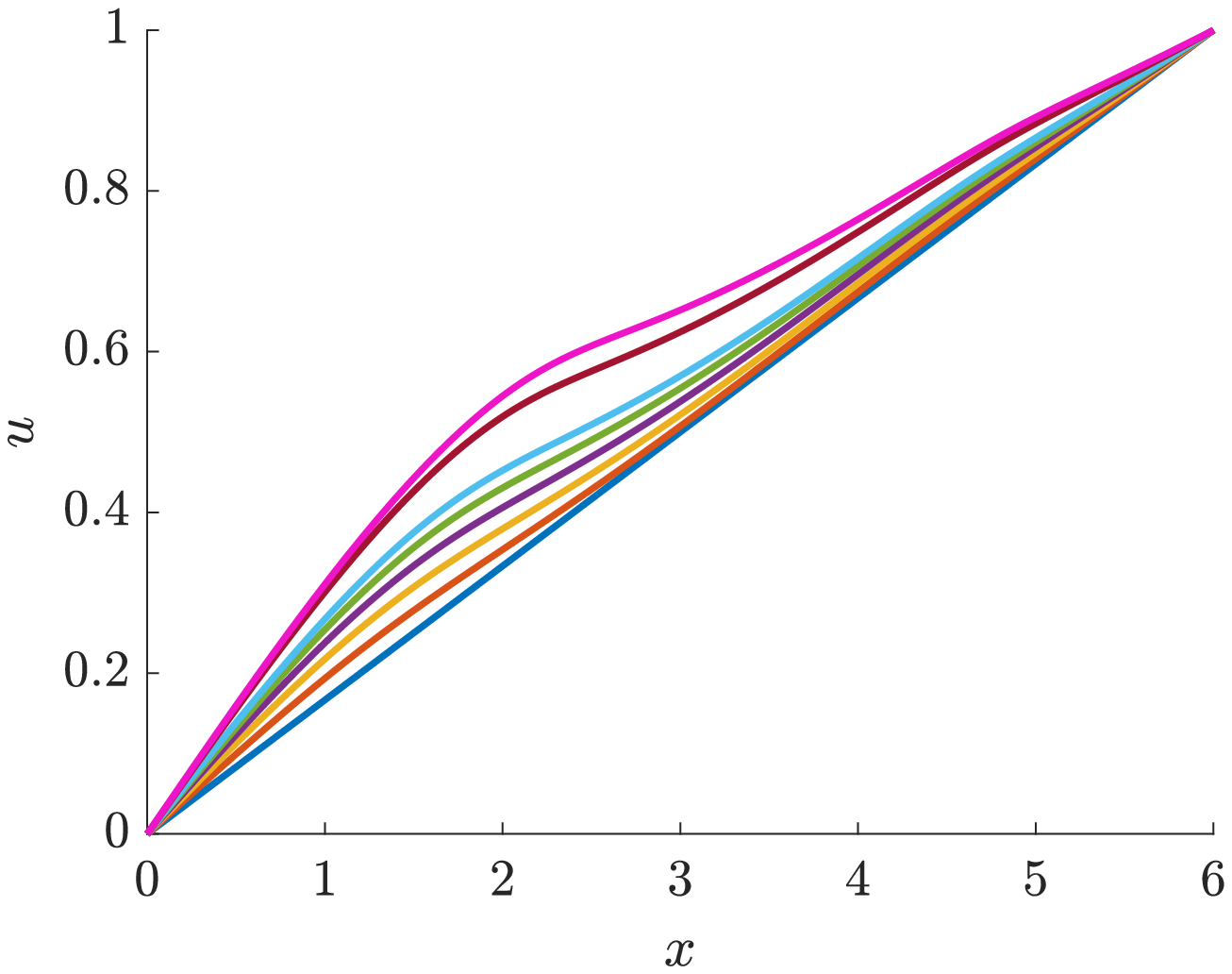}
      \put (3, 73) {b)}
   %   \linethickness{2pt}
\put(52,30){\vector(0,1){35}}
\put(50,67){$\mu_2$ increasing}
  \end{overpic}
  \\
 \centering
 \begin{overpic}[width=0.49\textwidth]{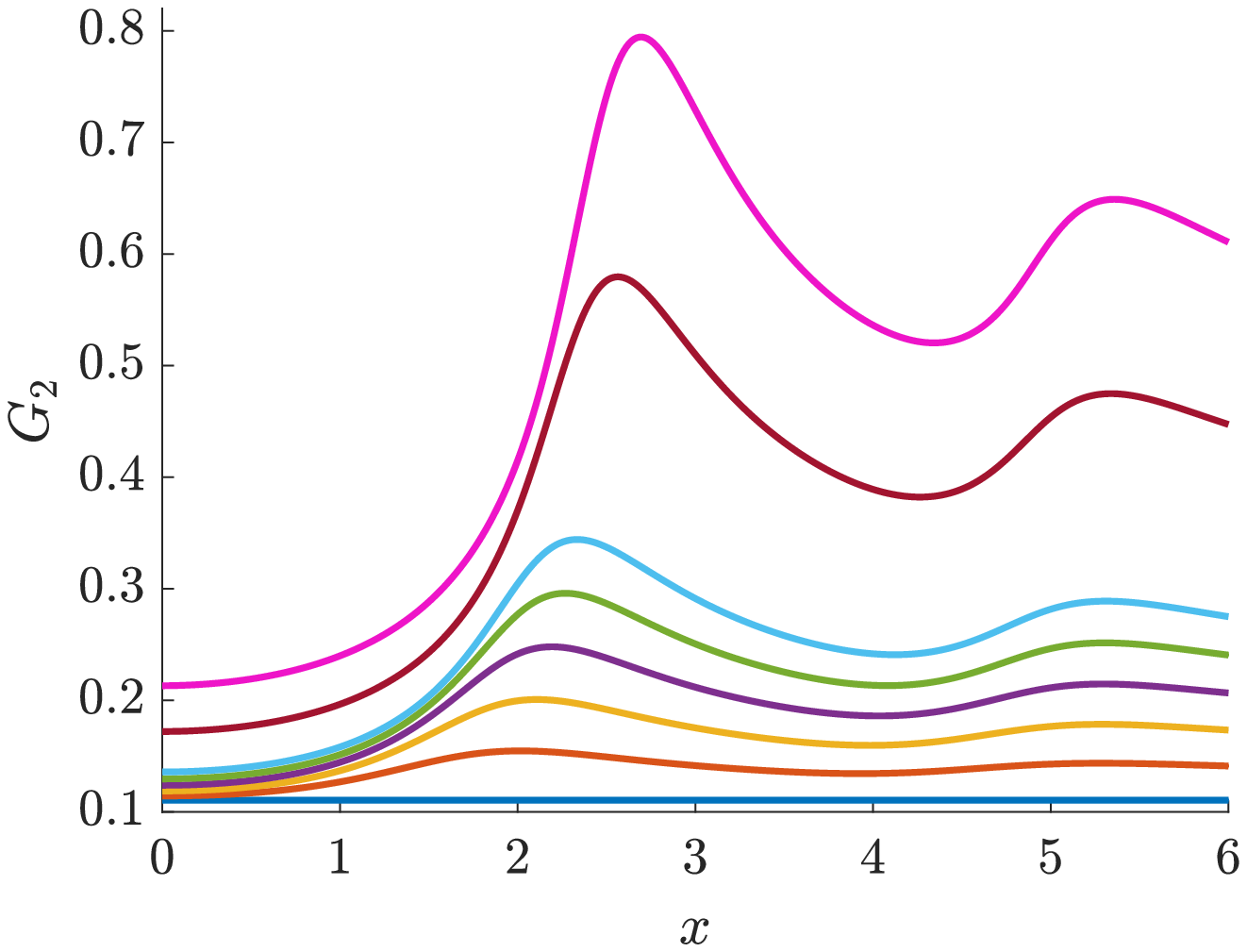}
      \put (3, 73) {c)}
     \put(70,10){\vector(0,1){55}}
\put(68,66){$\mu_2$ increasing}
  \end{overpic}
    \caption{Comparison of a) thickness, b) longitudinal velocity and c) $G_2$ at $t=5$, pulling with $L(t) = 1+t$, for $\mu_2=0,1,2,3,4,5,10,15$, with the conditions $h\left(x',0\right) = 1,\: \theta\left(x',y',0\right) = \cos\left(4\pi x' y'\right) - 0.1,\: \mu_1 =\mu_3 = 0$.  Notice that more extreme behaviour in $G_{2}$ correlates with greater change in the thickness and velocity profile across the sheet and increases with $\mu_2$.}
    \label{VariedCosine}
\end{figure}

In this section, we continue to use the initial conditions $h\left(x',0\right) = 1,\: \theta\left(x',y',0\right) = \cos\left(4\pi x' y'\right) - 0.1, \: \mu_1 = 0$, but now vary $\mu_2$ and $\mu_3$ and compare the state of the sheet at $t=5$. First, we note that varying $\mu_3$ with $\mu_2 = 0$ simply changes the value of the constant obtained from $G_{2}$.
We plot in Figure \ref{VariedCosine} the thickness and velocity $u$ in the sheet for varied $\mu_2$ and notice that for these choices of $\theta(x',y',0)$ and $h(x',0)$ that we see a global increase in the longitudinal velocity for increasing $\mu_2$. Additionally, we see that there are regions of the sheet than thin more quickly for increasing $\mu_2$, and other regions that thin more slowly. Intuition based upon the behaviour of a pipe flow would lead one to expect that in regions where the sheet is thicker, the velocity would be lower. This is not true here, and as in the previous subsection, this behaviour is linked to the behaviour of $G_{2}$ as we shall now demonstrate. %Taking $\mu_3 = 0$, we have
%\begin{align}
 %   G_{2}\left(x',\frac{1}{2},t\right) = \int \limits_{-\frac{1}{2}}^{\frac{1}{2}}   \frac{4\left(4+\mu_2\right)}{4+\mu_2\sin^2 2\theta}  \mathrm{d}y'.
%\end{align}
 Integrating \eqref{G2UX} and using that $u(L) = 1$, we may write
\begin{equation}
    1 = T \int_0^L \dfrac{1}{h G_{2}(x',\frac{1}{2},t)} \mathrm{d}x'. \label{Tension}
\end{equation}
\changes{We note that by \eqref{G2Defn} and \eqref{Tension}, as the fibres within the fluid sheet flatten out and $G_{2}$ increases everywhere and hence so does the tension, $T$, and that the tension will increase with increasing $\mu_2$. Eliminating tension from equation \eqref{Tension} gives a result for $u$,}
\begin{align}
    u = \dfrac{\int \limits_0^{x'} \dfrac{1}{h G_{2}(s,\frac{1}{2},t)} \mathrm{d}s}{\int \limits_0^L \dfrac{1}{h G_{2}(x',\frac{1}{2},t)} \mathrm{d}x'}, \label{Mu2LongVelo}
\end{align}
%We note that by \eqref{G2Defn} and \eqref{Tension}, as the fibres within the fluid sheet flatten out and $G_{2}$ increases everywhere and hence so does the tension, $T$, and that the tension will increase with increasing $\mu_2$.\\
\changes{and so $u$ behaves as a cumulative integral. If $G_2$ is fixed, the longitudinal velocity does behave as a pipe flow intuition would expect (in regions where $h$ is small, $1/h$ is large, and $u$ will increase). 
However, when $G_2$ is not fixed (i.e $\mu_2 \neq 0$), the behaviour of the longitudinal velocity is more complex. The derivative $u_x$ is dependent on the behaviour of $hG_2$. We see this in Figure \ref{VariedCosine}, where we give results for $\mu_3 = 0$ and see that for increasing $\mu_2$, $hG_2$ decreases (despite the increase in $G_2$), which leads to an greater $u_{x}$ on the left hand side of the domain. Where $hG_2$ is larger, around $x=3$, we see the gradient of $u$ decrease below that of $\mu_2 = 0$.\\
}

%\changes{Assuming $\mu_3 = 0, increasing $\mu_2$ has the effect of increasing $G_2$. Despite this, as we see in Figure \ref{VariedCosine}, we see that $hG_2$ decreases, which leads to a greater $u_{x}$ at the left hand side of the domain.
%In Figure \ref{VariedCosine}, we give results for $\mu_3 = 0$ and we see that for increasing $\mu_2$, $hG_2$ decreases (despite the increase in $G_2$), which leads to an greater $u_{x}$ on the left hand side of the domain. Where $hG_2$ is larger, around $x=3$, we see the gradient of $u$ decrease below that of $\mu_2 = 0$.
%We note that by \eqref{G2Defn} and \eqref{Tension}, as the fibres within the fluid sheet flatten out and $G_{2}$ increases everywhere and hence so does the tension, $T$, and that the tension will increase with increasing $\mu_2$.  \\

\begin{figure}
\begin{overpic}[width = 0.49\textwidth]{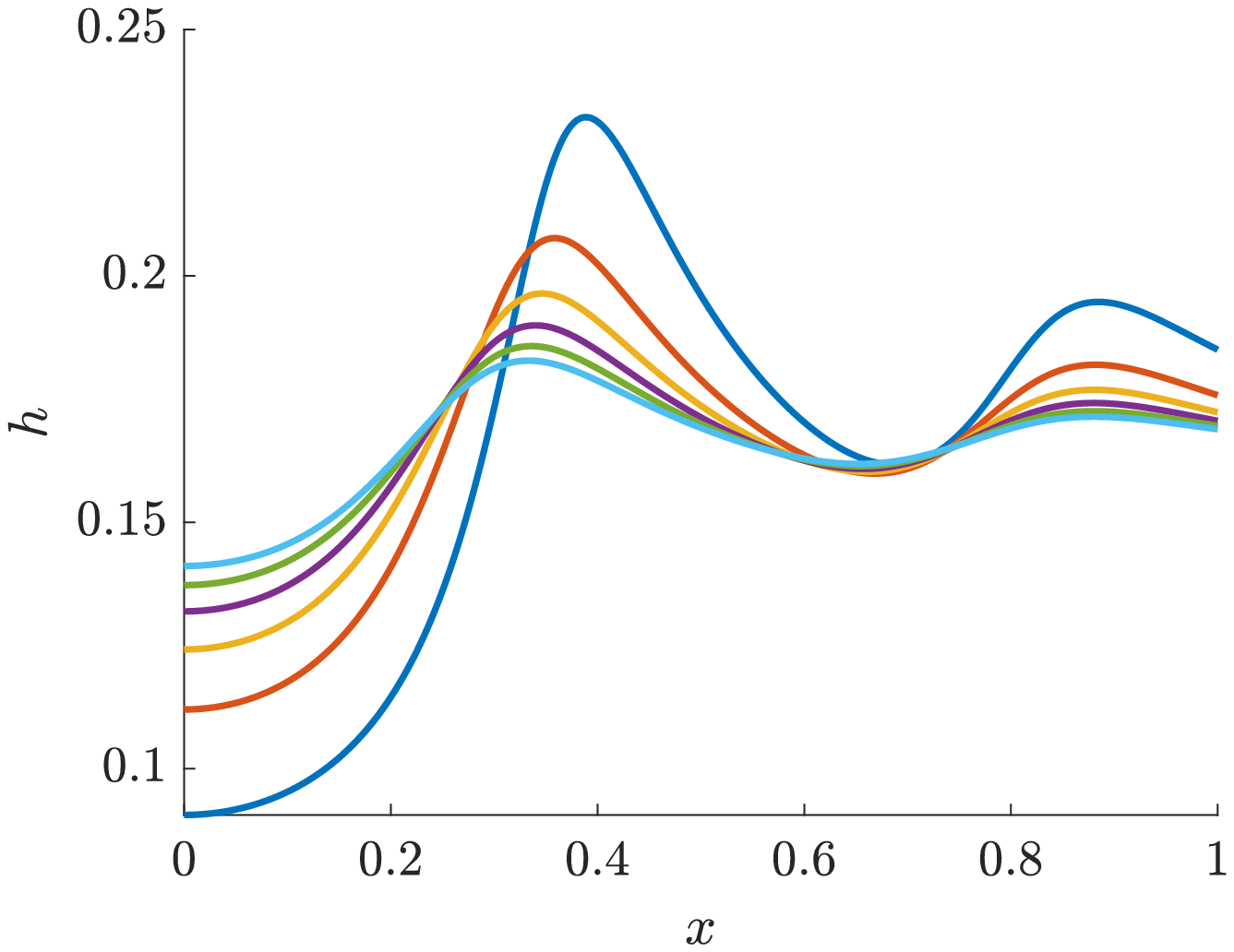}
\put (3,73) {a)}
\put(50,60) {\vector(0,-1){25}}
\put(47.5,32) {$\mu_3$ increasing}
\end{overpic}
\begin{overpic}[width = 0.49\textwidth]{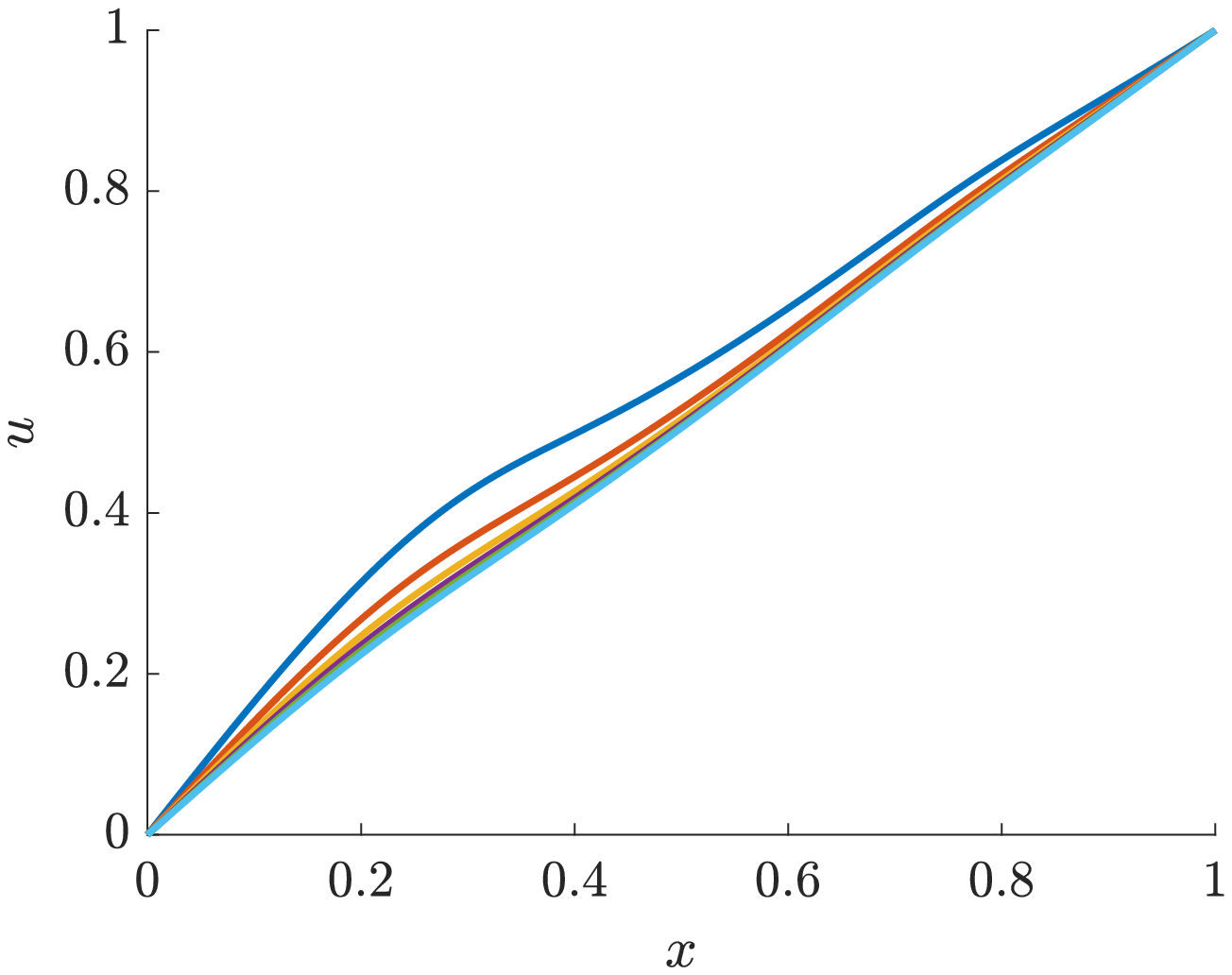}
\put (3,73) {b)}
\put(50,50) {\vector(0,-1){15}}
\put(47.5,32) {$\mu_3$ increasing}
\end{overpic}
\\
\centering
\begin{overpic}[width = 0.49\textwidth]{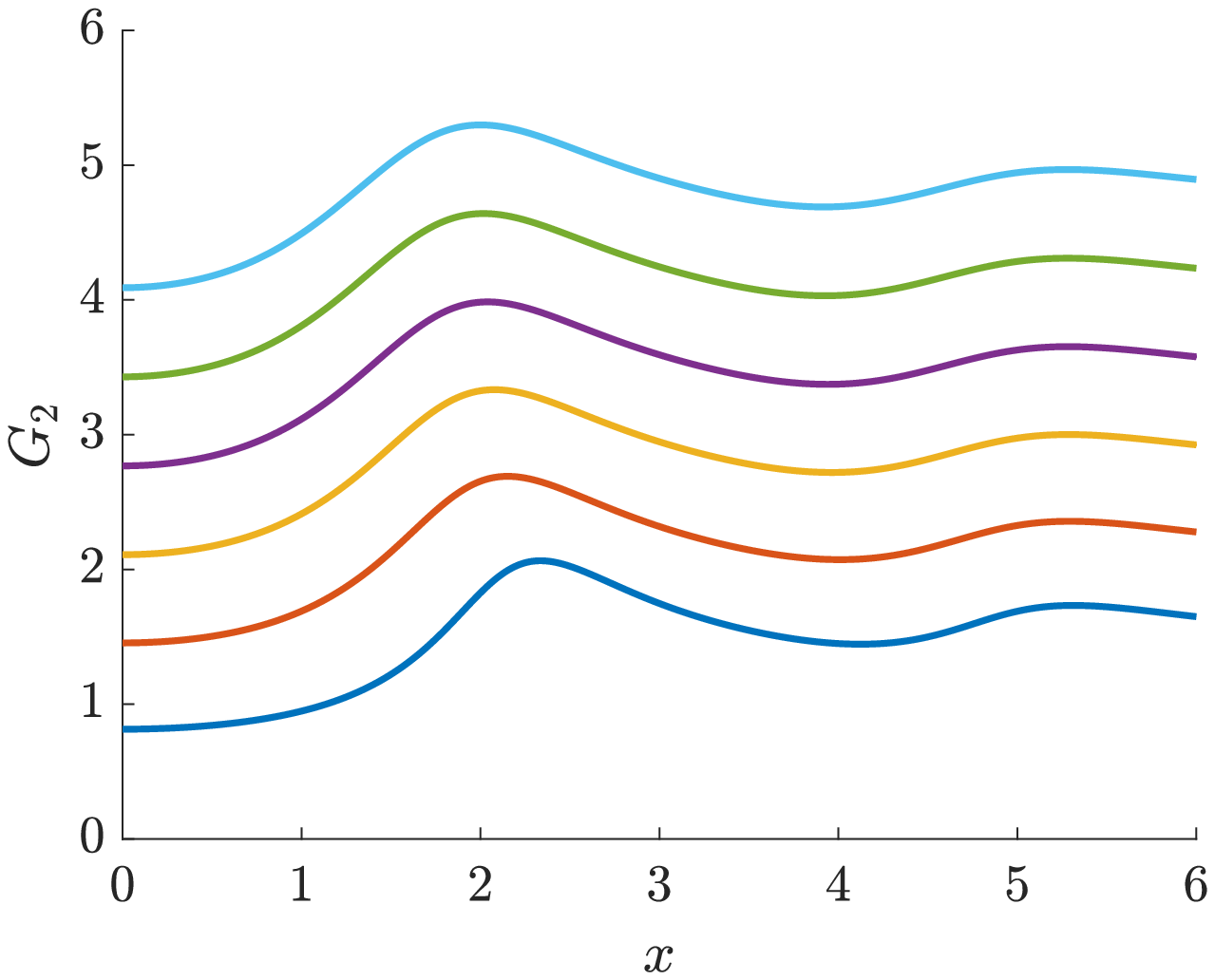}
\put (3,73) {c)}
\put(50,65) {\vector(0,-1){44}}
\put(47.5,18) {$\mu_3$ increasing}
\end{overpic}
\caption{Comparison of a) thickness of the sheet, and b) longitudinal velocity of the sheet at $t=5$ and c) $G_2$ at $t=5$, $\theta(x',y',0) = \cos \left(4\pi x' y'\right) -0.1$, for $\mu_1 = 0,\: \mu_2 =5$, and $\mu_3=0,1,2,3,4,5$.}
    \label{M3Varied}
\end{figure}

\indent If we allow $\mu_2, \mu_3 \neq 0 $, we find that increasing $\mu_3$ has the effect of globally increasing the value of $G_2$ and hence the tension applied to the ends of the sheet, and inhibits the uniformity-breaking behaviour of $\mu_2$.
%, since the parameter $\beta$ in \eqref{U ALE} is decreased. 
That is, increasing the $\mu_3$ term drives the fluid to behave in a `more Newtonian' manner, albeit with a higher tension, whilst increasing $\mu_2$ drives the non-Newtonian behaviour of breaking the uniformity of the sheet previously discussed. As an illustrative example, in Figure \ref{M3Varied} we plot the behaviour of the sheet at $t=5$ for varying $\mu_3$ with $\mu_2=5$ and initial conditions $h(x',0) = 1,\: \theta(x',y',0) = \cos \left(4 \pi x' y' \right) -0.1$. Notice that increasing $\mu_3$ causes the thickness of the sheet to exhibit less deviation from uniformity, and the longitudinal velocity to tend towards $u=x'$, the solution for a Newtonian fluid. 
\newpage 

\subsection{Behaviour of the fibres}
\subsubsection{Uniform initial fibre director field}
We now turn our attention to the \changes{fibre behaviour} within the sheet. In \cite{green2008extensional}, \changes{Green \& Friedman considered two special cases that validate our results.} It is found that for the case of an extensional flow with $\mu_1,\mu_2 = 0$ the fibres align with the direction of extension (i.e parallel to the $x$-axis) \changes{and} that for the special case of not extending the sheet ($\dot{L} = 0$) with $\mu_1,\mu_2,\mu_3 \neq0$, the fibres tend to align parallel to the $y$-axis. \\

\indent For $\mu_1 = 0$, we start by extending the results obtained for early-time with a constant initial fibre direction described in Section 6 of Green \& Friedman \cite{green2008extensional}. If $h_i, \theta_i$ are uniform, then we can see from equations \eqref{Theta Equation Rewrite Summary} and \eqref{U ALE} that $\dfrac{\p u}{\p x'} = 1$, and $\theta$ and $h$ must be functions of time only.  Moreover, \eqref{Theta Equation Rewrite Summary} yields
\begin{equation}
    \frac{\mathrm{d}\theta}{\mathrm{d}t} = \frac{ - \sin 2\theta\left(4+ 4\mu_3 +2 \mu_2 \sin^2\theta\right)\frac{1}{L}}{4+4\mu_3 + \mu_2\sin^2 2\theta}, \label{ThetaPassiveConstant}
\end{equation}
which implies that the behaviour of the fibres is determined by the sign of $\sin 2\theta$. For $0 < \theta < \dfrac{\pi}{2}$, we have $\dfrac{\p \theta}{\p t} < 0$, whilst for $\dfrac{\pi}{2} < \theta < \pi$, $\dfrac{\p \theta}{\p t} > 0$. Therefore, much like the case $\mu_1 = \mu_2 = 0$ studied in \cite{green2008extensional}, the fibres tend to orient themselves with the direction of extension, regardless of the value of $\mu_2$, given uniform initial conditions for $h$ and $\theta$.
\\
Additionally, we note that $\theta = \pm \frac{\pi}{2}$ are unstable fixed points of equation \eqref{ThetaPassiveConstant}, whilst $\theta = 0,\pi$ are stable fixed points.
\\

\begin{figure}
\begin{overpic}[width = 0.49\textwidth]{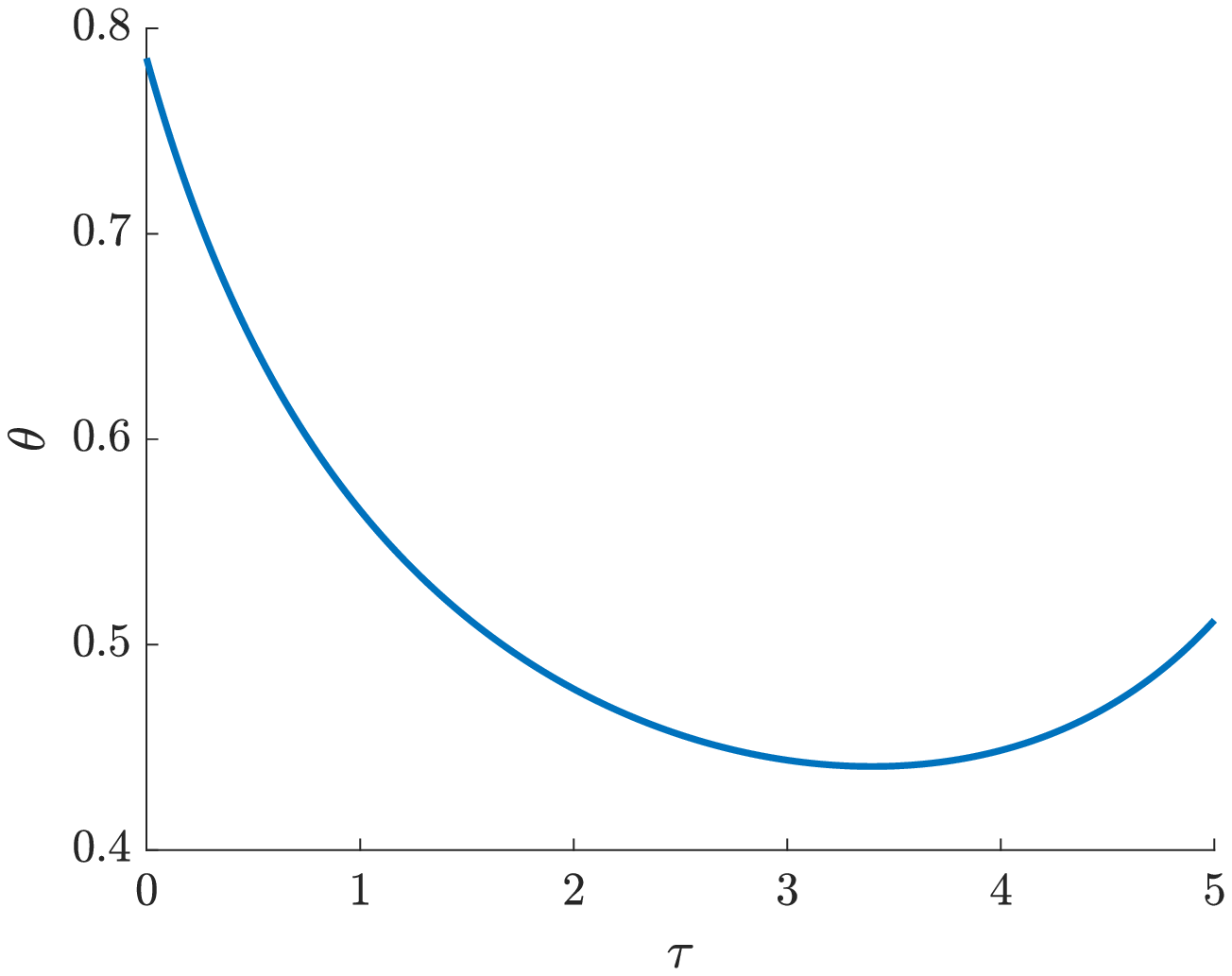}
\put (3,73) {a)}
\end{overpic}
\begin{overpic}[width = 0.49\textwidth]{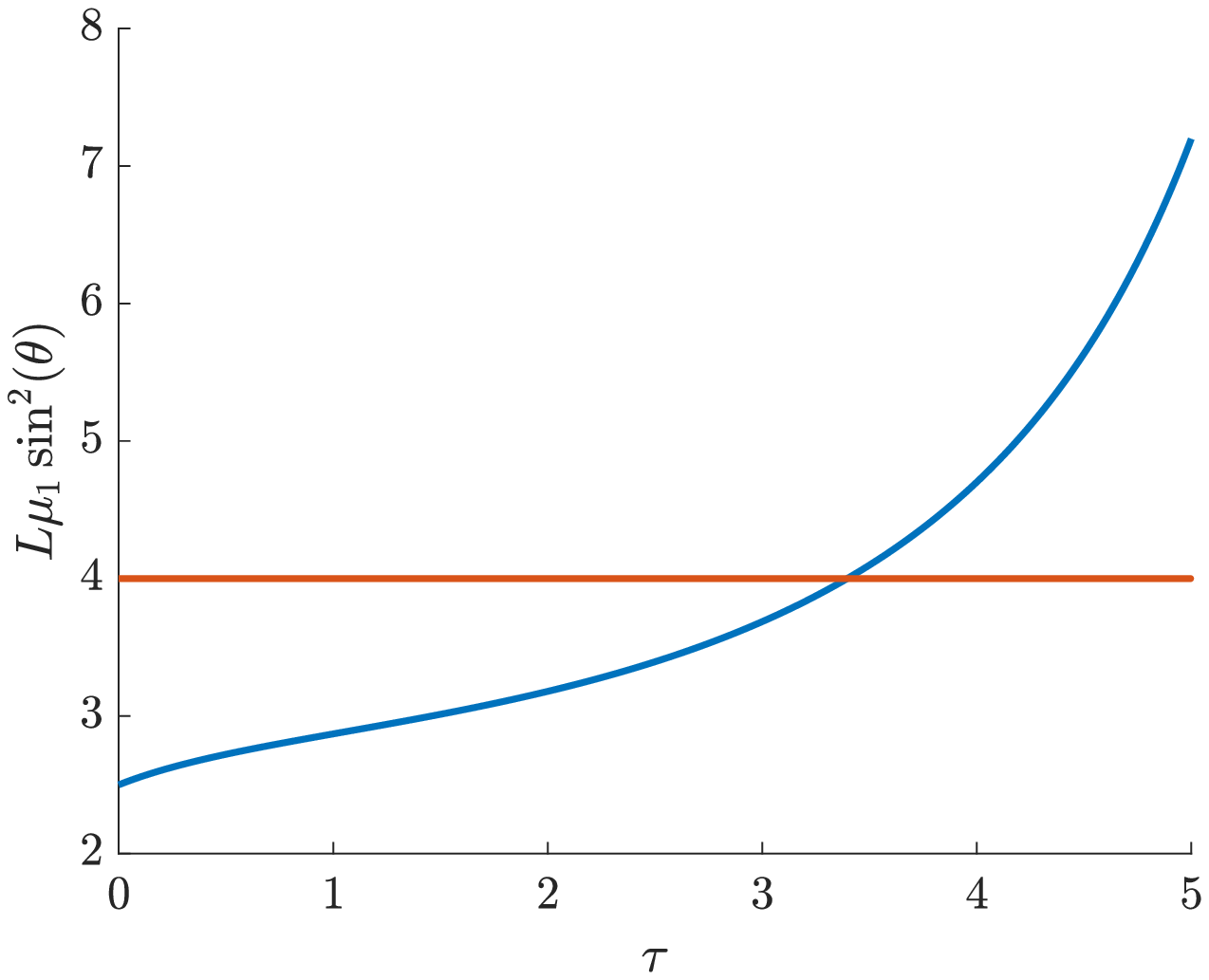}
\put (3,73) {b)}
\end{overpic}
    \caption{Evolution of a) the fibre angle $\theta$, and b) the left hand side of \eqref{DirectionCondition} (blue), with the threshold $2+2\mu_3,$ (red), for the choices of $\theta_i = \frac{\pi}{4}, L_i = h_i = 1, \mu_1 = 5, \mu_2 = 0, \mu_3 =1$. Note the reversal in the direction of rotation when $L\mu_1\sin^2\theta \geq 2+2\mu_3$. }
    \label{Sheet:Fig:Mu1ConstTheta}
\end{figure}

\indent We now consider the case $\mu_1 \neq 0$, for an initially constant $h_i$ and $\theta_i$. We again have that $\dfrac{\p u }{\p x'} = 1$ and $h,\theta$ remain uniform for all time. However, equation \eqref{Theta Equation Rewrite Summary} now yields  
\begin{align}
\frac{\mathrm{d}\theta}{\mathrm{d}t}= \frac{ \sin 2\theta \left( 2\mu_1\sin^2 \theta - \left(4+ 4\mu_3 +2 \mu_2 \sin^2\theta\right)\frac{1}{L}\right)}{4+4\mu_3 + \mu_2\sin^2 2\theta}, \label{ThetaActiveConstant}
\end{align}
and so the evolution of the fibre direction is less clear. By again considering the cases $0 < \theta < \dfrac{\pi}{2}$, $\dfrac{\pi}{2} < \theta < \pi$, $-\dfrac{\pi}{2} < \theta < 0 $, $-\pi < \theta < -\dfrac{\pi}{2}$
we find that in order for the fibres to align along \changes{the} $x$ axis with time, we require 
\begin{align}
    \sin^2 \theta \left(L\mu_1 - \mu_2\right) < 2+2\mu_3. \label{DirectionCondition}
\end{align}
If this condition is satisfied, fibres with angles between $-\dfrac{\pi}{2} < \theta < \dfrac{\pi}{2}$ will rotate towards the positive $x$-axis,  with fibres outside of this range rotating towards the negative $x$-axis, similarly to the above. However, since this expression includes $L(t)$, it is possible for fibres that initially rotate towards the longitudinal orientation can reverse their evolution as $L\mu_1$ grows with time to violate \eqref{DirectionCondition}. We illustrate this behaviour in Figure \ref{Sheet:Fig:Mu1ConstTheta}. For the choices of $\theta_i = \frac{\pi}{4}, L_i = h_i = 1, \mu_1 = 5, \mu_2 = 0, \mu_3 =1$, we give the evolution of the fibre angle and evolution of the left hand side \eqref{DirectionCondition}.\\% Initially, the fibres rotate towards the $x$-axis, until the inequality \eqref{DirectionCondition} is violated, at which time the fibres reverse their rotation and begin to rotate towards $\theta = \frac{\pi}{2}$. \\

\subsubsection{Non-uniform initial fibre angles}
\begin{figure}
\begin{overpic}[width = 0.49\textwidth]{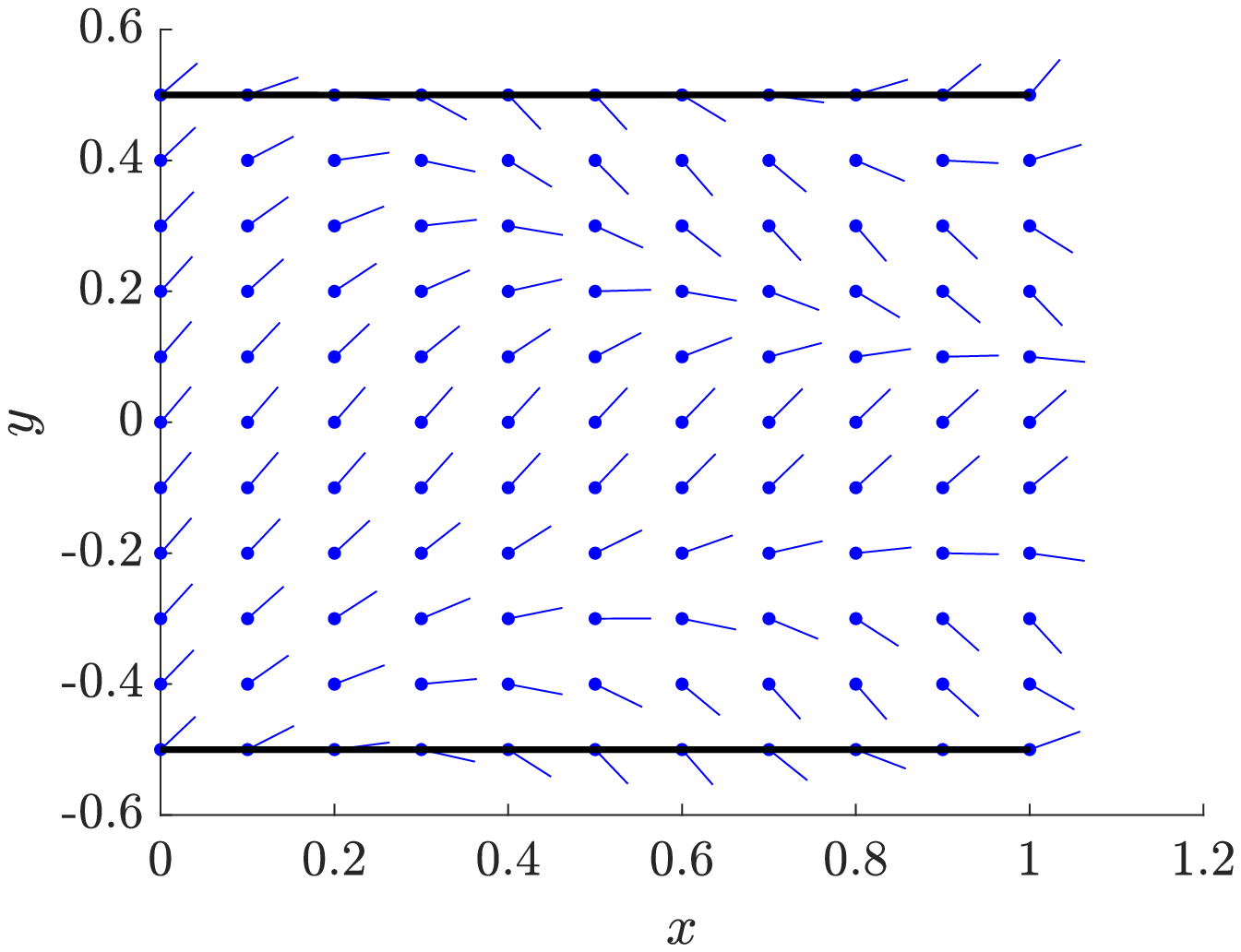}
\put (0,75){a)}
\end{overpic}
\begin{overpic}[width = 0.49\textwidth]{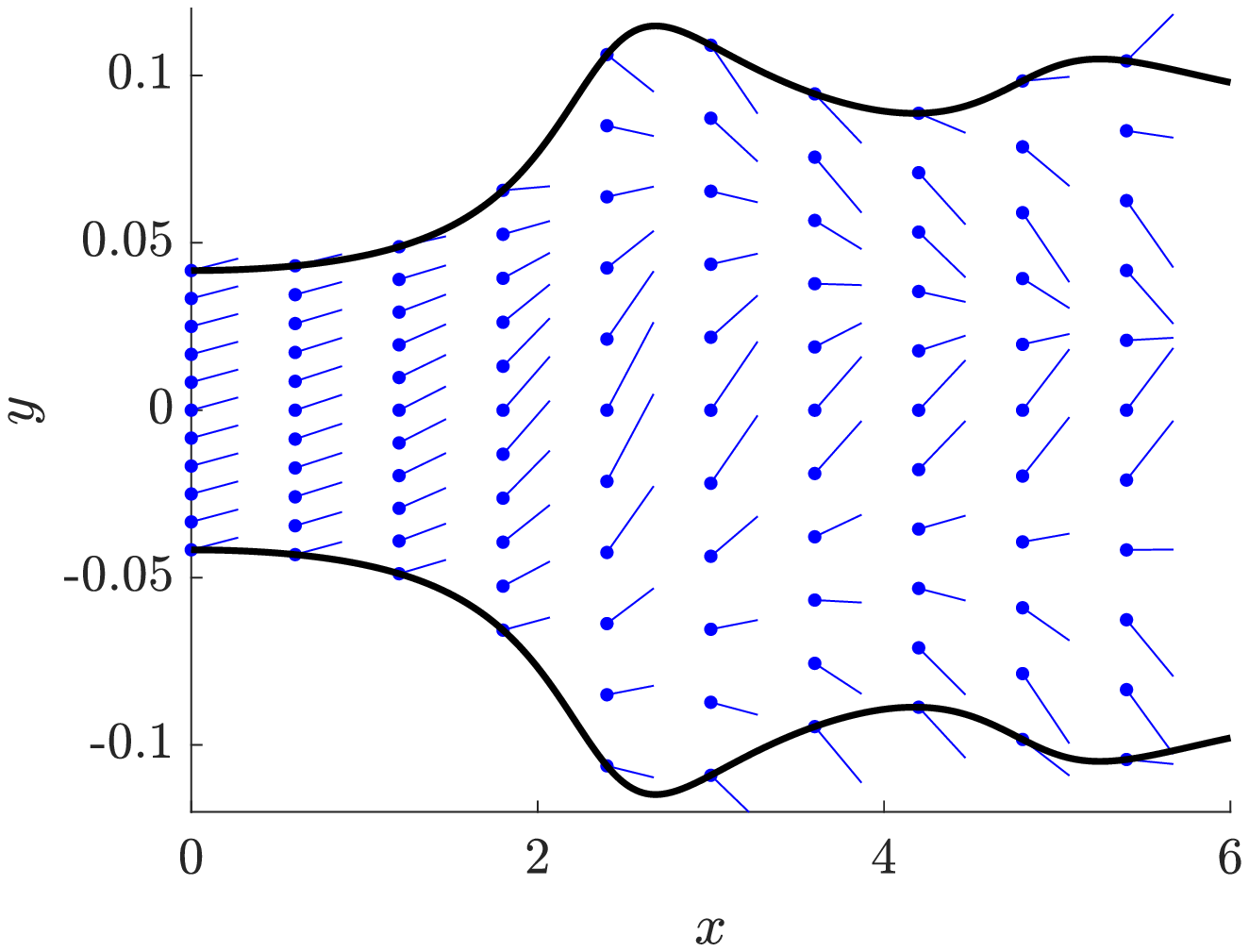}
\put (3,75){b)}
\end{overpic}
    \caption{a) The initial thickness of the sheet and orientation of the fibres, and b) the thickness of the sheet and orientation of the fibres at $t=5$, for the initial conditions $h(x',0) = 1,\theta(x',y',0) = \cos(4\pi x' y') - 0.1$ and choices of $\mu_1=\mu_3 = 0, \mu_2=5.$} 
    \label{FibreRealignment}
\end{figure}

Consider now the extension of the sheet with $\theta(x',y',0) = \cos(4\pi x' y') - 0.1$, $h(x',0)=1$ and $L(t) = 1+t$. We see in Figure \ref{FibreRealignment} that the fibres have a tendency to align in the direction of the sheet when $\mu_1 = 0$. We also note that the rate at which the fibres align in this direction is enhanced in the regions of the sheet that undergoes fastest thinning of the fluid. 

\subsection{\changes{Special cases for sheets possessing active behaviour}} % tension in the fibre direction, $\mu_1\neq 0$. }
\begin{figure}
\begin{overpic}[width = 0.49\textwidth]{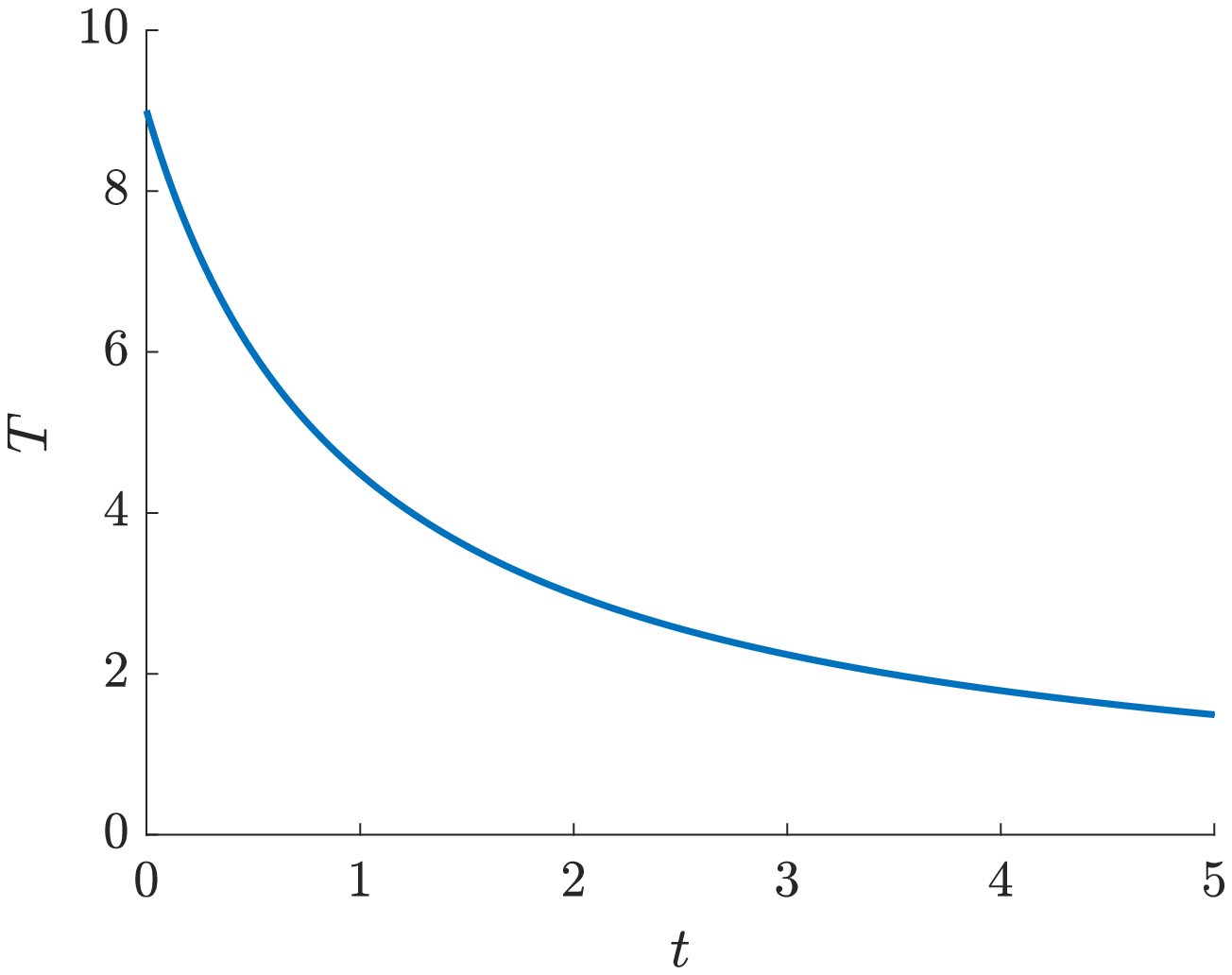}
\put (3,73) {a)}
\end{overpic}
\begin{overpic}[width = 0.49\textwidth]{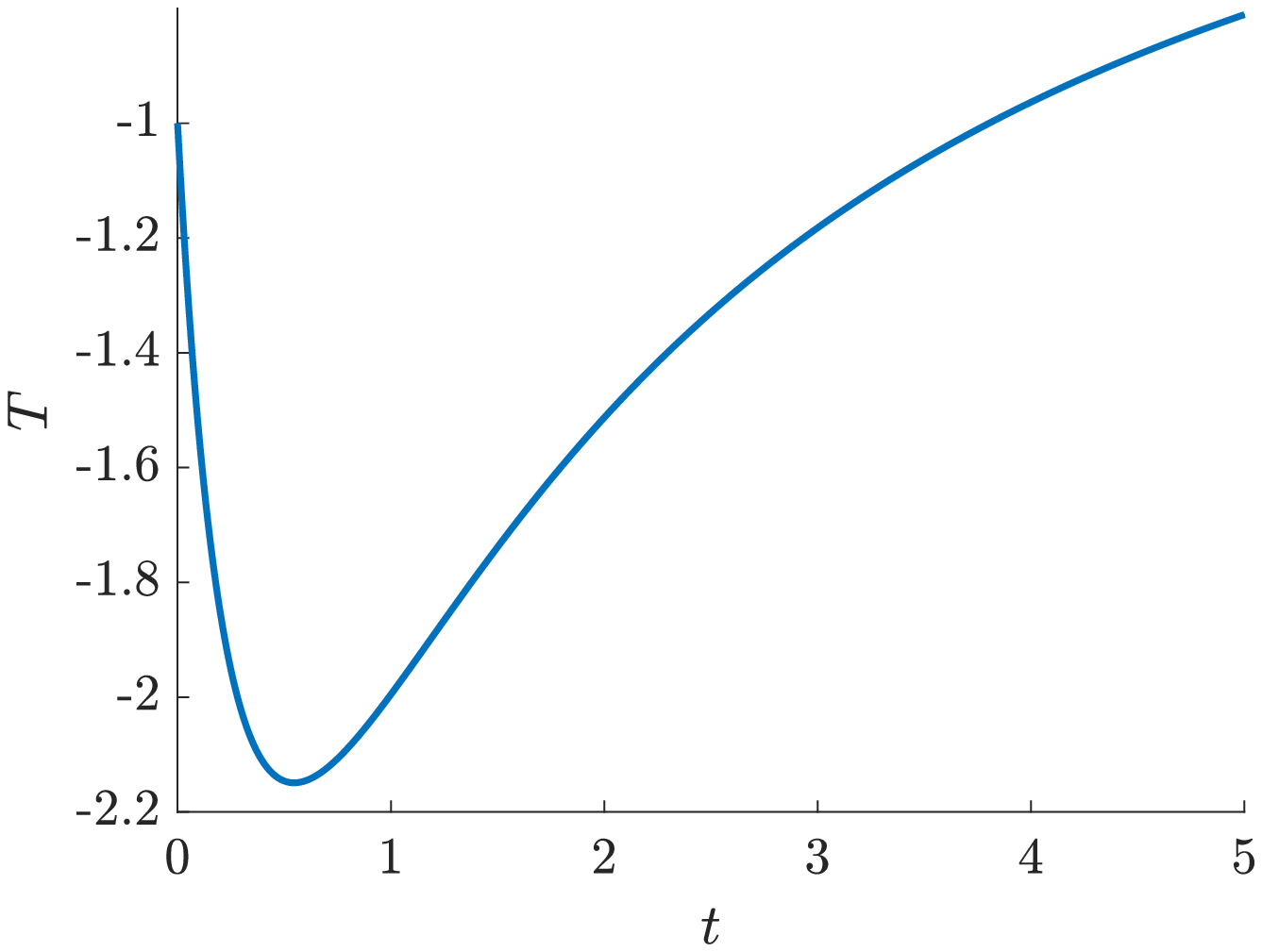}
\put (3,73) {b)}
    \end{overpic}
    \caption{Evolution of the tension applied to the sheet for two initial choices of fibre direction, a) $\theta_ i = 0$, and b) $\theta_ i = \dfrac{\pi}{2}$ up to $t=5$ with $h(x',0) = 1,\: \mu_1 = 5,\: \mu_2=\mu_3 = 0$, and prescribed pulling $L=1+t$.}
    \label{TensionTheta}
\end{figure}

%\subsubsection{Two special cases}
Here, we briefly examine two special cases for the sheet for which $\mu_1 > 0$. We start by prescribing $L(t) = 1+t$, $h_i = 1$, $\mu_1 = 5$, $\mu_2 = \mu_3 = 0$, and we test two \changes{choices for the initial fibre angle,} $\theta_i = 0$, corresponding to the fibres being aligned in the direction of extension, and $\theta_i = \dfrac{\pi}{2}$, corresponding to the fibres being aligned in the transverse direction in the sheet. In Figure \ref{TensionTheta} we plot the tension required to be applied to the sheet to achieve the prescribed rate of pulling for these two choices. First, we note that there is no evolution in $\theta$ with time. Next, we notice that for fibres arranged in the transverse direction of the sheet, that the tension applied to the sheet is negative.
To explain this we begin with the equation for $u$ in spatial variables,
\begin{align}
    T = \int_{H^{-}}^{H^{+}} \frac{\mu_1 \cos2\theta + \left(4 + 4\mu_3 + \mu_2 \right) \pd{u}{x}}{4+4\mu_3+ \mu_2 \sin^2 2\theta} \mathrm{d}y,
\end{align}
now since $\theta$ has no evolution for these particular choices of $\theta_i$ (as they are fixed points of equation \eqref{Theta Equation ALE}), then for $\theta_i = \dfrac{\pi}{2}$,
\begin{align}
    T = \frac{1}{4\left(1+\mu_3\right)}\left( \left(4+4\mu_3+\mu_2 \right) \pd{u}{x} - \mu_1\right) h.
\end{align}
It is possible that $\mu_1 > \left(4+4\mu_3+\mu_2 \right) u_x$ and hence $T<0$. We interpret this as being due to the fibres in the transverse direction attempting to contract the sheet, which due to mass \changes{conservation,} would generate a compression in the longitudinal direction, as the sheet attempts to extend longitudinally. If we define the total tension required to move the sheet at the prescribed speed as $T_L$, the tension caused by the fibres as $T_f$ and the tension applied to the sheet as $T$, then we expect that
\begin{equation}
    T_{L} = T+T_{f}.
\end{equation}
If the rate of extension is too slow to compensate for the compression generated by the fibres pulling in the transverse direction, then $T<0$. As discussed by Howell for the Newtonian case, \cite{howell1994extensional}, when the sheet is in compression we expect buckling to occur, and that the curvature of the centre--line will in time become significant. Thus, in this case, the nearly-straight centre--line scaling assumed in the Green \& Friedman model may be violated, and the model \changes{will no longer be valid.} 
%\\
%In this case the sheet is now undergoing a compression, and the Green and Friedman model is no longer valid, as the model contains the assumption that the centre–line is nearly straight. However, the argument above would lead us to expect that the overall tension would now become positive should the sheet be pulled fast enough. 

\subsection{Behaviour of the centre--line}
\subsubsection{Conditions for a flat centre--line}
In the Newtonian problem, it was shown by Howell \citep{howell1994extensional} that the centre--line of the sheet straightens on a timescale shorter than $\dfrac{L_{0}}{U}$ \cite{buckmaster1975buckling,howell1994extensional} (and is therefore generally taken to simply be $H =0$). This is not necessarily true for a transversely-isotropic fluid. It is noted in \citep{green2008extensional} that should $\theta = \theta(x',t)$ and $\mu_1=\mu_2=0$, then equations \eqref{U ALE} and \eqref{H ALE} together with the requirement that $H(0,t)=H(L(t),t)=0$, imply that the centre--line must be flat. We will now demonstrate that this is also true when $\mu_1, \mu_2 \neq 0$. Supposing $\theta = \theta(x',t)$, then the integrand of equations \eqref{U ALE} and \eqref{H ALE} can be evaluated explicitly, yielding
\begin{align}
    \pd{}{x'}\left( h f\right) &= 0, \label{CentrelineOne}
    \\
        \Pd{}{x'}\left( \frac{1}{2} h^2 f \right) &= \left(\Pd{H}{x'} + \frac{1}{2}\Pd{h}{x'}\right)hf,\label{CentrelineTwo}
\end{align}
where 
\begin{equation}
    f(x',t) = \frac{\mu_1\cos2\theta + \frac{1}{L}\left(4+4\mu_3 + \mu_2\right) \pd{u}{x'}}{4+4\mu_3 + \mu_2 \sin^2 2\theta}.
\end{equation}
Expanding the second derivative on the LHS of equation \eqref{CentrelineTwo} and using \eqref{CentrelineOne} yields that 
\begin{align}
    \Pd{H}{x'} = 0, \label{Hxx0}
\end{align}
and therefore $H$ must be a straight line. It is then possible to choose our co-ordinate system such that $H(x',t) \equiv 0$. \\
\\
\indent Now, let us consider what happens when $\theta(x',y',0) = \theta_{i}(y')$, and $\mu_1 = 0$. In this case the equation of the centre--line can be written as
\begin{align}
   \left(\Pd{H}{x'} + \frac{1}{2}\Pd{h}{x'}
    \right) h G_{2}\left(\frac{1}{2},t\right) = \Pd{}{x'} \left( h^2 \int\limits_{-\frac{1}{2}}^{\frac{1}{2}} G_{2}(y',t) \mathrm{d}y'\right). \label{CentrelinePassive}
\end{align}
We recall that, from Section \ref{Section:IUTI}, if we choose $\theta_i = \theta_i(y')$ only and the initial thickness $h_i$ to be uniform, then $G_2$ will not possess $x'$-dependence and $h$ will also remain uniform for all time. As a result we once again obtain \eqref{Hxx0}, so that asymmetry in the fibre angles over the centre--line is not sufficient to cause an initially uniform sheet with $\mu_1 = 0$ to deflect. We will now consider the conditions under which the centre--line of the sheet will not be straight.

%As we will see in Section \ref{Section:NZCentre}, the centre--line of a sheet will not (in general) be straight if either $\theta$ possesses $y'$-dependence with $h(x',0)$ that is non-uniform, or with $\mu_1 \neq 0$, or $\theta$ possessing both $x'$ and $y'$ dependencies.
\subsubsection{Conditions for a nonzero centre--line} \label{Section:NZCentre}
\begin{figure}
\centering
\begin{overpic}[width = 0.49\textwidth]{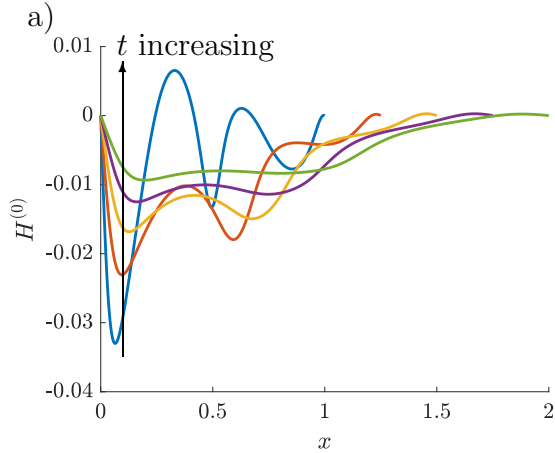}
\put (3,73){a)}
\put (19,17){\vector(0,1){50}}
\put (18,68){$t$ increasing}
\end{overpic}
    \caption{Evolution of the centre--line of the sheet under the initial conditions $h(x',0) = 1,\: \theta(x',y',0)=\sin\left(4\pi x' y'\right) -0.1$ with $\mu_2 = 5,\: \mu_1=\mu_3=0$. Plotted at $t=0, 0.25, 0.5, 0.75,1$.}
    \label{Fig:PassiveCentreline}
\end{figure}

For sheets possessing $\mu_1 = 0$ and a uniform condition for the thickness of the sheet, $h_i$, the choice of $\theta(x',y',0)=\theta_i(x',y')$ \changes{being a prescribed function that is non-symmetric (i.e. $\frac{\p \theta}{\p y'}\vert_{y'=0} \neq 0$) is required to obtain centre--line deflection.} In Figure \ref{Fig:PassiveCentreline} we give a plot of the centre--line evolution for the initial conditions $\theta(x',y',0) = \sin\left(4\pi x' y'\right) -0.1$ with $\mu_2=5,\: \mu_1=\mu_3 = 0$. \changes{We see that the centre--line initially possesses deflection and tends to $0$. This behaviour is driven by the right hand side of \eqref{CentrelinePassive} becoming small.} Once this occurs, the behaviour of the centre--line is then dominated by the $\Pd{h}{x'}$ term, which itself possesses an implicit dependence on $G_{2}$ through $u$. Additionally, for a transversely isotropic fluid with $\mu_1 = 0$, the deflection is small and decays quickly. Since the centre--line does not instantaneously collapse to $0$, this suggests that considering the behaviour of the fluid on a short timescale may yield some interesting behaviour that is markedly different from a Newtonian fluid. 
\\

%If we consider a condition for $h(x',0)$ that is not uniform, we find that there will exist a small deflection when $\theta(x',y',0)= \theta_i(y')$ only, since endowing $h$ with $x'$-dependence will generate a non-linear $u$, hence $u_{x}$ will gain $x'$-dependence through equation \eqref{U ALE} and hence $\theta$ will gain $x'$-dependence after the first time step through equation \eqref{Theta Equation Rewrite Summary}.

If we consider a condition for $h(x',0)$ that is not uniform, we find that there will exist a small deflection when $\theta(x',y',0)= \theta_i(y')$ only, since endowing $h$ with $x'$-dependence will impart $x'$-dependence on $\pd{u}{x}$ through equation \eqref{U ALE}, and hence $\theta$ will gain $x'$-dependence after the first time step through equation \eqref{Theta Equation Rewrite Summary}.

\section{Results \changes{on the short timescale}} \label{Section:TIBNTResults}
\changes{We now examine the behaviours of the sheet over a short timescale of $t \sim \varepsilon^2 \dfrac{L_0}{U}$. For a Newtonian fluid, it was demonstrated by Howell \cite{howell1994extensional} that the leading order equations for the extensional flow of a slender, viscous, Newtonian sheet predict that the centre--line of the sheet is straight. Hence, the model cannot satisfy an initial condition in which the centre--line is not straight. In order to study the behaviour of initially curved sheets, a short timescale analysis is performed. The required timescale must be $\eps^2 \frac{L_0}{U}$ \cite{buckmaster1975buckling}. In the transversely isotropic problem it is similarly impossible to satisfy an arbitrary initial condition for the centre--line and, as we have demonstrated, there exist choices of the key parameters and initial distribution of fibre angles that give rise to a centre--line that is non-zero on the flow timescale. This indicates that there may exist interesting behaviours, different from the Newtonian case, over a short timescale.}

\subsection{Short-time evolution of the centre--line}
%We have now established that equation \eqref{H ALE} permits a curved centre--line on the $\dfrac{L_0}{U}$ timescale, unlike for the Newtonian case. In order to study the behaviour of a film with an arbitrary initial centre--line, we consider a short timescale of $\varepsilon^2 \dfrac{L_0}{U}$.\\ 

\begin{figure}
\begin{overpic}[width = 0.49\textwidth]{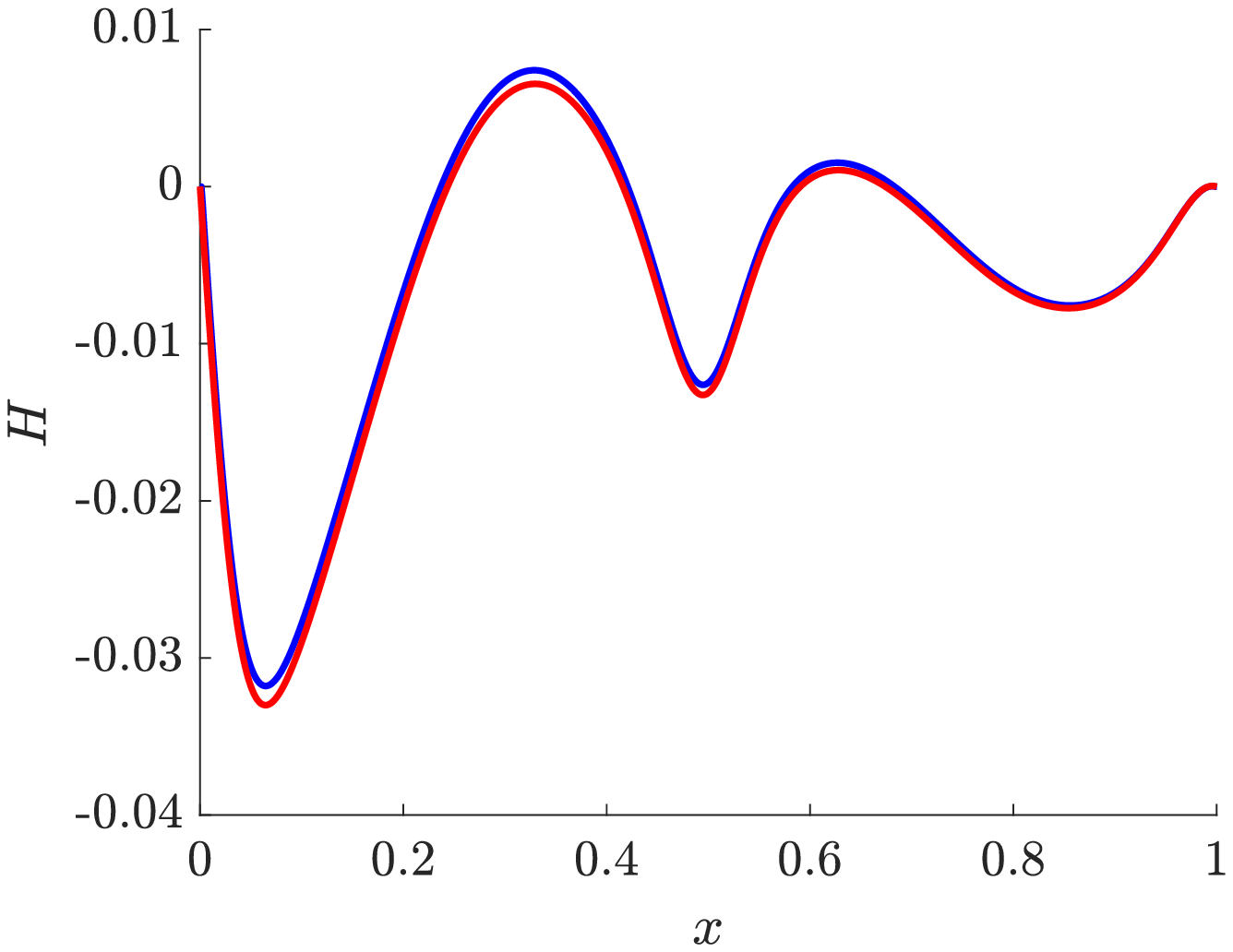}
\put (3,73) {a)}
\end{overpic}
\begin{overpic}[width = 0.49\textwidth]{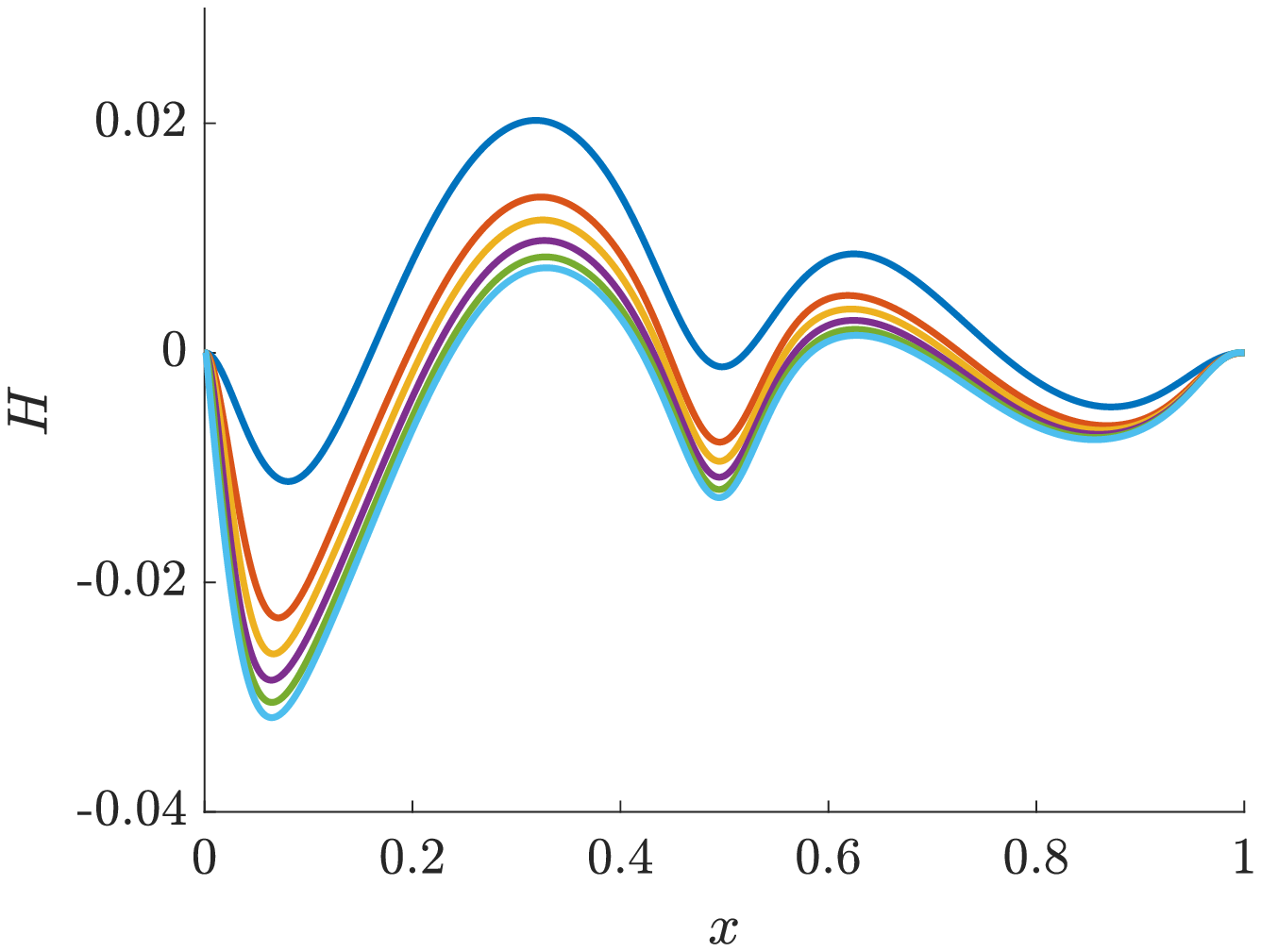}
\put (3,73) {b)}
\put (52.5,55) {\vector(0,-1){25}}
\put (51.5,56){$\tau$ increasing}
    \end{overpic}
    \\
\centering
\begin{overpic}[width = 0.49\linewidth]{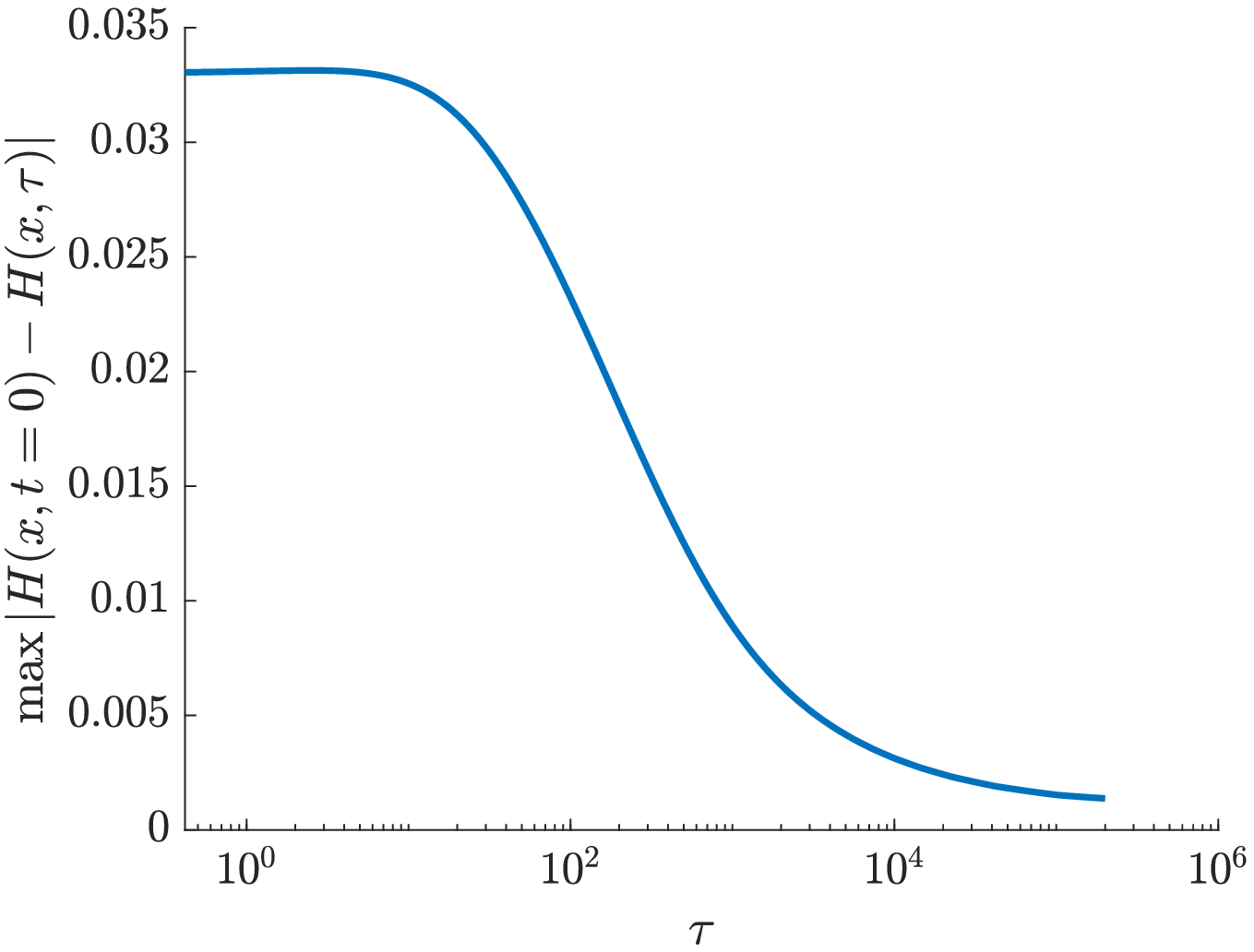}
    \put (3,73) {c)}
\end{overpic}
\begin{overpic}[width = 0.49\linewidth]{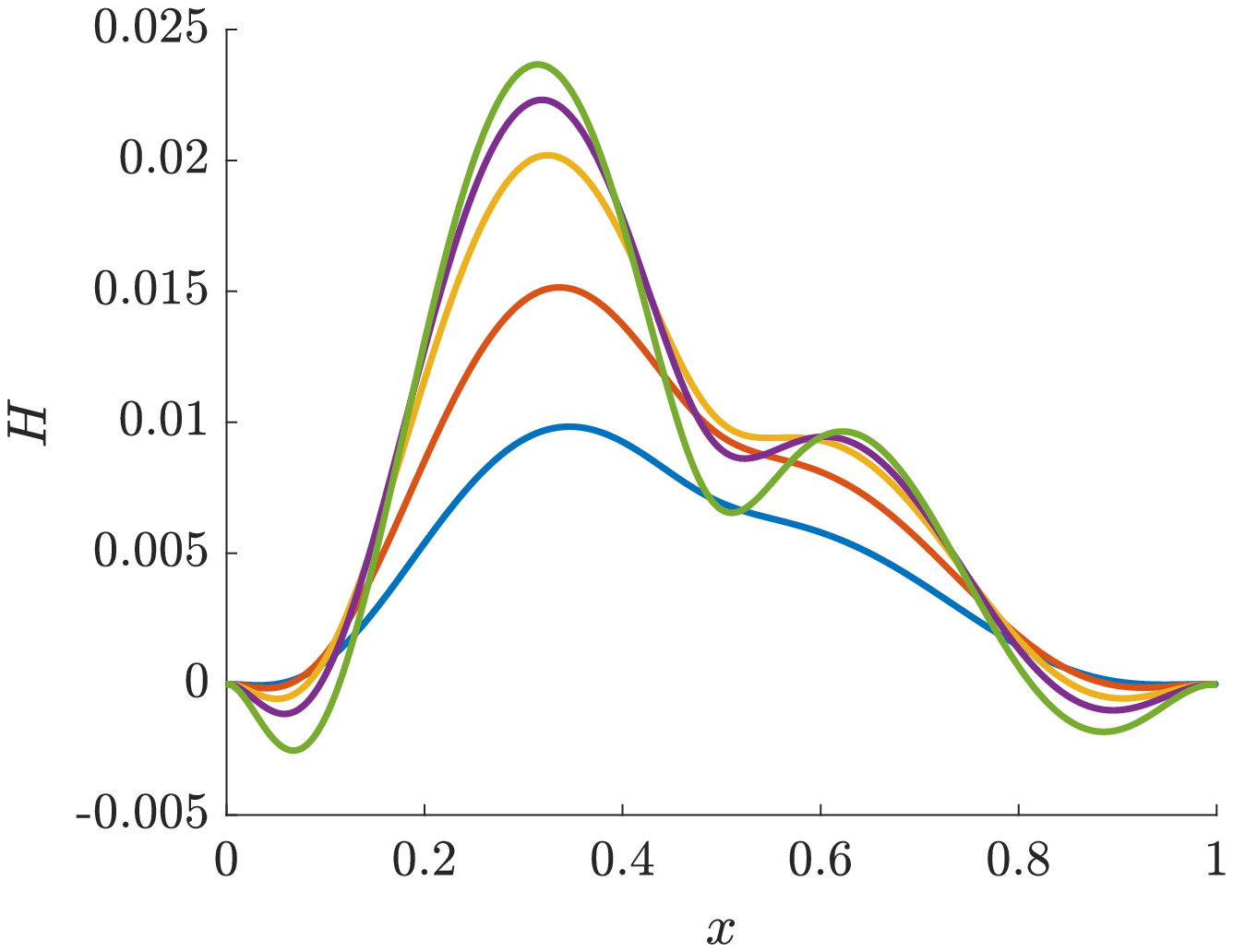}
    \put (3,73) {d)}
    \put (40,30) {\vector(0,1){40}}
\put (39,27.5){$\tau$ increasing}
\end{overpic}
    \caption{Position of the centre--line a) at $t=0$ (red) and $\tau=200000$ (blue) and b) at $\tau=100,500,1000,2500,10000,200000$, where $\tau =0$ is omitted as $H(x,0) = 0$, In c) we plot the maximum of the absolute difference between the evolving centre--line on the short timescale and the Green \& Friedman result and in d) we give the centre--line at $\tau = 2.5,5,10,15,25$.
    The conditions for all of the above are $H(x,0) = 0,\: h(x,0) = 1,\: \theta(x,y,0) = \sin(4\pi x y ) - 0.1,\: \mu_1 = \mu_3 = 0,\: \mu_2 = 5$,\: $L(0) = 1$.}
    \label{Fig:BNTFigures}
\end{figure}

First, we check that the long-time behaviour of the short time model is consistent with the Green \& Friedman model. In Figure \ref{Fig:BNTFigures}, we give the evolution of the centre--line over the short timescale for the initial conditions of $\mu_1 = \mu_3 = 0,\: \mu_2 = 5$, $L(0) = 1$, $\theta(x,y,0) = \sin(4\pi x y) - 0.1$, ${H^{(0)}(0,\tau) = H^{(0)}(L,\tau) = \dfrac{\p H^{(0)}}{\p x}(L,\tau) = \dfrac{\p H^{(0)}}{\p x}H(L,\tau) = 0}$, up to $\tau = 200000$, and compare the values of $H(x,\tau = 200000)$ with $H(x,t=0)$. We see that the short time-scale result closely matches the result produced by the solver for the Green \& Friedman \changes{model.} In Figure \ref{Fig:BNTFigures}b we give the evolution of the centre--line over $\tau$. Notice in Figure \ref{Fig:BNTFigures}b the centre--line converges to the Green \& Friedman model fairly quickly, and there is a small absolute difference between the centre--line at $\tau=10000$, and $\tau = 200000$.\\

\indent For a Newtonian fluid, Howell was able to obtain a analytical expression for the decay of the centre--line of an initially curved sheet undergoing stretching by use of eigenfunction expansions \cite{howell1994extensional}, assuming the same boundary conditions we use in this section, $H(0,\tau) = H(1,\tau) = \dfrac{\p H}{\p x}(0,\tau) =  \dfrac{\p H}{\p x}(1,\tau) =0$. It is found that the centre--line decays exponentially to $H=0$. The behaviour for a transversely isotropic fluid is more complex. 
The results plotted in Figure \ref{Fig:BNTFigures} began with the initial condition $H(x,\tau = 0) = 0$, and we immediately see from Figure \ref{Fig:BNTFigures}c that the convergence to the Green \& Friedman centre--line is not exponential for all time. Indeed, there is an initial lag, as the centre--line evolves away from the flat initial condition to adopt a similar shape of the centre--line predicted by the Green \& Friedman model, before decaying to the expected result. As an illustrative example, we include Figure \ref{Fig:BNTFigures}d. This figure shows the formation of the peaks and troughs of the general shape of the centre--line given by the Green \& Friedman model. 

%\\

\subsection{Effect of key parameters upon convergence}
\begin{figure}
\centering
\begin{overpic}[width = 0.49\linewidth]{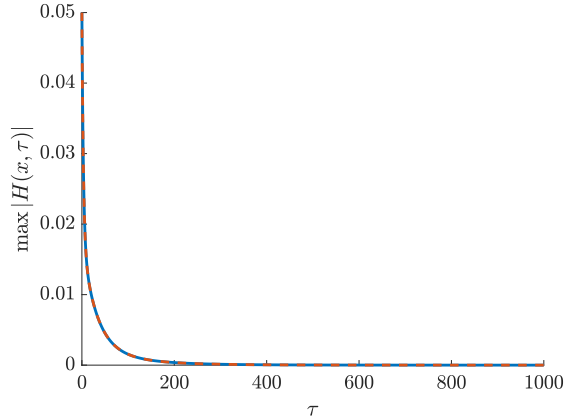}
%\put (3,73) {a)}
%\put (40,70){\vector(0,-1){30}}
%\put (38.5,37){$\tau$}
\end{overpic}
    \caption{The decay of the maximum value of $H(x,\tau)$ for $\mu_3 = 0$ (blue) and $\mu_3 = 10$ (red) with the conditions $H(x,0) = \frac{1}{20} e^{-\frac{\left(x-0.5\right)^2}{0.01}},\: \mu_1 = \mu_2 = 0,\: h(x,0) = 1,\: L =1$.}
    \label{Fig:M3GaussianDecay}
\end{figure}

In this subsection we discuss the effect of increasing $\mu_2,\mu_3$ upon the convergence of the short timescale centre--line to the result from the Green \& Friedman model. First we note that if $\mu_2 = 0$, $\mu_3$ has no effect upon convergence. In Figure \ref{Fig:M3GaussianDecay} we give a plot of the decay of the maximum value of $H(x,\tau)$ for the initial conditions of $H(x,0) = \frac{1}{20} e^{-\frac{\left(x-0.5\right)^2}{0.01}}, \theta(x,y,0) = \sin(4\pi x y), h(x,0) = 1, L =1$. There is no difference in the decay of the centre--line to flat between the Newtonian case and $\mu_3 = 10$.\\

\begin{figure}
\begin{overpic}[width = 0.49\linewidth]{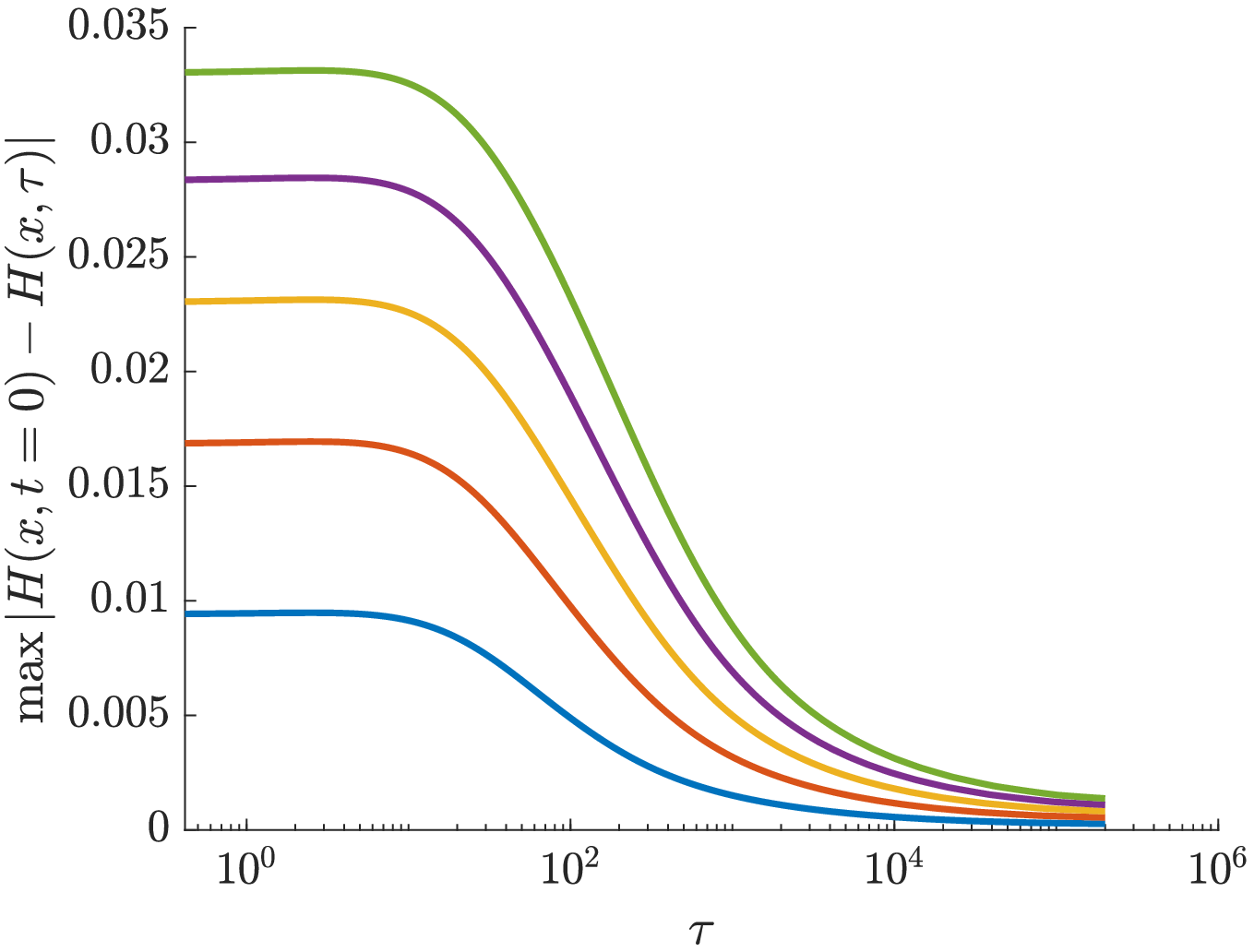}
    \put(3,73){a)}
    \put(20,20){\vector(0,1){50}}
       \put(18,72){$\mu_2$ increasing}
\end{overpic}
\begin{overpic}[width =0.49\linewidth]{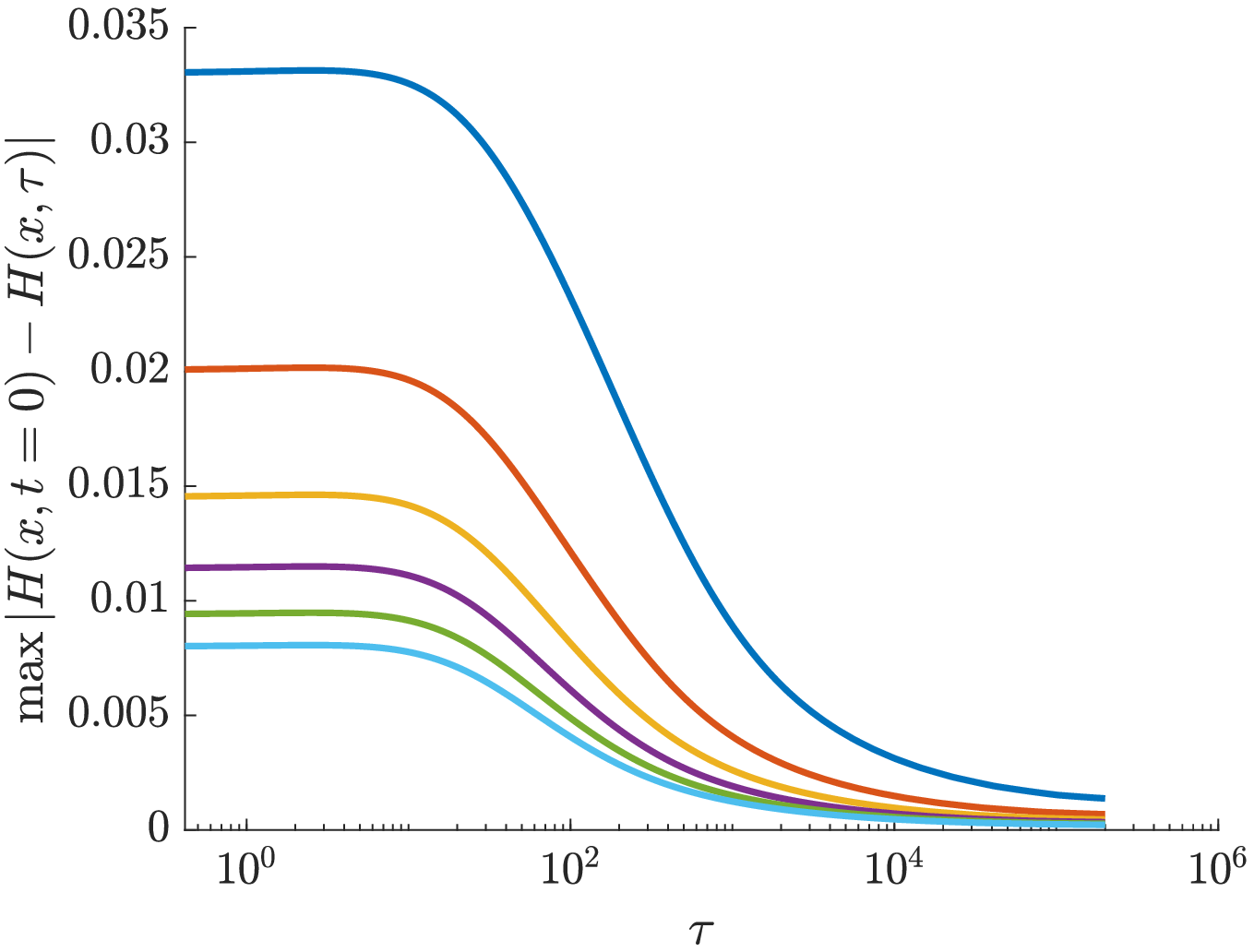}
    \put(3,73){b)}
      \put(20,63){\vector(0,-1){46}}
      \put(15.5,14.5){$\mu_3$ increasing}
\end{overpic} 
    \caption{Maximum absolute difference across the sheet between the centre--line on the short timescale and the result obtained from the Green \& Friedman model \eqref{H ALE} for a) $\mu_3=0$,\: $\mu_2 = 1,2,3,4,5$ and b) $\mu_2=5$,\: $\mu_3 = 0,1,2,3,4,5$, with the conditions $H(x,0) = 0,\: h(x,0) = 1,\: \theta(x,y,0) = \sin(4\pi x y ) - 0.1, \: \mu_1 = 0,\: L(0) = 1$. }
    \label{Fig:BNT:VariedMu2Mu3}
\end{figure}

\indent Now choosing $\mu_2 > 0$, in Figure \ref{Fig:BNT:VariedMu2Mu3} we plot the maximum absolute difference across the sheet between the centre--line on the short timescale and the result obtained by solving equation \eqref{H ALE} for varied values of $\mu_2,\mu_3$ with the conditions $h(x,0) = 1, \theta(x,y,0) = \sin(4\pi x y ) - 0.1, \mu_1 = 0, \mu_2 = 5,L(0) = 1$. First, in Figure \ref{Fig:BNT:VariedMu2Mu3}a, we fix $\mu_3 = 0$ and vary $\mu_2$ (note the case of $\mu_2 =5$ corresponds to the example given above). We see that as $\mu_2$ increases, the initial difference between the flat initial condition of the centre--line and the result of the Green \& Friedman model also increases. This is due to the deepening of the trough around $x=0.5$ seen in Figure \ref{Fig:BNTFigures}a. We note that changing $\mu_2$ does not appear to affect the length of the lag as the centre--line is adopting the correct general shape before decaying, but as $\mu_2$ increases, the decay is faster.
In Figure \ref{Fig:BNT:VariedMu2Mu3}b, we fix $\mu_2 = 5$ and vary $\mu_3$ (as before, the case of $\mu_3 = 0$ corresponds to the example given above). Once again, we see that the effect of increasing $\mu_3$ has little affect upon the convergence of the centre--line, and that the behaviour we see here is a consequence of the role of $\mu_3$ in moderating the effects of $\mu_2$ in the Green \& Friedman model as discussed in Section \ref{Section:GreenFriedman}.

\newpage
\section{Discussion} \label{Section:Discussion}
%--------------------------\section{Discussion two}
In this paper we have constructed \changes{and employed} a numerical strategy to solve the model proposed by Green \& Friedman for the extensional flow of a thin two-dimensional sheet of a fibre-reinforced fluid, first reducing the model by eliminating $u^{(1)}$ and then employing a arbitrary Lagrangian--Eulerian method. We have shown how the distribution of fibres within the fluid can cause interesting non--Newtonian behaviours such as driving non-uniformity in the development of the thickness of an initially uniform sheet and deflection of the centre--line even with the implicit assumption that the centre--line is nearly straight. Our results also show that the bulk properties of a passive transversely isotropic fluid sheet are controlled largely by the behaviour of a derived `effective viscosity'.\\

\indent As far as the behaviour of an active transversely isotropic fluid is concerned, preliminarily we have seen that allowing $\mu_1 \neq 0$ allows the fibres to develop towards alignments that are not in the direction of extension of the fluid. However, if the fibres are aligned in the longitudinal direction of the sheet and possesses active behaviour, the tension within the sheet is increased. Active behaviour giving rise to greater tension has been observed in the seeding of hydrogels with a suspension of self-aligning cells \cite{antman2017mechanically}. Future work in this area could include constructing more biologically realistic multiphase model that incorporates the work in this paper as a fibrous extracellular matrix or hydrogel, with the cells exhibiting active behaviour instead of the fibres. This could result in a model of how different experimental setups lead to different alignment patterns of cells and could determine the best conditions to grow neural tissue.
    \\
    
\indent There are a number of other avenues for further work related to this paper. We model sheets that are nearly straight, with the employment of a Cartesian co-ordinate system restricting the model to examining sheets which are initially slightly curved, i.e $\dfrac{H}{L_0}$ is small. Where this is not the case, future work could entail the use of a curvilinear co-ordinate system to approach sheets with curvature in the centre--line, in works similar to Ribe \cite{ribe2002general}. As the sheet starts to become very thin, there may be a new regime where the $\mu_1$ term in \eqref{U ALE} dominates the $\dfrac{\p u}{\p x'}$ term. Perhaps the behaviour of the sheet in this regime may shed light upon how the active behaviour of the fluid may drive breakup of the sheet.
Simpler modifications could include prescribing the tension applied to the ends of the sheet, rather than prescribing the length. Furthermore we could also modify the model to include the effects of surface tension, inertia, and body forces.
\\
\\
\textbf{Declaration of competing interests}
\\
None.

    \appendix
\section{Model equations in full} \label{AppendixEqns}
To briefly summarise, the dimensionless equations are
\begin{equation}
    \pd{u}{x} + \pd{v}{y} = 0,
\end{equation}
for conservation of mass, with the momentum equation \eqref{DimensionFullMomentum} yielding in the $x$-direction:
\begin{align*}
- \varepsilon^2 \pd{p}{x} &+ \Pd{u}{y} + \varepsilon^2 \Pd{u}{x} + \varepsilon^2 \mu_1 \pd{}{x} (\cos^2{\theta}) 
+ \varepsilon \mu_1 \pd{}{y} ( \cos{\theta} \sin{\theta} ) \\
&+ \mu_2 \pd{}{x} \left[ \varepsilon^2 \cos^4{\theta} \pd{u}{x} + \cos^3{\theta} \sin{\theta} \left( \varepsilon \pd{u}{y} + \varepsilon^3 \pd{v}{x} \right) + \varepsilon^2 \cos^2{\theta} \sin^2{\theta} \pd{v}{y} \right] \\
&+ 2 \mu_3 \pd{}{x} \left[ 2 \varepsilon^2 \cos^2{\theta} \pd{u}{x} + \cos{\theta} \sin{\theta} \left( \varepsilon\pd{u}{y} +  
\varepsilon^3 \pd{v}{x} \right)  \right] \\
&+  \mu_2 \pd{}{y} \left[\cos{\theta} \sin{\theta}  \left(\varepsilon\cos^2{\theta} \pd{u}{x} + \cos{\theta} \sin{\theta} \left( \pd{u}{y} + \varepsilon^2 \pd{v}{x}  \right) + \varepsilon\sin^2{\theta} \pd{v}{y}  \right) \right] \\ 
&+  \mu_3 \pd{}{y} \left[  \pd{u}{y} +  \varepsilon^2 \pd{v}{x} \right] =0, \numberthis
\label{inc nondim x mom}
\end{align*}
\noindent whilst in the $y$ direction we have:
\begin{align*}
-  \varepsilon \pd{p}{y} &+ \varepsilon^3 \Pd{v}{x} +  \varepsilon \Pd{v}{y} 
+  \varepsilon\mu_1 \pd{}{y} (\sin^2{\theta}) +   \varepsilon^2 \mu_1 \pd{}{x} (\cos{\theta} \sin{\theta}) \\
&+ \mu_2 \pd{}{y} \left[ \varepsilon\sin^2{\theta} \cos^2{\theta} \pd{u}{x} + \cos{\theta} \sin^3{\theta} \left( \pd{u}{y} + \varepsilon^2 \pd{v}{x} \right) + \varepsilon \sin^4{\theta} \pd{v}{y} \right] \\
&+ 2   \mu_3 \pd{}{y} \left[ 2\varepsilon \sin^2{\theta} \pd{v}{y} + \cos{\theta} \sin{\theta} \left(  \pd{u}{y} +  \varepsilon^2 \pd{v}{x} \right)  \right] \\
&+ \mu_2 \pd{}{x} \left[ \varepsilon^2 \sin{\theta} \cos^3{\theta} \pd{u}{x} + \cos^2{\theta} \sin^2{\theta} \left(  \varepsilon \pd{u}{y} + \varepsilon^3 \pd{v}{x} \right) +  \varepsilon^2 \cos{\theta} \sin^3{\theta} \pd{v}{y} \right] \\
&+  \mu_3 \pd{}{x} \left[ \varepsilon \pd{u}{y} +  \varepsilon^3 \pd{v}{x} \right] =0,\numberthis
\label{inc nondim y mom}
\end{align*}
with the fibre director field being given by
\begin{align}
    \varepsilon\pd{\theta}{t} + \varepsilon u \pd{\theta}{x} + \varepsilon v \pd{\theta}{y} = -\varepsilon\sin{\theta} \cos{\theta} \pd{u}{x} 
- \sin^2{\theta} \pd{u}{y} + \varepsilon^2 \cos^2{\theta} \pd{v}{x} + \varepsilon \sin{\theta} \cos{\theta} \pd{v}{y} .\label{nondim theta}
\end{align}

 \section{Simplification of the equation for $\theta$} \label{AppendixTheta}
 In the main text, we claimed that equation \eqref{Theta Equation ALE} permitted great simplification by noting that the equation corresponded only to advection in a purely horizontal direction the reference domain. To demonstrate this simplification, suppose $\tilde\theta\left(x',y',t\right)$ is a function defined over $D_{\text{ref}}$ that satisfies the advection equation
\begin{equation}
    \pd{\tilde{\theta}}{t} + \tilde{u}\pd{\tilde\theta}{x'} = \tilde{f}, \label{A1}
\end{equation}
where $\tilde{u}\left(x',t\right)$ and $\tilde{f}\left(x',t\right)$ are a horizontal advection velocity and forcing term respectively. We can relate $\theta(x,y,t) = \tilde\theta(x(x',t),y(x',y',t),t)$, and using the mapping $\pmb\Phi$, which gives
\begin{align}
    \pd{\tilde{\theta}}{t} = \pd{\theta}{t} + \frac{\dot{L}x}{L}\pd{\theta}{x} +\left(\pd{H}{t} + \left(\frac{y-H}{h}\right) \pd{h}{t} + \frac{\dot{L}x}{L}\left(\pd{H}{x} + \left(\frac{y-H}{h}\right) \pd{h}{x} \right) \right) \pd{\theta}{y}, \label{TildeThetaT}\\
    \pd{\tilde{\theta}}{x'} = L\pd{\theta}{x} + L\left(\pd{H}{x} + \left(\frac{y-H}{h}\right)\pd{h}{x}\right)\pd{\theta}{y}.  \label{TildeThetaX}
\end{align}
Substituting \eqref{TildeThetaT}-\eqref{TildeThetaX} into \eqref{A1}, then yields
\begin{equation}
    \pd{\theta}{t} + \pd{\theta}{x} \left(L\tilde{u} + \frac{\dot{L}x}{L}\right) + \pd{\theta}{y}\left(\pd{H}{t} + \left(\frac{y-H}{h}\right)\pd{h}{t} + \left(\pd{H}{x} + \left(\frac{y-H}{h}\right)\pd{h}{x}\right)\left(L\tilde{u} + \frac{\dot{L}x}{L}\right) \right) = \tilde{f},
\end{equation}
we now choose $u\left(x,t\right) = L\tilde{u} + \dfrac{\dot{L}x}{L}$, in order to recover the correct coefficient of $\dfrac{\p \theta}{\p x}$. Examining the coefficient of the $\dfrac{\p \theta}{\p y}$ term we note that
\begin{align}
    \pd{H}{t} + \left(\frac{y-H}{h}\right)\pd{h}{t} + u\left(\pd{H}{x} + \left(\frac{y-H}{h}\right)\pd{h}{x}\right)
    \\
    =\pd{H}{t} + u \pd{H}{x} + \left(\frac{y-H}{h}\right)\left(-h \pd{u}{x}\right) \label{A7}
    \\
    =\pd{H}{t} + \pd{}{x}\left(uH\right)  - y \pd{u}{x} = v
\end{align}
where we have used the equation for conservation of mass, \eqref{Cons Mass}, to obtain \eqref{A7}. Here, we have demonstrated that the coefficient of $\theta_y$ is precisely $v$ when mapping back from $D_{ref}$ to the original domain. Therefore, we have shown that the advection of $\theta$ is purely horizontal upon the reference domain, with velocity $\tilde{u}\left(x',t\right)= \dfrac{u - \dot{L} x'}{L}  = \dfrac{u-u_{\text{mesh}}}{L}$.

\section{Discretisation of the Green and Friedman integral equations} \label{AppendixDiscretisation}
In what follows, the treatment of the integral equations is in the Eulerian framework. The integral equations \eqref{U ALE},\eqref{H ALE} require further treatment before being discretised and solved.  Introduce
\begin{align}
    F\left(x,y,t\right) = \frac{\mu_1 \cos 2\theta + \left(4+4\mu_3+\mu_2\right) u_{x}}{4+4\mu_3+\mu_2\sin^2 2\theta},
\end{align}
it will be convenient to write $F=F_{1} + u_{x}F_2$, where
\begin{align}
    &F_1 = \frac{\mu_1 \cos 2\theta }{4+4\mu_3+\mu_2\sin^2 2\theta}, &F_2 = \frac{ 4+4\mu_3+\mu_2 }{4+4\mu_3+\mu_2 \sin^2 2\theta},
\end{align}
in much the same way, we also introduce notation for the integrals of $F$, by defining $G=G_1 + u_{x}G_2,$ where
\begin{equation}
    G_{m}\left(x,y,t\right) = \int\limits_{H^{-}}^{y} F_{m}\left(x,s,t\right) \mathrm{d}s,
\end{equation}
for $m=1,2$. Equation \eqref{U ALE} can now be written as 
\begin{align}
    0 = \pd{}{x} \left( G_{1}\left(x,H^{+},t\right)+ u_{x}\left(x,t\right)G_{2}\left(x,H^{+},t\right)\right), \label{U4}
\end{align}
using the trapezoidal rule, 
\begin{equation}
    G_{m}\left(x,H^{+},t\right) = \frac{h\left(x,t\right)}{N-1}\left(\frac{ F_{m}\left(x,y_{0},t\right) + F_{m}\left(x,y_{N-1},t\right)}{2} + \sum_{i=1}^{N-2} F_{m}\left(x,y_{i},t\right) \right),
\end{equation}
where $N$ is the number of nodes in the $y$-direction. We note that upon substituting the trapezoidal rule into \eqref{U4}, the result does not depend on $H$. That is, its presence in the integration limits is redundant and effectively just describes a vertical translation. Therefore, $H$ is decoupled from the rest of the system and we can solve the equations for $u$ and $\theta$ and then compute $H$ as required at a desired time. For completeness, we include the discretisation for equation \eqref{U4}. First, introduce the notation
\begin{align}
    \left[G_{m} \right]_{i,N-1}^{k} = G_{m}\left(x_{i},H_{i}^{k}+ h_{i}^{k}/2, t_{k} \right),
\end{align}
now \eqref{U4} gives us, through centred finite differences,  
\begin{multline}
    \frac{\left[G_{1}\right]^{k}_{i-1,N-1} - \left[G_{1}\right]^{k}_{i+1,N-1}}{2L/\left(M-1\right)} = \frac{}{}\frac{\left[G_{2}\right]^{k}_{i+1,N-1} - \left[G_{2}\right]^{k}_{i-1,N-1}}{2L/\left(M-1\right)} \frac{u_{i+1}^{k} - u_{i}^{k} }{2L/\left(M-1\right)} \\
    + \frac{u_{i}^{k} - 2u_{i}^{k} + u_{i-1}^{k}}{\left(L/\left(M-1\right)\right)^2}\left[G_{2}\right]_{i,N-1}^{k},
\end{multline}
where $N,M$ are the number of nodes in the vertical and horizontal directions respectively so that $i = 0:M-1, j=0:N-1$. Noting that $u_{0}^{k} = 0$, $u_{M-1}^{k} = \dot{L}\left(t_{k}\right)$, and that $G_{m}$ is readily precomputed at each time-step $k$, yields a tri-diagonal system for $u^{k}$. If we now consider the equation for $H$, \eqref{h ALE}, this is the only equation in the model that is indeed easier to treat in the Eulerian framework than the ALE. Using equation \eqref{UOne} and the Leibniz rule we may write equation \eqref{Original H integral} as 
\begin{equation}
    0 = -\left(H_{xx} + \frac{h_{xx}}{2}
    \right) G\left(x,H^{+},t\right) + \Pd{}{x} \int\limits_{H^{-}}^{H^{+}} G(x,y,t) \mathrm{d}y, \label{AppendixH}
\end{equation}
in order to proceed, one must apply the trapezoidal rule twice to each $G_{m}$. Applying it once yields
\begin{equation}
    \int\limits_{H^{-}}^{H^{+}} G_{m}\left(x,y,t\right)\mathrm{d}y  = \frac{h\left(x,t\right)}{N-1}\left(\frac{ G_{m}\left(x,y_{0},t\right) + G_{m}\left(x,y_{N-1},t\right)}{2} + \sum_{j=1}^{N-2} G_{m}\left(x,y_{j},t\right) \right), \label{Simpsons1}
\end{equation}
then, for each $j > 0$, 
\begin{equation}
    G_{m}\left(x,y_{j},t\right) = \frac{h\left(x,t\right)}{N-1}\left(\frac{ F_{m}\left(x,y_{0},t\right) + F_{m}\left(x,y_{j},t\right)}{2} + \sum_{i=1}^{j-1} F_{m}\left(x,y_{i},t\right) \right), \label{Simpsons2}
\end{equation}
for the case $j=0$, $G_{m}\left(x,y_{0},t\right) = 0$. Substitution of \eqref{Simpsons2} into \eqref{Simpsons1} yields
\begin{multline}
    \int\limits_{H^{-}}^{H^{+}} G_{m}\left(x,y,t\right)\mathrm{d}y  = \left( \frac{h}{N-1}\right)^2 \left(\frac{2N-3}{4} F_{m}\left(x,y_0,t\right) + \frac{1}{4}F_{m}\left(x,y_{N-1},t\right) \right.
    \\
    \left. + \sum_{j=1}^{N-2} \left(N-1-j\right)F_{m}\left(x,y_j,t\right) \right). \label{GMSimpsons}
\end{multline}
Finally, we require the introduction of the notation
\begin{align}
    \left[IG_{m}\right]^{k}_{i} = \int\limits_{H^{-}}^{H^{+}} G_{m}\left(x_{i},y,t_{k}\right)\mathrm{d}y,
\end{align}
for $m=1,2$. Clearly, we use \eqref{GMSimpsons} to precompute $[IG_{m}]$ at the required nodes as necessary. The discretisation of equation \eqref{AppendixH} is then 
\begin{multline}
    \frac{H_{i-1}^{k} - 2H_{i}^{k} + H_{i+1}^{k}}{\left(L(t_{k})/\left(M-1\right) \right)^2}\left( \left[G_1\right]_i^k + \frac{u_{i+1}^k - u_{i-1}^k}{2L\left(t_k\right)/\left(M-1\right)}\left[G_2\right]_i^{k} \right)
    \\
    =  -\frac{h_{i-1}^{k} - 2h_{i}^{k} + h_{i+1}^{k}}{2\left(L(t_{k})/\left(M-1\right) \right)^2}\left( \left[G_1\right]_i^k + \frac{u_{i+1}^k - u_{i-1}^k}{2L\left(t_k\right)/\left(M-1\right)}\left[G_2\right]_i^{k} \right) + \frac{\left[IG_1\right]_{i-1}^{k} - 2\left[IG_1\right]_{i}^{k} + \left[IG_1\right]_{i+1}^{k} }{\left(L\left(t_k\right)/\left(M-1\right)\right)^2}
    \\
    + \frac{u_{i+2}^k - 2u_{i+1}^k + 2u_{i-1}^k -u_{i-2}^k }{2\left(L\left(t_k\right)/\left(M-1\right)\right)^3}\left[IG_{2}\right]_{i}^{k+1} + \frac{u_{i+1}^k - 2u_{i}^{k} + u_{i-1}^{k}}{\left(L\left(t_k\right)/\left(M-1\right)\right)^2}\frac{\left[IG_2\right]_{i+1}^{k} -\left[IG_2\right]_{i-1}^k}{2L\left(t_k\right)/\left(M-1\right)}
    \\
    + \frac{\left[IG_{2}\right]_{i+1}^k - 2\left[IG_{2}\right]_{i}^{k} + \left[IG_{2}\right]_{i-1}^{k}}{\left(L\left(t_k\right)/\left(M-1\right)\right)^2}\frac{u_{i+1}^{k} -u_{i-1}^k}{2L\left(t_k\right)/\left(M-1\right)}.  \label{HDiscretised}
\end{multline}
Equation \eqref{HDiscretised} contains wholly precomputable quantities on the RHS. Therefore, similar to the equation for $u$, this creates a tri-diagonal system to be solved for $H$. For the specific cases of $i=0,M-1$, we have the boundary condition that $H_0^k = H_{M-1}^k = 0$. For the cases of $i=1,M-2$, the discretisation must be modified slightly as the stencil for $u_{xxx}$ is too wide. This can be done in a number of ways and is omitted.

\section{Derivation of the short timescale integral equations} \label{Appendix:BNTDeriv}
\indent In order to close the model, we must go to yet higher orders in order to obtain equations for $u^{(0)},H^{(0)}$. Our approach is to integrate the relevant equations over the depth of the sheet and apply the no-stress boundary conditions at $y=H^{(0)^{\pm}}$. We find that a number of higher order terms in the boundary conditions will be eliminated by substitution of previously obtained quantities. The $\mathcal{O}\left(\varepsilon^3\right)$ $x$-momentum equation is
\begin{multline}
    \pd{}{x}\left[ -p^{(0)} + 2\pd{u^{(0)}}{x} + \mu_1 \cos^2\theta^{(0)} \right. 
 + \mu_2 \left( \cos^4\theta^{(0)}  \pd{u^{(0)}}{x} + \cos^3\theta^{(0)}\sin\theta^{(0)}\left(\pd{u^{(1)}}{y} + \pd{V^{(1)}}{x}\right) \right.
 \\
 \left. + \cos^2\theta^{(0)}\sin^2\theta^{(0)}\pd{V^{(2)}}{y} \right) + \left. 2\mu_3\left( 2\cos^2\theta^{(0)}\pd{u^{(0)}}{x} + \cos\theta^{(0)}\sin\theta^{(0)}\left( \pd{u^{(1)}}{y} +\pd{V^{(1)}}{x}\right)\right) \right] - \Pd{u^{(0)}}{x} 
 \\
 = -\pd{}{y} \left[ \pd{u^{(2)}}{y} + \f{\p V^{(2)}}{\p x} +  \mu_1 \left( \theta^{(1)}\cos2\theta^{(0)}\right) +  \mu_2  \left(\cos^3\theta^{(0)}\sin\theta^{(0)}\pd{u^{(1)}}{x} \right.  \right.
 \\
  + \theta^{(1)}\left(\cos^4\theta^{(0)} - 3\sin^2\theta^{(0)}\cos^2\theta^{(0)}\right)\pd{u^{(0)}}{x}  + \cos^2\theta^{(0)}\sin^2\theta^{(0)}\left(\pd{u^{(2)}}{y} + \pd{V^{(2)}}{x}\right)
  \\  + \f{1}{2} \theta^{(1)} \sin 4\theta^{(0)} \left(\pd{u^{(1)}}{y} + \pd{V^{(1)}}{x}\right) + \cos\theta^{(0)}\sin^3\theta^{(0)}\pd{V^{(3)}}{y}  
   \\
\left. \left.   + \theta^{(1)}\left(3\sin^2\theta^{(0)}\cos^2\theta^{(0)} - \sin^4\theta^{(0)}\right)\pd{V^{(2)}}{y} \right) \colour{black} + \mu_3 \left(\pd{u^{(2)}}{y} + \pd{V^{(2)}}{x} \right) \right] \colour{black} + \f{\p^2 V^{(2)}}{\p x \p y}, \label{RewrittenOE3XM}
\end{multline}
and its associated no-stress boundary condition is
  \begin{multline} 
  \pd{u^{(2)}}{y} + \pd{V^{(2)}}{x} + \mu_1 \theta^{(1)} \cos2\theta^{(0)} + \mu_2\left( \cos^3\theta^{(0)}\sin\theta^{(0)} \pd{u^{(1)}}{x} + \theta^{(1)} \left(\cos^4\theta^{(0)} - 3\sin^2\theta^{(0)}\cos^2\theta^{(0)}\right)\pd{u^{(0)}}{x} \right.  \\
 \left.  + \cos^2\theta^{(0)}\sin^2\theta^{(0)}\left( \pd{u^{(2)}}{y} + \pd{V^{(2)}}{x} \right) + \f{1}{2} \theta^{(1)} \sin 4\theta^{(0)} \left(\pd{u^{(1)}}{y} + \pd{V^{(1)}}{x}\right) + \cos\theta^{(0)}\sin^3\theta^{(0)}\pd{V^{(3)}}{y} \right. 
   \\
 \left.   + \theta^{(1)}\left(3\sin^2\theta^{(0)}\cos^2\theta^{(0)} - \sin^4\theta^{(0)}\right)\pd{V^{(2)}}{y} \right) +\mu_3\left(\pd{V^{(2)}}{x} + \pd{u^{(2)}}{y}\right) 
\\
= \left( \pd{H^{(0)}}{x} \pm \f{1}{2}\pd{h^{(0)}}{x}\right)\left[ -p^{(0)} + 2\pd{u^{(0)}}{x} + \mu_1 \cos^2\theta^{(0)} \right.  \\
 + \mu_2 \left( \cos^4\theta^{(0)}  \pd{u^{(0)}}{x} + \cos^3\theta^{(0)}\sin\theta^{(0)}\left(\pd{u^{(1)}}{y} + \pd{V^{(1)}}{x}\right) + \cos^2\theta^{(0)}\sin^2\theta^{(0)}\pd{V^{(2)}}{y} \right)  \\
 \left. 2\mu_3\left( 2\cos^2\theta^{(0)}\pd{u^{(0)}}{x} + \cos\theta^{(0)}\sin\theta^{(0)}\left( \pd{u^{(1)}}{y} +\pd{V^{(1)}}{x}\right)\right) \right]; \text{ on } y=H^{(0)^{\pm}}. \label{OE3STRESSXM}
 \end{multline}
 We note that the remaining terms outside of the derivatives in \eqref{RewrittenOE3XM} cancel due to the continuity equation at $\mathcal{O}(\eps^2)$. Integrating equation \eqref{RewrittenOE3XM} over the depth of the sheet yields
 \begin{align}
     \int\limits_{H^{(0)^{-}}}^{H^{(0)^{+}}} \pd{}{x}\left( -p^{(0)} + g_1(x,\tau)\right) \mathrm{d}y = - \left[ \pd{u^{(2)}}{y} + \pd{V^{(2)}}{x} + g_{2}(x,\tau) \right]_{y=H^{(0)^{-}}}^{y=H^{(0)^{+}}}, 
 \end{align}
 where $g_1,g_2$ are functions containing the collected $\mu_1,\mu_2,\mu_3$ terms from \eqref{RewrittenOE3XM} and are readily obtained by inspection. Application of \eqref{OE3STRESSXM} now yields
 \begin{multline}
     \int\limits_{H^{(0)^{-}}}^{H^{(0)^{+}}} \pd{}{x}\left( -p^{(0)} + g_1\right) \mathrm{d}y = \left(\pd{H^{(0)}}{x} - \frac{1}{2}\pd{h^{(0)}}{x}\right)\left( -p^{(0)} + g_1\right)_{y=H^{(0)^{-}}} \\
     - \left(\pd{H^{(0)}}{x} + \frac{1}{2}\pd{h^{(0)}}{x}\right)\left( -p^{(0)} + g_1\right)_{y=H^{(0)^{+}}},
     \end{multline}
 and hence by use of the Leibniz rule, we now obtain an equation for $\bar{u}$
\begin{multline}
    \frac{\p}{\p x} \int \limits_{H^{(0)^{-}}}^{H^{(0)^{+}}}  \left( 4\pd{u^{(0)}}{x} + \mu_1 \cos2\theta^{(0)} +\mu_2\left( \cos^2 2\theta^{(0)} \pd{u^{(0)}}{x} + \f{1}{4}\sin4\theta^{(0)}\left(\pd{u^{(1)}}{y} + \pd{V^{(1)}}{x} \right)  \right) 
     \right. \\ \left. + 4\mu_3 \pd{u^{(0)}}{x} \right)\mathrm{d}y = 0. \label{EquationUBar}
\end{multline}
A similar process at $\mathcal{O}\left(\eps^4\right)$ yields an equation for $H$, 
\begin{multline}
       \pd{}{x} \int \limits_{H^{(0)^{-}}}^{H^{(0)^{+}}} \pd{}{x} \int\limits_{H^{(0)^{-}}}^{y} \left( 4\pd{u^{(0)}}{x} + \mu_1\cos2\theta^{(0)} + \mu_2 \cos^2 2\theta^{(0)}\pd{u^{(0)}}{x}
    +  \f{\mu_2}{4}\sin4\theta^{(0)}\left( \pd{u^{(1)}}{y} + \pd{V^{(1)}}{x}\right) \right. 
    \\
    \left.+ 4\mu_3\pd{u^{(0)}}{x} \right) \mathrm{d}y'  \mathrm{d}y  = 0. \numberthis \label{OriginalH}
\end{multline}
We note that these equations are of the same form as equations \eqref{Original u integral} and  \eqref{Original H integral}, with the difference being that $u^{(0)}$ now possesses $y$-dependence. Substitution of \eqref{UOne} and \eqref{Compat} into both \eqref{EquationUBar} and \eqref{OriginalH} and use of the Liebniz rule upon \eqref{OriginalH} yield equations \eqref{UTransformedIntro} and \eqref{HTransformedIntro}.
\section{Discretisation of the short timescale model}\label{AppendixBNTDisc}
In this section, we give the discretisation of the integral equations in the system of the short timescale equations, namely \eqref{UTransformedBNTSec}, \eqref{HTransformedBNTSec}. As before, we drop the superscript notation for leading-order quantities.
We discretise with
\begin{equation}
    \theta(x_{i},y_{j},\tau_{k}) = \theta_{i,j}^{k}, \text{ etc}
\end{equation}
and for functions that contain subscripts:
\begin{equation}
    Z_1(x_i,y_j,\tau_k)  = [Z_1]_{i,j}^k, \text{ etc}
\end{equation}
where $i=1:M-1,j=1:N-1,$ where $M,N$ are the number of nodes in the $x,y$-directions.%Using equation \eqref{BarUBNT} and the linear transformation $y=H+h\tilde{y}$ upon equation \eqref{UBar}, we obtain the following equation for $\bar{u}$:
Similarly to the discretisation of the Green and Friedman model equations, we define
\begin{align}
   Z_{1}(x,y,\tau) = \int \limits_{-\frac{1}{2}}^{y} \frac{\mu_1 \cos 2\theta}{4+4\mu_3 + \mu_2 \sin^2 2\theta} h\mathrm{d}\tilde{y}, \\ 
    Z_{2}(x,y,\tau) = \int \limits_{-\frac{1}{2}}^{y} \frac{4+4\mu_3 + \mu_2}{4+4\mu_3+\mu_2 \sin^2 2\theta}h \mathrm{d}\tilde{y},  \\
    J(x,t) = \int \limits_{-\frac{1}{2}}^{\frac{1}{2}} \frac{\left(4+4\mu_3+\mu_2\right) h^2 y' }{4+4\mu_3+\mu_2 \sin^2 2\theta} \mathrm{d}\tilde{y},
\end{align}
we note that the definitions of $Z_{1},Z_{2}$ here are similar to $G_{1},G_{2}$ in appendix \ref{AppendixDiscretisation}, but are not precisely the same. It is possible to obtain $G_{1},G_{2}$ from $Z_{1},Z_{2}$ by undoing both the transformation $y=H+h\tilde{y}$ and the short-timescale. Since the ALE transformation in Section \ref{Section:SectionALE} acts as a linear transformation on the integral equations, one may treat $Z_1,Z_2$ as the integrals in ALE form on the short-timescale. 
We may now rewrite equation \eqref{UTransformedBNTSec} as
\begin{multline}
   Z_{2}(x,\frac{1}{2},\tau)\Pd{\bar{u}}{x}  + \pd{Z_{2}(x,\frac{1}{2},\tau)}{x} \pd{\bar{u}}{x} + \frac{\p^2 H}{\p x \p \tau}\left(\pd{Z_{2}(x,\frac{1}{2},\tau)}{x} \pd{H}{x} + Z_{2}(x,\frac{1}{2},\tau) \Pd{H}{x}\right) \\
        + \frac{\p^3 H}{\p x^2 \p \tau}\left(Z_{2}(x,\frac{1}{2},\tau) \pd{H}{x} - \pd{J}{x}\right) -J\frac{\p^4 H}{\p x^3 \p \tau} = -\pd{Z_{1}(x,\frac{1}{2},\tau)}{x}. \label{UDisc}
\end{multline}
We choose to use a FTCS finite difference method, hence the discretisation of \eqref{UDisc} is 
\begin{multline}
    [Z_2]_{i, N-1}^k \FD{\bar{u}}{k}{2} + \frac{[Z_2]_{i+1, N-1}^k -[Z_2]_{i-1, N-1}^k}{2\Delta x}\FD{\bar{u}}{k}{1} +
    \\
    \frac{\left(H_{i+1}^{k+1} -  H_{i-1}^{k+1} \right) - \left(H_{i+1}^{k} - H_{i-1}^{k} \right)}{ 2 \Delta \tau \Delta x}\left( \frac{[Z_2]_{i+1, N-1}^k -[Z_2]_{i-1, N-1}^k}{2\Delta x}\FD{H}{k}{1} \right. 
    \\
    \left. + [Z_2]_{i, N-1}^k \FD{H}{k}{2} \right) 
    \\
     +  \frac{\left(H_{i+1}^{k+1} - 2 H_{i}^{k+1} + H_{i-1}^{k+1} \right) - \left(H_{i+1}^{k} - 2 H_{i}^{k} + H_{i-1}^{k} \right)}{\Delta \tau \Delta x^2}\left([Z_2]_{i, N-1}^k \FD{H}{k}{1} \right. 
     \\
     \left. - \FD{J}{k}{1}  \right) 
       -J_{i}^{k}\frac{\left(H_{i+2}^{k+1} -2H_{i+1}^{k+1} + 2 H_{i-1}^{k+1} - H_{i-2}^{k+1} \right) - \left(H_{i+2}^{k} -2H_{i+1}^{k} + 2 H_{i-1}^{k} - H_{i-2}^{k} \right)}{2 \Delta \tau \Delta x^3}
       \\
       = -\frac{[Z_1]_{i+1, N-1}^k -[Z_1]_{i-1, N-1}^k}{2\Delta x},
\end{multline}
This discretisation can be written in the matrix form
\begin{equation}
    \Delta \tau \Delta x^2 \mathbf{b} + \mathbf{M_H}\mathbf{H}^{k} = \mathbf{M_H}\mathbf{H}^{k+1} + \Delta \tau \Delta x \mathbf{M_{\bar{U}}} \mathbf{\bar{u}}^{k}, \label{MatrixEquationBNT1}
\end{equation}
where $\boldsymbol{H}^{k} = \left(H_{1}^{k},H_{2}^{k},\dots,H_{N+1}^{k}\right)^{T}$, $\boldsymbol{\bar{u}}^{k} = \left(\bar{u}_{1}^{k},\bar{u}_{2}^{k},\dots,\bar{u}_{N+1}^{k}\right)^{T}$, and $\boldsymbol{M_{H}}, \boldsymbol{M_{\bar{U}}}$ are matrices whose entries are the coefficients of the $H^{k+1}$ and $\bar{u}^k$ terms respectively, and are dependent upon the choice of discretisation of the $x$-derivatives of $H,\bar{u}$. We note that the left hand side of \eqref{MatrixEquationBNT1} is known and precomputable at each time step. Due to the $\dfrac{ \p H^4}{\p x^3 \p \tau}$ term, the stencil must be adjusted at the ends of the domain, by using biased finite differences.
\\
\indent Next, we define more functions for notational convenience:
\begin{align}
    IZ_{1}(x,\tau) = \int \limits_{-\frac{1}{2}}^{\frac{1}{2}} \int \limits_{-\frac{1}{2}}^{\tilde{y}} \frac{\mu_1 \cos 2\theta}{4+4\mu_3+\mu_2\sin^2 2\theta} h^2\mathrm{d}\tilde{y}'\mathrm{d}\tilde{y}, \\ 
    IZ_{2}(x,\tau) = \int \limits_{-\frac{1}{2}}^{\frac{1}{2}}\int \limits_{-\frac{1}{2}}^{\tilde{y}} \frac{4+4\mu_3 + \mu_2}{4+4\mu_3+\mu_2 \sin^2 2\theta}h^2\mathrm{d}\tilde{y}'\mathrm{d}\tilde{y},  \\
    K(x,\tau) = \int \limits_{-\frac{1}{2}}^{\frac{1}{2}} \int \limits_{-\frac{1}{2}}^{\tilde{y}} \frac{\left(4+4\mu_3+\mu_2\right) h^3 \tilde{y}' }{4+4\mu_3+\mu_2\sin^2 2\theta}\mathrm{d}\tilde{y}'\mathrm{d}\tilde{y}.
\end{align}
With the introduced functions, we may express \eqref{HTransformedBNTSec} as
\begin{multline}
   IZ_{2} \frac{\p^3 \bar{u}}{\p x^3} + 2\pd{IZ_2}{x} \Pd{\bar{u}}{x} + \pd{\bar{u}}{x} \left( \Pd{IZ_2}{x} -Z_2\left(\Pd{H}{x} +\frac{1}{2}\Pd{h}{x}\right)\right)
   \\
   - K \frac{\p^5 H}{\p x^4 \p \tau} + \frac{\p^4 H}{\p x^3 \p \tau} \left(IZ_2 \pd{H}{x} - 2 \pd{K}{x}\right) 
    \\
    +\frac{\p ^3 H}{\p x^2 \p \tau} \left( 2 IZ_2 \Pd{H}{x} + 2 \pd{IZ_2}{x}\pd{H}{x}+J\left(\Pd{H}{x} +\frac{1}{2}\Pd{h}{x}\right) - \Pd{K}{x}\right)
    \\
    + \frac{\p^2 H}{\p x \p \tau} \left( IZ_2 \frac{\p ^3 H}{\p x^3}  + 2 \pd{IZ_2}{x}\Pd{H}{x} + \Pd{IZ_2}{x}\pd{H}{x} - Z_{2}\pd{H}{x} \left(\Pd{H}{x} +\frac{1}{2}\Pd{h}{x}\right) \right) 
    \\
      = Z_{1} \left(\Pd{H}{x} +\frac{1}{2}\Pd{h}{x}\right) -  \Pd{IZ_1}{x}. \label{HDisc}
\end{multline}
\begin{landscape}
The discretisation of \eqref{HDisc} is: 
\begin{multline}
[IZ_2]_{i}^{k} \FD{\bar{u}}{k}{3} + \frac{[IZ_2]_{i+1}^{k} - [IZ_2]_{i-1}^{k}}{\Delta x}\FD{\bar{u}}{k}{2} 
\\
+ \FD{\bar{u}}{k}{1}\left( \frac{[IZ_2]_{i+1}^{k} -2[IZ_2]_{i}^{k} +[IZ_2]_{i-1}^{k}}{\Delta x^2} - Z_{2_{i}}\left(\FD{H}{k}{2} + \frac{1}{2}\FD{h}{k}{2} \right)\right)
\\
+ \frac{\left(H_{i+1}^{k+1} -  H_{i-1}^{k+1} \right) - \left(H_{i+1}^{k} - H_{i-1}^{k} \right)}{ 2 \Delta \tau \Delta x}\left(\frac{[IZ_2]_{i+1}^{k}-2[IZ_2]_{i}^{k}+[IZ_2]_{i-1}^{k}}{\Delta x^2} \FD{H}{k}{1} \right. \\
\left.
+ \frac{[IZ_2]_{i+1}^{k} - [IZ_2]_{i-1}^{k}}{\Delta x}\FD{H}{k}{2} \right.
\\
\left. +[IZ_2]_{i}^{k}\FD{H}{k}{3} - Z_{2_i}^{k}\FD{H}{k}{1}\left(\FD{H}{k}{2} + \frac{1}{2}\FD{h}{k}{2} \right) \right)
\\
+\frac{\left(H_{i+1}^{k+1} -2H_i^{k+1}  H_{i-1}^{k+1} \right) - \left(H_{i+1}^{k} -2H_i^{k}  + H_{i-1}^{k}  \right)}{\Delta \tau \Delta x^2}\left(2[IZ_2]_{i}^{k} \FD{H}{k}{2} + \frac{IZ_{2_{i+1}}-IZ_{2_{i-1}}}{\Delta x}\FD{H}{k}{1} \right.
\\
\left. - \FD{K}{k}{2} + J_{i}^{k}\left( \FD{H}{k}{2} + \frac{1}{2}\FD{h}{k}{2} \right) \right)
\\
+ \frac{\left(H_{i+2}^{k+1} -2H_{i+1}^{k+1} + 2 H_{i-1}^{k+1} - H_{i-2}^{k+1} \right) - \left(H_{i+2}^{k} -2H_{i+1}^{k} + 2 H_{i-1}^{k} - H_{i-2}^{k} \right)}{2 \Delta \tau \Delta x^3}  \left(IZ_{2_i}^k \FD{H}{k}{1} - \FD{K}{k}{1} \right)
\\
-K_i^k \frac{\left(H_{i+2}^{k+1} -4 H_{i+1}^{k+1} + 6H_{i}^{k+1} -4 H_{i-1}^{k+1} + H_{i-2}^{k+1} \right) - \left(H_{i+2}^{k} -4 H_{i+1}^{k} + 6H_{i}^{k} -4 H_{i-1}^{k} + H_{i-2}^{k} \right)}{\Delta \tau \Delta x^4} 
\\= Z_{1_i}^{k} \left(\FD{H}{k}{2} + \frac{1}{2}\FD{h}{k}{2} \right) - \frac{IZ_{2_{i+1}}^k - 2 IZ_{2_{i}}^k + IZ_{2_{i-1}}^k}{\Delta x^2}. \label{HALEDisc}
\end{multline}
\end{landscape}
As before, we may write \eqref{HALEDisc} in the form
\begin{equation}
    \Delta \tau \Delta x^2 \mathbf{c} + \mathbf{M_H'}\mathbf{H}^{k} = \mathbf{M_H'}\mathbf{H}^{k+1} + \Delta \tau \Delta x \mathbf{M_{\bar{U}}'} \mathbf{\bar{u}}^{k}, 
\end{equation}
where the coefficients of all $H^{k+1},\bar{u}^{k}$ terms are entries within $\boldsymbol{M'_{H}},\boldsymbol{M'_{\bar{U}}}$ respectively. \\
\indent In our implementation, the matrices $\boldsymbol{M'_{\bar{U}}}, \boldsymbol{M'_{H}}$ are a quintuple banded matrices. The additional derivative in $H$ does not change the size of the stencil. We include the boundary conditions for $H$ in the first two and final two lines of both matrices, so that it is not necessary to adjust the stencil near the endpoints of the domain. Hence, we can construct the linear system
\begin{align}
    \begin{pmatrix} 
 \mathbf{M_H}& \Delta \tau \Delta x \mathbf{M_{\bar{U}}} \\
\mathbf{M_H'} &\Delta \tau \Delta x  \mathbf{M_{\bar{U}}'} 
\end{pmatrix}
  \begin{pmatrix} 
 \mathbf{H}^{k+1}\\\mathbf{\bar{u}}^{k}
\end{pmatrix}
=
  \begin{pmatrix} 
   \Delta \tau \Delta x^2 \mathbf{b} + \mathbf{M_H}\mathbf{H}^{k} \\
  \Delta \tau \Delta x^2 \mathbf{c} + \mathbf{M_H'}\mathbf{H}^{k}
\end{pmatrix}.
\end{align}
This linear system is solved at each $k$, with equation $\eqref{FirstTheta}$ requring $H^{k+1}$ in order to update $\theta^{k} \rightarrow \theta^{k+1}$.

\bibliographystyle{plain}
\bibliography{biblio}

\end{document}